\documentclass[longauth]{aa} 
\usepackage[utf8]{inputenc}
\usepackage[varg]{txfonts}
\usepackage[breaklinks, colorlinks, citecolor=blue, linkcolor=blue]{hyperref}
\usepackage[usenames, dvipsnames]{color}
\usepackage{amsmath}
\usepackage{graphicx}
\usepackage{siunitx}
\usepackage{float}
\usepackage{url}
\usepackage{longtable}
\usepackage{fancyhdr}
\usepackage{natbib}
\usepackage[normalem]{ulem}
\usepackage[version=4]{mhchem}
\usepackage{tabularx} 
\makeatletter

\def\instrefs#1{{\def\scsep{\def\scsep{,}}\@for\w:=#1\do{\scsep\ref{inst:\w}}}}
\renewcommand{\inst}[1]{\unskip$^{\instrefs{#1}}$}

\makeatother
\begin{document} 

\title{A detailed analysis of the Gl~486 planetary system}

   
\titlerunning{The Gl~486 planetary system}

\authorrunning{Caballero et al.}

\author{J.\,A.~Caballero\inst{cabesac}              
\and E.~Gonz\'alez-\'Alvarez\inst{cabinta}          
\and M.~Brady\inst{unichicago}                      
\and T.~Trifonov\inst{mpia,sofia}                   
\and T.\,G.~Ellis\inst{louisiana}                   
\and C.~Dorn\inst{unizuerich}                       
\and C.~Cifuentes\inst{cabesac}                     
\and K.~Molaverdikhani\inst{lmu,origins,mpia,lsw}   
\and J.\,L.~Bean\inst{unichicago}                   
\and T.~Boyajian\inst{louisiana}                    
\and E.~Rodr\'iguez\inst{iaa}                       
\and J.~Sanz-Forcada\inst{cabesac}                  
\and M.\,R.~Zapatero~Osorio\inst{cabinta}           
\and C.~Abia\inst{ugr}                              
\and P.\,J.~Amado\inst{iaa}                         
\and N.~Anugu\inst{mountwilson}                     
\and V.\,J.\,S.~B\'ejar\inst{iac,ull}               
\and C.\,L.~Davies\inst{exeter}                     
\and S.~Dreizler\inst{iag}                          
\and F.~Dubois\inst{astrolab}                       
\and J.~Ennis\inst{annarbor}                        
\and N.~Espinoza\inst{stsci}                        
\and C.\,D.~Farrington\inst{mountwilson}            
\and A.~Garc\'ia~L\'opez\inst{cabesac}              
\and T.~Gardner\inst{annarbor}                      
\and A.\,P.~Hatzes\inst{tls}                        
\and Th.~Henning\inst{mpia}                         
\and E.~Herrero\inst{ice,ieec}                      
\and E.~Herrero-Cisneros\inst{cabinta}              
\and A.~Kaminski\inst{lsw}                          
\and D.~Kasper\inst{unichicago}                     
\and R.~Klement\inst{mountwilson}                   
\and S.~Kraus\inst{exeter}                          
\and A.~Labdon\inst{esochile}                       
\and C.~Lanthermann\inst{mountwilson}               
\and J.-B.~Le\,Bouquin\inst{grenoble}               
\and M.\,J.~L\'opez~Gonz\'alez\inst{iaa}            
\and R.~Luque\inst{iaa,unichicago}                  
\and A.\,W.~Mann\inst{carolina}                     
\and E.~Marfil\inst{cabesac}                        
\and J.\,D.~Monnier\inst{annarbor}                  
\and D.~Montes\inst{ucm}                            
\and J.\,C.~Morales\inst{ice,ieec}                  
\and E.~Pall\'e\inst{iac,ull}                       
\and S.~Pedraz\inst{caha}                           
\and A.~Quirrenbach\inst{lsw}                       
\and S.~Reffert\inst{lsw}                           
\and A.~Reiners\inst{iag}                           
\and I.~Ribas\inst{ice,ieec}                        
\and C.~Rodr\'iguez-L\'opez\inst{iaa}               
\and G.~Schaefer\inst{mountwilson}                  
\and A.~Schweitzer\inst{hs}                         
\and A.~Seifahrt\inst{unichicago}                   
\and B.\,R.~Setterholm\inst{annarbor}               
\and Y.~Shan\inst{oslo,iag}                         
\and D.~Shulyak\inst{iaa}                           
\and E.~Solano\inst{cabesac}                        
\and K.\,R.~Sreenivas\inst{ariel}                   
\and G.\,Stef\'ansson\inst{princeton,russell}       
\and J.~St\"urmer\inst{lsw}                         
\and H.\,M.~Tabernero\inst{cabinta}                 
\and L.~Tal-Or\inst{ariel}                          
\and T.~ten\,Brummelaar\inst{mountwilson}           
\and S.~Vanaverbeke\inst{astrolab,vvs,leuven}       
\and K.~von\,Braun\inst{lowell}                     
\and A.~Youngblood\inst{greenbelt}                  
\and M.~Zechmeister\inst{iag}                       
}

\institute{
\label{inst:cabesac}Centro de Astrobiolog\'ia (CSIC-INTA), ESAC, Camino bajo del castillo s/n, 28692 Villanueva de la Ca\~nada, Madrid, Spain \\ 
\email{caballero@cab.inta-csic.es}
\and
\label{inst:cabinta}Centro de Astrobiolog\'ia (CSIC-INTA), Carretera de Ajalvir km 4, 28850 Torrej\'on de Ardoz, Madrid, Spain 
\and
\label{inst:unichicago}Department of Astronomy and Astrophysics, University of Chicago, 5640 South Ellis Avenue, Chicago, IL 60637, USA 
\and
\label{inst:mpia}Max-Planck-Institut f\"ur Astronomie, K\"onigstuhl 17, 69117 Heidelberg, Germany 
\and
\label{inst:sofia}Departament of Astronomy, Sofijski universitet ,,Sv. Kliment Ohridski'', 5 James Bourchier Boulevard, 1164 Sofia, Bulgaria 
\and
\label{inst:louisiana}Louisiana State University, 202 Nicholson Hall, Baton Rouge, LA 70803, USA 
\and
\label{inst:unizuerich}Universit\"at Z\"urich, Institute for Computational Science, Winterthurerstrasse 190, CH-8057, Z\"urich, Switzerland 
\and
\label{inst:lmu}Universit\"ats-Sternwarte, Ludwig-Maximilians-Universit\"at M\"unchen, Scheinerstrasse 1, 81679 M\"unchen, Germany 
\and
\label{inst:origins}Exzellenzcluster Origins, Boltzmannstrasse 2, 85748 Garching, Germany 
\and
\label{inst:lsw}Landessternwarte, Zentrum f\"ur Astronomie der Universit\"at Heidelberg, K\"onigstuhl 12, 69117 Heidelberg, Germany 
\and
\label{inst:iaa}Instituto de Astrof\'isica de Andaluc\'ia (CSIC), Glorieta de la Astronom\'ia s/n, 18008 Granada, Spain 
\and
\label{inst:ugr}Departamento de F\'isica Te\'orica y del Cosmos, Universidad de Granada, 18071 Granada, Spain 
\and
\label{inst:mountwilson}The CHARA Array of Georgia State University, Mount Wilson Observatory, Mount Wilson, CA 91203, USA 
\and 
\label{inst:iac}Instituto de Astrof\'isica de Canarias (IAC), 38200 La Laguna, Te\-ne\-ri\-fe, Spain 
\and 
\label{inst:ull}Departamento de Astrof\'isica, Universidad de La Laguna, 38206 La Laguna, Te\-ne\-ri\-fe, Spain 
\and
\label{inst:exeter}Astrophysics Group, Department of Physics \& Astronomy, University of Exeter, Stocker Road, Exeter, EX4 4QL, UK 
\and
\label{inst:iag}Institut f\"ur Astrophysik und Geophysik, Georg-August-Universit\"at G\"ottingen, Friedrich-Hund-Platz 1, 37077 G\"ottingen, Germany 
\and
\label{inst:astrolab}AstroLAB IRIS, Provinciaal Domein ``De Palingbeek'', Verbrandemolenstraat 5, 8902 Zillebeke, Ieper, Belgium 
\and
\label{inst:annarbor}Astronomy Department, University of Michigan, Ann Arbor, MI 48109, USA 
\and
\label{inst:stsci}Space Telescope Science Institute, 3700 San Martin Drive, Baltimore, MD 21218, USA 
\and
\label{inst:tls}Th\"uringer Landessternwarte Tautenburg, Sternwarte 5, 07778 Tautenburg, Germany 
\and
\label{inst:ice}Institut de Ci\`encies de l’Espai (ICE, CSIC), Campus UAB, Can Magrans s/n, 08193 Bellaterra, Barcelona, Spain 
\and 
\label{inst:ieec}Institut d’Estudis Espacials de Catalunya (IEEC), 08034 Barcelona, Spain 
\and
\label{inst:esochile}European Southern Observatory, casilla 19001, Santiago 19, Chile 
\and
\label{inst:grenoble}Institut de Planetologie et d'Astrophysique de Grenoble, Grenoble 38058, France 
\and
\label{inst:carolina}Department of Physics and Astronomy, The University of North Carolina at Chapel Hill, Chapel Hill, NC 27599, USA 
\and
\label{inst:ucm}Departamento de F\'{i}sica de la Tierra y Astrof\'{i}sica and IPARCOS-UCM (Instituto de F\'{i}sica de Part\'{i}culas y del Cosmos de la UCM), Facultad de Ciencias F\'{i}sicas, Universidad Complutense de Madrid, 28040 Madrid, Spain 
\and
\label{inst:caha}Centro Astron\'omico Hispano en Andaluc\'ia, Observatorio de Calar Alto, Sierra de los Filabres, 04550 G\'ergal, Almer\'ia, Spain 
\and
\label{inst:hs}Hamburger Sternwarte, Universit\"at Hamburg, Gojenbergsweg 112, 21029 Hamburg, Germany 
\and
\label{inst:oslo}Centre for Earth Evolution and Dynamics, Department of Geosciences, Universitetet i Oslo, Sem S{\ae}lands vei 2b, 0315 Oslo, Norway 
\and
\label{inst:ariel}Department of Physics, Ariel University, Ariel 40700, Israel 
\and
\label{inst:princeton}Department of Astrophysical Sciences, Princeton University, 4 Ivy Lane, Princeton, NJ 08540, USA 
\and
\label{inst:russell}Henry Norris Russell Fellow 
\and
\label{inst:vvs}Vereniging Voor Sterrenkunde, Oude Bleken 12, 2400 Mol, Belgium 
\and
\label{inst:leuven}Centre for Mathematical Plasma Astrophysics, Katholieke Universiteit Leuven, Celestijnenlaan 200B, bus 2400, 3001 Leuven, Belgium 
\and
\label{inst:greenbelt}Exoplanets and Stellar Astrophysics Laboratory, NASA Goddard Space Flight Center, Greenbelt, MD 20771, USA 
\and
\label{inst:lowell}Lowell Observatory, 1400 W. Mars Hill Road, Flagstaff, AZ 86001, USA 
}

\date{Received 14 March 2022 / Accepted 07 June 2022}

 
\abstract
{The Gl~486 system consists of a very nearby, relatively bright, weakly active M3.5\,V star at just 8\,pc with a warm transiting rocky planet of about 1.3\,$R_\oplus$ and 3.0\,$M_\oplus$ that is ideal for both transmission and emission spectroscopy and for testing interior models of telluric planets.}
{To prepare for future studies, we thoroughly characterise the planetary system with new accurate and precise data collected with state-of-the-art photometers from space and spectrometers and interferometers from the ground.}
{We collected light curves of seven new transits observed with the {\em CHEOPS} space mission and new radial velocities obtained with MAROON-X at the 8.1\,m Gemini North and CARMENES at the 3.5\,m Calar Alto telescopes, together with previously published spectroscopic and photometric data from the two spectrographs and {\em TESS}. 
We also performed near-infrared interferometric observations with the CHARA Array and new photometric monitoring with a suite of smaller telescopes (AstroLAB, LCOGT, OSN, TJO).
This extraordinary and rich data set was the input for our comprehensive analysis.}
{From interferometry, we measure a limb-darkened disc angular size of the star Gl~486 at $\theta_{\rm LDD}$ = $0.390 \pm 0.018$\,mas.
Together with a corrected {\em Gaia} EDR3 parallax, we obtain a stellar radius $R_\star$ = $0.339 \pm 0.015$\,$R_\odot$.
We also measure a stellar rotation period at $P_{\rm rot}$ = 49.9 $\pm$ 5.5\,d, an upper limit to its XUV (5--920\,\AA) flux with new {\em Hubble}/STIS data, and, for the first time, a variety of element abundances (Fe, Mg, Si, V, Sr, Zr, Rb) and C/O ratio.
Besides, we impose restrictive constraints on the presence of additional components, either stellar or substellar, in the system.
With the input stellar parameters and the radial-velocity and transit data, we determine the radius and mass of the planet Gl~486\,b at $R_{\rm p}$ = $1.343 ^{+0.063} _{-0.062}$\,$R_\oplus$ and $M_{\rm p}$ = $3.00 ^{+0.13} _{-0.13}$\,$M_\oplus$, with relative uncertainties in planet radius and mass of 4.7\,\% and 4.2\,\%, respectively.
From the planet parameters and the stellar element abundances, we infer the most probable models of planet internal structure and composition, which are consistent with a relatively small metallic core with respect to the Earth, a deep silicate mantle, and a thin volatile upper layer.
With all these ingredients, we outline prospects for Gl~486\,b atmospheric studies, especially with forthcoming {\em James Webb Space Telescope} observations.}
{}


\maketitle
%

\section{Introduction}
\label{sec:intro}

Over the twenty-seven years of discoveries since the seminal work by \citet{MayorQueloz1995}, exoplanet searches have resulted in almost 5000 candidate detections. 
Statistical analyses of large samples of surveyed stars show that planets are ubiquitous, with occurrence rates greater than 0.5 planets per FGK-type star for orbital periods between one day and a few hundred days, based on estimates using radial velocity (RV) data \citep{Howard2010, Mayor2011} and transits \citep{Fressin2013, Petigura2013, KunimotoMatthews2020}. 
Occurrence rates for planets with low-mass M-dwarf hosts are even higher, with values exceeding one planet per star \citep{Cassan2012, Bonfils2013, DressingCharbonneau2015, Gaidos2016, Sabotta2021, Mulders2021}, and possibly further increasing from early to mid M-type dwarfs \citep[][but see \citealt{BradyBean2022} for the opposite]{HardegreeUllman2019}.

\begin{table*}
\centering
\small
\caption{Transiting planets with radius and mass determination at less than 10\,pc$^a$.} 
\label{tab:10pc}
\begin{tabular}{lcccc lccc l}
\hline
\hline
\noalign{\smallskip}
Star                    & $d$   & $J$   & Sp.       & $L_\star$                       & Planet    & $R_{\rm p}$           & $M_{\rm p}$           & $S_{\rm p}$           & References \\
                        & [pc]  & [mag] & type      & [$10^{-5}$\,$L_\odot$]    &           & [$R_\oplus$]  & [$M_\oplus$]  & [$S_\oplus$]  &           \\
\noalign{\smallskip}
\hline
\noalign{\smallskip}
\object{HD 219134}      & 6.53  & $\sim$3.9 & K3\,V & $28200 \pm 790$             & b         & $1.602 ^{+0.055} _{-0.055}$    & $4.74 ^{+0.19} _{-0.19}$       & $187.8 ^{+7.0} _{-7.0}$    & Mot15, Gil17 \\
\noalign{\smallskip}
    &   &   &   &   & c & $1.511 ^{+0.047} _{-0.047}$ & $4.36 ^{+0.22} _{-0.22}$ & $66.2 ^{+2.5} _{-2.5}$    & \\
\noalign{\smallskip}
\object{LTT 1445 A}$^b$ & 6.86  & 7.29  & M4.0\,V   & $794.8 \pm 8.0$              & b         & $1.305 ^{+0.066} _{-0.061}$    & $2.87 ^{+0.26} _{-0.25}$  & $5.47 ^{+0.20} _{-0.21}$  & Win19, Win22 \\
\noalign{\smallskip}
\object{Gl 486}$^c$     & 8.08  & 7.20  & M3.5\,V   & $1213 \pm 8$              & b         & $1.305 ^{+0.063} _{-0.067}$   & $2.82 ^{+0.11} _{-0.12}$  & $43.3^{+2.2}_{-2.4}$ & Tri21 \\
\noalign{\smallskip}
\object{Gl 367}         & 9.42  & 7.83  & M1.0\,V   & $3036 \pm 23$            & b         & $0.718 ^{+0.054} _{-0.054}$   & $0.546 ^{+0.078} _{-0.078}$   & $602 ^{+34} _{-34}$& Lam21 \\
\noalign{\smallskip}
\object{Gl 357}$^d$     & 9.44  & 7.34  & M2.5\,V   & $1612 \pm 13$             & b         & $1.217 ^{+0.084} _{-0.083}$   & $1.84 ^{+0.31} _{-0.31}$  & $13.2 ^{+1.5} _{-1.5}$ & Luq19 \\
\noalign{\smallskip}
\object{AU Mic}$^e$     & 9.71  & 5.44  & M0.5\,V   & $9875 \pm 86$             & b         & $4.38^{+0.18} _{-0.18}$      & $20.1 ^{+1.7} _{-1.6}$   & $8.15 ^{+0.30} _{-0.30}$ & Pla20, Cal21 \\
\noalign{\smallskip}
\object{Gl 436}         & 9.76  & 6.90  & M2.5\,V   & $2408 \pm 12$             & b         & $4.10 ^{+0.16} _{-0.16}$      & $21.36 ^{+0.20} _{-0.21}$    & $30.7 ^{+2.2} _{-2.2}$ & But04, Lan14, Tri18 \\
\noalign{\smallskip}
\object{HD 260655}  & 10.00  & 6.67  & M0.0\,V   & $3631 \pm 18$             & b         & $1.240 ^{+0.023} _{-0.023}$   & $2.14 ^{+0.34} _{-0.34}$  & $42.21 ^{+0.72} _{-0.72}$ & Luq22 \\
\noalign{\smallskip}
    &   &   &   &   & c & $1.533 ^{+0.051} _{-0.046}$ & $3.09 ^{+0.48} _{-0.48}$ & $16.10 ^{+0.28} _{-0.28}$    & \\
\noalign{\smallskip}
\hline
\end{tabular}
\tablebib{
But04: \citet{Butler2004};
Cal21: \citet{Cale2021};
Gil17: \citet{Gillon2017a};
Lam21: \cite{Lam2021};
Lan14: \citet{Lanotte2014};
Luq19: \citet{Luque2019};
Luq22: \citet{Luque2022};
Mot15: \citet{Motalebi2015};
Pla20: \citet{Plavchan2020};
Tri18: \citet{Trifonov2018};
Tri21: \citet{Trifonov2021};
Win19: \citet{Winters2019};
Win22: \citet{Winters2022}.
}
\tablefoot{
\tablefoottext{a}{Stellar bolometric luminosities, $L_\star$, and planet instellation, $S_{\rm p}$, computed by us as in \citet{Cifuentes2020} and \citet{MartinezRodriguez2019}, respectively ($L_\odot = 3.828 \cdot 10^{26}$\,W, $S_\oplus$ = 1361\,W\,m$^{-2}$).
The remaining star and planet parameters were taken from \citet[][{\em Gaia} EDR3 $d$]{GaiaBrown2021}, \citet[][2MASS $J$]{Skrutskie2006}, \citet[][and references therein, spectral type]{AlonsoFloriano2015},
and the references listed in the last column.}
\tablefoottext{b}{LTT~1445 A has a second transiting planet with precise mass determination of $1.54^{+0.20}_{-0.19}$\,$M_\oplus$ and a minimum radius of 1.15\,$R_\oplus$. 
\citet{Winters2022} could not determine the radius directly as the signal-to-noise ratio of their light curve permits both grazing and non-grazing configurations.}
\tablefoottext{c}{See this work for new planet parameters of Gl~486\,b.}
\tablefoottext{d}{Gl~357 has at least two more non-transiting planets detected via RV with approximate minimum masses of 3.40\,$M_\oplus$ and 6.1\,$M_\oplus$ \citep{Luque2019}.}
\tablefoottext{e}{AU~Mic, a member of the young $\beta$~Pictoris moving group, has a second transiting planet with precise radius determination of $3.51^{+0.16}_{-0.16}$\,$R_\oplus$ and a $5 \sigma$ upper limit on the mass of 20.3\,$M_\oplus$. 
\citet{Cale2021} could not determine the mass directly as the stellar activity amplitude is one order of magnitude greater than the planet semi-amplitude.}
}
\end{table*}

Our solar neighbourhood is the prime hunting ground for exoplanets around M dwarfs because of the relative abundance of such stars and the brightness limitations of observing them at farther distances. 
Generally, nearby planets offer the bonus of better perspectives for follow-up characterisation because of their relatively brighter hosts (i.e., higher signal-to-noise ratio, S/N) and greater star-planet angular separation (inversely proportional to the distance) for astrometric measurement and direct imaging. 
\citet{Reyle2021} determined that $61.3 \pm 5.9$\,\% of the reported stars and brown dwarfs in the 339 known systems within 10\,pc of the Sun have M spectral type \citep[see also:][]{Reid2002,Henry2006}.
This abundance is not only due to the peak of the mass function, but also to the span of the M-star spectral classification, which covers a wide range of properties (e.g. $\Delta L \approx$ 0.08--0.0004\,$L_\odot$, $\Delta M \approx$ 0.6--0.08\,$M_\odot$; \citealt{Cifuentes2020}).
From the estimated planet occurrence rates above, the immediate vicinity of the Sun should be populated by several hundred planets. 
As a result, many RV planet searches have focused on nearby M dwarfs, particularly the UVES \citep{Kuerster2003, Zechmeister2009}, HRS/HET \citep{Endl2003}, HARPS \citep{Bonfils2013, AstudilloDefru2017b}, RedDots \citep{AngladaEscude2016, Dreizler2020, Jeffers2020}, and the CARMENES survey \citep{Quirrenbach2014, Reiners2018, Zechmeister2019}. 
A total of {97} planet candidates 
in {46} stellar systems 
with distances shorter than 10\,pc have been found so far, with {74} planet candidates 
in {37} systems 
with M dwarf hosts\footnote{Data from \url{https://exoplanetarchive.ipac.caltech.edu/}, \url{https://exoplanet.eu}, and \url{https://gruze.org/10pc/}, all accessed on {3~May~2022}.
The list does not contain \object{Barnard's Star}~b (a contested cool super-Earth candidate around the second closest stellar system -- \citealt{Ribas2018, Lubin2021}), but instead contains other questioned exoplanet candidates (\citealt{Reyle2021} and references therein).}. 

The relative bonanza of nearby exoplanets diminishes greatly when considering only those that experience transits because of the relatively low geometric probability of eclipse. 
Assuming the same rates as above, one could expect a dozen transiting planets within 10\,pc.
These relatively scarce nearby transiting planets are, therefore, highly valuable and of great interest, especially for atmospheric studies, which at present mostly rely on emission and transmission spectroscopy of transiting planets \citep{VidalMadjar2003, Charbonneau2009}.

The measurement of rocky planet atmospheres has proven very challenging with today's instrumentation because of their expected small scale height and large contrast with the host star. 
A particularly favourable example is \object{55~Cnc}\,e, whose short orbital separation and luminous host lead to an equilibrium temperature, $T_{\rm eq}$, of $\sim$2400\,K. 
Such a combination has allowed for observations of the phase variation and has pointed at inefficient heat transfer, casting doubt on the existence of an atmosphere \citep{Demory2016}. 
Another example is \object{LHS~3844}\,b \citep{Vanderspek2019}, with a much lower $T_{\rm eq}$ of $\sim$800\,K. 
A phase curve was also obtained, but the results were also compatible with the planet having no atmosphere \citep{Kreidberg2019}. 
A case such as LHS~3844\,b is valuable, but the host star is relatively faint ($V \approx 15.3$\,mag), making the planet properties difficult to measure. 
For example, no dynamical mass is yet available for this planet.
Other potentially interesting nearby systems for atmosphere characterisation of rocky planets with dynamical mass determination are 
\object{Gl~357} \citep{Luque2019}, 
\object{Gl~367} \citep{Lam2021}, 
\object{Gl~1132} \citep{BertaThompson2015}, 
\object{L~98--59} \citep{Kostov2019}, 
\object{L~231--32} \citep[TOI--270,][]{Guenther2020}, 
\object{LHS~1140} \citep{Ment2019}, 
and \object{TRAPPIST--1} \citep[2MUCD~12171,][]{Gillon2017b}.
Of them, the planets most probed for the existence of atmospheres have probably been the seven in the TRAPPIST-1 system \citep{deWit2016,deWit2018,Bourrier2017a,Bourrier2017b,Zhang2018,Wakeford2019,Gressier2022}.
However, none of their atmospheres have been successfully detected because of the observational difficulties (faint primaries and low $T_{\rm eq}$).
The two transiting rocky planets that have been analysed so far, 55~Cnc\,e and LHS~3844\,b, seem to point at the absence of thick atmospheres around close-in hot rocky planets
\citep[e.g.,][]{RiddenHarper2016, Jindal2020, Deibert2021}, but the very limited statistics do not permit any general conclusions.

Transiting rocky exoplanets around nearby M dwarfs are also the key to comparative geology and geochemistry. 
Until recently, the only rocky bodies for which we could study and model their interiors were Mercury, Venus, Earth, Mars, and the largest Solar System moons and dwarf planets.
However, with the advent of very precise photometry and RV and the discovery of nearby transiting telluric planets, mostly with the {\em Transiting Exoplanet Survey Satellite} ({\em TESS}; \citealt{Ricker2015}), now we can compare the structure and composition of Solar System bodies and of exoplanets.
For example, \citet{Lam2021} inferred that Gl~367\,b, a dense, ultrashort-period sub-Earth planet transiting a nearby M dwarf, has an iron core radius fraction of $86 \pm 5$\,\%, similar to that of Mercury’s interior.
On the other hand, \citet{Demangeon2021} reported iron cores of 12\,\% and 14\,\% in total mass of \object{L~98--59}\,b and~c, for which there is no counterpart in our Solar System. 
Planets c and d of \object{$\nu^{02}$~Lup}, a very bright Sun-like star, seem to have retained small hydrogen-helium envelopes and a possibly large water fraction, but planet b probably has a rocky, mostly dry composition \citep{Delrez2021}.
Additional analyses of internal structure of rocky exoplanets are more theoretical  \citep{Schulze2021, Adibekyan2021} or oriented towards non-transiting planets, such as \object{Proxima Centauri}\,b \citep{Brugger2016, Herath2021, Noack2021, Acuna2022}.

In Table~\ref{tab:10pc} we compile the {ten} transiting planets (in {eight} systems) with precise radius and mass determination at less than 10\,pc, which are expected to be cornerstones for atmospheric studies with the {\em James Webb Space Telescope}, being commissioned at present.
Among them, there are two Neptune-mass planets, {seven} super- and exo-Earths, and one sub-Earth with a wide range of instellation (insolation) from $S \sim$ 5.5\,$S_\oplus$ to 600\,$S_\oplus$.
Not tabulated, in orbit to the seven stellar hosts there are another four planet candidates missing precise radius or mass determinations (see notes).
Table~\ref{tab:10pc} does not list L~98--59\,b and~c, also expected to be cornerstone rocky transiting planets around relatively bright early M dwarfs, but at slightly over 10\,pc.

On the one hand, HD~219134 stands out against the other stars in Table~\ref{tab:10pc} because of its closeness, apparent brightness, and possession of two well-investigated planets.
On the other hand, it also stands because of its luminosity and spectral type, as it is the only host with a spectral type other than M.
However, being a K3\,V star, the planet-to-star radius ratio is not as good for planet investigation as for the other six early- and mid-M dwarfs, which are smaller.
Besides, the large instellation on HD~219134\,b (and, to a lesser degree, on HD~219134\,c) leads to a situation similar to 55~Cnc\,e, with very hot surfaces and, probably, evaporated atmospheres.
The second closest star in Table~\ref{tab:10pc} is the M4.0\,V star LTT~1445\,A, which is the primary of a hierarchical triple stellar system with a fainter double companion at an average separation of 5\,arcsec \citep{Rossiter1937} and two rocky planets.

The third closest star with a transiting planet with precise radius and mass determination is Gl~486, which is the second brightest (in the $J$ band) M dwarf with a transiting rocky planet.
The host star is also a photometrically and RV-quiet M3.5\,V star, which helps reducing the impact of stellar activity on both RV and transit observations.
Its planet, Gl~486\,b, had at the time of discovery the greatest emission spectroscopic metric and second greatest transmission spectroscopic metric of all known transiting planets \citep{Kempton2018, Trifonov2021}.
The planet is warm ($T_{\rm eq} \sim$ 700\,K), but below the limit for a molten surface at about 880\,K \citep[][and references therein]{Mansfield2019}, and has a short orbital period of $\sim$1.47\,d that allows observing transits every three nights with a good time sampling.
Besides, because of its declination, it is observable from both hemispheres.
All these parameters make Gl~486\,b a nearby transiting rocky planet ideal for atmospheric and internal structure investigations. 
However, key exoplanet parameters, such as the scale height, which quantifies the extension of an atmosphere, or the core-to-mantle ratio, which quantifies the amounts of silicates and iron of an interior (if the planet is differentiated into core and mantle), strongly depend on the mass and radius of the exoplanet.

Here, we improve the mass and radius determination of the exoearth Gl~486\,b in terms of both accuracy (closeness of the measurements to the true value of the quantity) and precision (closeness of the measurements to each other) based on a large and varied collection of data sets and analyses.
The data sets include new {\em CHEOPS} transit observations that complement public {\em TESS} space photometry, high-resolution spectroscopy collected with MAROON-X and CARMENES, near-infrared interferometry with the CHARA Array, ultraviolet spectroscopy with the {\em Hubble Space Telescope}, and multi-site photometric follow-up from the ground with a number of small telescopes.
Using state-of-the-art techniques and tools, we measure a nearly-model-independent stellar radius, put limits on the presence of additional companions, measure a stellar rotation period shorter than previously considered, determine a suite of photospheric abundances, and determine planet mass and radius with uncertainties of 4.2\,\% and 4.7\,\%, respectively. 
From these inputs, we compute different models of Gl~486\,b internal structure and atmospheric composition useful for forthcoming observations with {\em Webb}.

\section{Star and planet}
\label{sec:star+planet}

\subsection{Gl 486}

The star Gl~486 was discovered by \citet{Wolf1919} using a proper motion survey
of low-luminosity stars with photographic plates collected with the Bruce double astrograph on K\"onigstuhl, Heidelberg.
Due to its proximity, Gl~486 is a well-studied star with more than one hundred refereed publications on topics ranging from photometry \citep{Leggett1992} through spectroscopy \citep{Wright2004} to planet searches \citep{Bonfils2013}.
Table~\ref{tab:star} summarises the stellar parameters of Gl~486.

\begin{table}
\centering
\small
\caption{Stellar parameters of Gl~486$^a$.} 
\label{tab:star}
\begin{tabular}{lcr}
\hline
\hline
\noalign{\smallskip}
Parameter & Value & Reference \\ 
\noalign{\smallskip}
\hline
\noalign{\smallskip}
\multicolumn{3}{c}{\em Basic identifiers and data}\\
\noalign{\smallskip}
Wolf                                & 437                   & Wol19 \\
Gl                                  & 486                   & Gli69 \\
Karmn                               & J12479+097            & AF15, Cab16a \\
Sp. type                            & M3.5\,V      & PMSU \\
$T$ [mag]                           & $8.8223 \pm 0.0073$ & ExoFOP-{\em TESS}$^b$ \\
\noalign{\smallskip}
\multicolumn{3}{c}{\em Astrometry and kinematics}\\
\noalign{\smallskip}
$\alpha$ (J2016.0)                  & 12:47:55.53  & {\it Gaia} EDR3 \\
$\delta$ (J2016.0)                  & +09:44:57.7  & {\it Gaia} EDR3 \\
$\mu_{\alpha}\cos\delta$ [$\mathrm{mas\,a^{-1}}$]  & $-1008.267 \pm 0.040$ & {\it Gaia} EDR3 \\
$\mu_{\delta}$ [$\mathrm{mas\,a^{-1}}$] & $-460.034 \pm 0.033$ & {\it Gaia} EDR3 \\
$\varpi$ [mas]                      & $123.722 \pm 0.033$ & {\it Gaia} EDR3, Lin21 \\
$d$ [pc]                            & $8.0827 \pm 0.0021$ & {\it Gaia} EDR3, Lin21 \\
$\gamma$ [$\mathrm{km\,s^{-1}}$]    & $+19.106 \pm 0.013$ & Sou18 \\ 
$\dot{\gamma}$ [$\mathrm{m\,s^{-1}\,a^{-1}}$] & $+0.2274 \pm 0.0011$ & This work \\
$U$ [$\mathrm{km\,s^{-1}}$]         & $-20.6015 \pm 0.0093$ & This work \\
$V$ [$\mathrm{km\,s^{-1}}$]         & $-39.8626 \pm 0.0076$ & This work \\
$W$ [$\mathrm{km\,s^{-1}}$]         & $+12.440 \pm 0.012$ & This work \\
Galactic population                 & Thin disc & This work \\
\noalign{\smallskip}
\multicolumn{3}{c}{\em Fundamental parameters}\\
\noalign{\smallskip}
$\theta_{\rm LDD}$ [mas]            & $0.390 \pm 0.018$ & This work \\
$R_\star$ [$R_{\odot}$]             & $0.339 \pm 0.015$ & This work \\
$M_\star$ [$M_{\odot}$]             & $0.333 \pm 0.019$ & This work \\
$L_\star$ [$10^{-6}\,L_\odot$]      & $12120 \pm 82$    & This work \\
$T_{\mathrm{eff}}$ [K]              & $3291 \pm 75$     & This work \\ 
$\log{g_{\rm spec}}$                & $4.82 \pm 0.12$ & Mar21 \\
{[Fe/H]}                            & $-0.15 \pm 0.13$ & Mar21$^{c}$ \\
\noalign{\smallskip}
\multicolumn{3}{c}{\em Activity and age}\\
\noalign{\smallskip}
$v \sin i_\star$ [$\mathrm{km\,s^{-1}}$] & $<2.0$ & Rei18 \\
$P_{\rm rot,phot}$ [d]              & 49.9 $\pm$ 5.5 & This work$^{d}$ \\ 
pEW(He~{\sc i} D$_3$) [\AA]         & $+0.098 \pm 0.007$ & Fuh20 \\
pEW(H$\alpha$) [\AA]                & $+0.163 \pm 0.016$ & Fuh20 \\
pEW(Ca~{\sc ii} IRT$_1$) [\AA]      & $+0.609 \pm 0.003$ & Fuh20 \\
pEW(He~{\sc i} IR) [\AA]            & $+0.046 \pm 0.013$ & Fuh20 \\
$\log R'_{\rm HK}$                  & $-5.461^{+0.067}_{-0.079}$ & This work$^{e}$ \\
$\langle B \rangle$ [G]             & $<$ 240 & Rei22 \\ 
$\log{L_{\rm X}}$ [erg\,s$^{-1}$]   & $<$ 26.62 & Ste13 \\ 
Age [Ga]                            & 1--8 & This work$^{f}$ \\
\noalign{\smallskip}
\hline
\end{tabular}
\tablebib{
AF15: \citet{AlonsoFloriano2015};
Cab16a: \citet{Caballero2016a};
ExoFOP-{\em TESS}: \url{https://exofop.ipac.caltech.edu/tess/};
Fuh20: \citet{Fuhrmeister2020};
{\em Gaia} EDR3: \citet{GaiaBrown2021};
Gli69: \citet{Gliese1969};
Lin21: \citet{Lindegren2021};
Mar21: \citet{Marfil2021};
PMSU: \citet{Reid1995};
Rei18: \citet{Reiners2018};
Rei22: \citet{Reiners2022};
Ste13: \citet{Stelzer2013};
Sou18: \citet{Soubiran2018};
Wol19: \citet{Wolf1919}.
}
\tablefoot{
\tablefoottext{a}{Thoughout the paper, we use the symbol ``a'' for {\em annus} (year in Latin), the unit of time that is exactly 365.25\,d (86\,400\,s): \url{http://exoterrae.eu/annus.html}.}
\tablefoottext{b}{See Table~\ref{tab:phot} for multiband photometry different from {\em TESS} $T$.}
\tablefoottext{c}{See Sect.~\ref{sec:abundances} for an element abundance analysis.}
\tablefoottext{d}{See Sect.~\ref{sec:Prot} for the $P_{\rm rot}$ determination from ground photometry.}
\tablefoottext{e}{From data compiled by \citet{Perdelwitz2021}.}
\tablefoottext{f}{\citet{Passegger2019} assumed a mean age of 5\,Ga.}
}
\end{table}

Spectral typing of Gl~486 has varied in the narrow interval between M3.0\,V \citep{Bidelman1985} and M4.0\,V \citep{Newton2014}, consistent with the M-dwarf spectral typing uncertainty of 0.5 subtypes \citep{AlonsoFloriano2015}.  
We used the {\em Gaia} EDR3 \citep{GaiaBrown2021} equatorial coordinates, proper motions, and the magnitude-, colour, and ecliptic latitude-corrected parallax \citep{Lindegren2021} of Gl~486, 
together with the absolute RV, $\gamma$, of \citet{Soubiran2018}, which is similar to other determinations in the literature (see Table~\ref{tab:gamma}), for determining the components of the Galactocentric space velocity, $UVW$, and assigning the star to the Galactic thin disc kinematic population as in \citet{CortesContreras2016}.
As in \citet{Kuerster2003}, we also computed the secular radial acceleration, $\dot{\gamma}$, which must be taken into account in long-term monitoring of nearby stars \citep{vandeKamp1977}.

\begin{table*}
\caption{Data sets of Gl~486 used in this work.}
\label{tab:data}
\centering
\begin{tabular}{lc cc cc l}
\hline \hline
\noalign{\smallskip}
Facility & Run, visit & \multicolumn{2}{c}{Observing dates (UT)} & Filter, {instrument,} & $N_{\rm obs (used)}^{a}$ & Reference$^{b}$ \\
 & or sector & Start & End & or channel &  &  \\ 
\noalign{\smallskip}
\hline 
\noalign{\smallskip}
\multicolumn{7}{c}{\em Space photometry}\\
\noalign{\smallskip}
{\em TESS}      & 23 & 18 March 2020 & 16 April 2020 & $T$ & 13\,167 (13\,167) & Tri21 \\
\noalign{\smallskip}
{\em CHEOPS}    & 1 &  \multicolumn{2}{c}{05 April 2021} & Open & 470 (429) & This work \\
                & 2 & \multicolumn{2}{c}{07 April 2021} &  & 436 (398) & This work \\
                & 3 & \multicolumn{2}{c}{12 April 2021} &  & 406 (370) & This work \\
                & 4 & \multicolumn{2}{c}{15 April 2021} &  & 414 (372) & This work \\
                & 5 & \multicolumn{2}{c}{18 April 2021} &  & 396 (366) & This work \\
                & 6 & \multicolumn{2}{c}{10 June 2021} &  & 341 (288) & This work \\
                & 7 & \multicolumn{2}{c}{26 June 2021} &  & 284 (232) & This work \\
\noalign{\smallskip}
\hline 
\noalign{\smallskip}
\multicolumn{7}{c}{\em High-resolution spectroscopy}\\
\noalign{\smallskip}
CARMENES        & 1 & 13 January 2016 & 10 June 2020 & VIS & 80 (76) & Tri21 \\
                & 2 & 01 May 2021 & 07 May 2021 &  & 5 (5) & This work \\
\noalign{\smallskip}
MAROON-X        & 1 & 20 May 2020 & 02 June 2020 & Blue, Red & 65 (65) & Tri21 \\
                & 2 & 16 April 2021 & 30 April 2021 &  & 8 (8) & This work \\
                & 3 & 25 May 2021 & 02 June 2021 & & 8 (8) & This work \\
\noalign{\smallskip}
\hline 
\noalign{\smallskip}
\multicolumn{7}{c}{\em Interferometry}\\
\noalign{\smallskip}
CHARA           & 1 & \multicolumn{2}{c}{24 May 2021} & MIRC-X & 126 & This work \\
                & 2 & \multicolumn{2}{c}{27 May 2021} & MIRC-X & 402 & This work \\
\noalign{\smallskip}
\hline 
\noalign{\smallskip}
\multicolumn{7}{c}{\em Space spectroscopy}\\
\noalign{\smallskip}
{\em Hubble}$^c$    & {1} & \multicolumn{2}{c}{15 March 2022} & {STIS G140L} & {1} & {This work} \\
                & {2} & \multicolumn{2}{c}{16 March 2022} & {STIS G140M} & {1} & {This work} \\
\noalign{\smallskip}
\hline 
\noalign{\smallskip}
\multicolumn{7}{c}{\em Ground photometry}\\
\noalign{\smallskip}
ASAS-SN & 1 & 14 February 2012 & 26 November 2018 & $V$ & 972 (958) & Tri21 \\ 
                & 2 & 04 December 2017 & 10 May 2020 & $g'$ & 1064 (1054) & Tri21 \\ 
\noalign{\smallskip}
AstroLAB & 1 & 19 May 2021 & 27 June 2021 & $V$ & 39 (39) & This work \\ 
\noalign{\smallskip}
LCOGT     & 1 & 22 April 2021 & 27 July 2021 & $B$ & 440 (429) & This work \\ 
\noalign{\smallskip}
OSN     & 1 & 17 May 2021 & {30 April 2022} & T90 $V$ & {1729 (1729)} & This work \\ 
\noalign{\smallskip}
SuperWASP & 1 & 5 February 2008 & 29 March 2011 & North & 182 (181) & Tri21 \\
                & 2 & 30 January 2013 & 15 July 2014 & South & 184 (178) & Tri21 \\ 
\noalign{\smallskip}
TJO             & 1 & 31 March 2021 & {24 April 2022} & LAIA $R$ & {610 (594)} & This work \\
\noalign{\smallskip}
\hline 
\end{tabular}
\tablefoot{
    \tablefoottext{a}{Number of eventually used data points in parenthesis.
    For SuperWASP North and South, we tabulate the $N_{\rm obs (used)}$ binned per night; the actual total number of SuperWASP data amounts to 51\,720.}
    \tablefoottext{b}{Data sets presented for the first time in this work or used already by \cite{Trifonov2021} [Tri21].}    
    \tablefoottext{c}{Under GO~16701, there are also {\em Hubble} STIS spectra with CCD/G430L (2900--5700\,\AA) and NUV/G230L (1700--3200\,\AA).}     
    }
\end{table*}

As described in detail in Sects.~\ref{sec:chara} and~\ref{sec:Rstar}, from the corrected {\em Gaia} parallax and the limb darkening-corrected stellar angular diameter, $\theta_{\rm LDD}$, measured by us with near-infrared interferometric observations, we derived a precise, model-independent, stellar radius, $R_\star$.
We integrated the spectral energy distribution of Gl~486 from Johnson $B$ to {\em WISE} $W4$ as in \cite{Cifuentes2020} and got the stellar bolometric luminosity, $L_\star$ ($L_{\rm bol}$).
The multiband photometry of the star is listed in Table~\ref{tab:phot} and its spectral energy distribution is shown in Fig.~\ref{fig:SED}.
With the stellar radius, bolometric luminosity, and the Stefan-Boltzmann law we set the effective temperature, $T_{\rm eff}$, which is similar to previous determinations (see Table~\ref{tab:teff}).
In particular, our $T_{\rm eff}$ agrees within $1\sigma$ with the values of \citet{Passegger2019} and \citet{Marfil2021} computed via spectral synthesis on a number of regions of the high-S/N, high-resolution, optical and near-infrared CARMENES template spectrum around atomic and molecular lines sensitive to changes in stellar parameters, but insensitive to Zeeman broadening caused by magnetic activity.
Finally, from the stellar radius and the empirical mass-radius relation of \cite{Schweitzer2019}, we determined the stellar mass, $M_\star$.
\citet{Trifonov2021} instead derived $R_\star$ from the Stefan-Boltzmann law, $L_\star$ from \citet{Cifuentes2020}, who integrated the star's spectral energy distribution in the same wavelength region but with the deprecated {\em Gaia} DR2 parallax, and $T_{\rm eff}$ from \citet{Passegger2019}.

Apart from $T_{\rm eff}$, \citet{Marfil2021} also determined the stellar surface gravity, $\log{g}$, and iron abundance, [Fe/H], which is the most frequently used proxy for metallicity in stellar astrophysics \citep[][]{Wheeler1989, Baraffe1998, Nordstroem2004, Ammons2006}. 
Additional element abundances are presented in Sect.~\ref{sec:abundances}.

Gl~486 is a very weakly active M dwarf \citep{StaufferHartmann1986,WalkowiczHawley2009,Browning2010,BoroSaikia2018,Fuhrmeister2018,Fuhrmeister2019,Schoefer2019,Lafarga2021}. 
The very low projected rotational velocity as measured by \citet{Reiners2018} agrees with previous determinations by \cite{Delfosse1998}, \cite{Jenkins2009}, \cite{Reiners2012}, or \cite{Moutou2017}, and with the long rotation period, $P_{\rm rot}$, of about {50}\,d (Sects.~\ref{sec:Prot} and~\ref{sec:planetradiusandmass}).
Following \citet{Fuhrmeister2020}, the lines of He~{\sc i}~D$_3$, H$\alpha$, Ca~{\sc ii}~IRT, and He~{\sc i}~$\lambda$10\,830\,\AA, which are robust spectroscopic activity indicators, are all in absorption (see their Table~1 for the line wavelengths).
Uncertainties in pseudo-equivalent widths (pEWs) of the lines were estimated from the standard deviation, which is 1.4826 times the median absolute deviation about the median (``MAD'') tabulated by \citet{Fuhrmeister2020} in absence of outliers.
As expected from its weak activity, the Ca~{\sc ii} H\&K indicator $\log R'_{\rm HK}$ is also very low.
For Table~\ref{tab:star}, we computed the logarithm of mean $R'_{\rm HK}$ of 8 HIRES, 2 ESPaDOnS, 2 UVES, 1 FEROS, and 1 HARPS measurements collected by \citet{Perdelwitz2021}, and propagated uncertainties from the standard deviation of the mean \citep[see also:][]{AstudilloDefru2017a, Houdebine2017, Hojjatpanah2019}.
\citet{Reiners2022} investigated Zeeman-sensitive Ti~{\sc i} and FeH lines and estimated an upper limit of the stellar average magnetic field strength at $\langle B \rangle =$ 240\,G as in \citet{Shulyak2019}.
We also tabulate an upper limit on the X-ray luminosity from the limit on observed flux of \citet{Stelzer2013} and the {\em Gaia} EDR3 distance.
Besides, in Sect.~\ref{sec:stellarcoronalemission} we evaluate the stellar coronal emission from X-ray and extreme ultraviolet (EUV) data.
Finally, because of the weak activity and potential kinematics membership in the Galactic thin disc, the age of Gl~486 is rather unconstrained.


\subsection{Gl 486\,b}

The warm terrestrial planet Gl 486\,b was discovered by \citet{Trifonov2021}.
With a set of methods and tools including Markov chain Monte Carlo, nested sampling, and Gaussian process (GP) regression, they performed a joint Keplerian 
parameter optimisation analysis of proprietary CARMENES, MAROON-X, and public {\em TESS} data.
For the planet Gl~486\,b, \citet{Trifonov2021} determined an orbital period of $P = 1.467119^{+0.000031}_{-0.000030}$\,d and an orbital inclination of $i_b = 88.4^{+1.1}_{-1.4}$\,deg. 
Together with the RV semi-amplitude of $K = 3.37^{+0.08}_{-0.08}$\,m\,s$^{-1}$, their stellar parameters of Gl~486, and the rest of the joint fit estimates, they obtained a dynamical mass of $M_b = 2.82^{+0.11}_{-0.12}$\,M$_\oplus$, a semi-major axis of $a_b = 0.01732^{+0.00027}_{-0.00027}$\,au, and a planet radius of $R_b = 1.305^{+0.063}_{-0.067}$\,R$_\oplus$. 
They concluded that the Gl~486\,b orbit is circular with a maximum possible eccentricity of $e_b < 0.05$ with a 68.3\,\% confidence level, which is expected given the short orbital period and the probable star-planet tides that circularise the orbit. 
They also performed a series of star-planet tidal simulations of the Gl~486 system and found that Gl~486\,b very quickly reached synchronous rotation.

From the planet mass and radius calculated in the joint RV and transit analysis, \citet{Trifonov2021} derived the planet bulk density and surface gravity at  $\rho_b \sim 1.3 \, \rho_\oplus$ and $g_b \sim 1.7 \, g_\oplus$ with relative errors of 17\,\% and 12\,\%, respectively. 
From the location of Gl~486\,b in a planet mass-radius diagram, its iron-to-silicate ratio matches that for an Earth-like internal composition. 
The inferred mass and radius of about 2.82\,M$_\oplus$ and 1.30\,R$_\oplus$ put Gl~486\,b at the boundary between Earth and super-Earth planets, but with a relatively high bulk density.
They also pointed towards a massive terrestrial planet rather than an ocean planet. 
Besides, with these data, the escape velocity at 1\,$R_b$ resulted into $v_e = 16.4^{+0.6}_{-0.5}$\,km\,s$^{-1}$ that, together with an energy-limited escape model and its X-ray flux upper limit, suggested a very small photo-evaporation ratio of $\dot{M} < 10^7$\,kg\,s$^{-1}$. 
From the stellar bolometric luminosity $L_\star$ and the planet semi-major axis, they inferred a planet instellation of $S_b = 40.3^{+1.5}_{-1.4}$\,$S_\oplus$ and, together with an assumed Bond albedo $A_{\rm Bond} = 0$, an equilibrium temperature of $T_{\rm eq} = 701^{+13}_{-13}$\,K. 
Planets with $T_{\rm eq}$ above 880\,K, such as 55~Cnc\,e and LHS~3844\,b, are expected to have molten surfaces and no atmospheres except for vapourised rocks (Sect.~\ref{sec:intro}). 
In contrast, Gl~486\,b is too cold to be a lava world, and its high temperature, while being below the 880\,K boundary, makes it one of the most suitable known rocky planet for emission and transmission spectroscopy and phase curve studies in search for an atmosphere.

\section{Data}
\label{sec:data}

Table~\ref{tab:data} summarises all the data sets of Gl~486 used in this work.
For each run, visit, or sector, it tabulates (start and end) observing date, filter, instrument, or channel, number of observations, $N_{\rm obs}$, and if the data set was already used by \citet{Trifonov2021}.
Table~\ref{tab:data} contains data sets of space photometry, high-resolution spectroscopy, interferometry, space spectroscopy, and ground photometry, which are detailed below.

\subsection{Space photometry}

\subsubsection{\em CHEOPS}
\label{sec:cheops}

\begin{figure}
    \centering
    \includegraphics[width=0.49\textwidth]{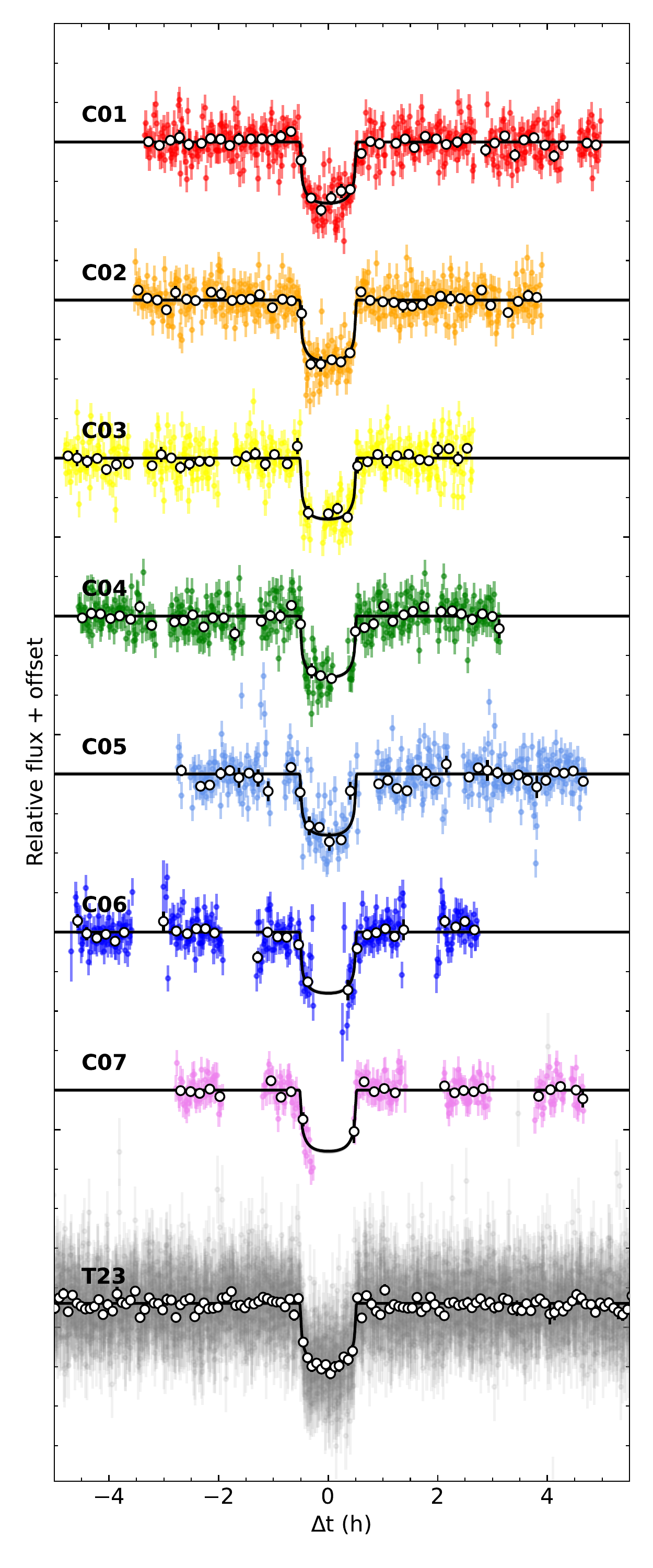}
        \caption{Post-processed light curves of the seven {\em CHEOPS} visits and around null phase of the 13 {\em TESS} transits in sector 23.
        From top to bottom: {\em CHEOPS} 1 (red), 2 (orange), 3 (yellow), 4 (green), 5 (light blue), 6 (dark blue), 7 (pink), and {\em TESS} (grey).
        Open circles denote binned data ({\em CHEOPS}: 10 points, {\em TESS}: 30 points), while the solid black lines denote the best model in the joint RV+transit fit (Sect.~\ref{sec:planetradiusandmass}).}
        \label{fig:CHEOPS+TESS}
\end{figure} 

Precise exoplanet radius measurements are among the main science goals of the ESA {\em CHEOPS} space mission. 
We refer to \citet{Futyan2020} and \citet{Benz2021} for general descriptions of the mission, and \citet{Hoyer2020},  \citet{Lendl2020}, and, especially, \citet{Maxted2022} for the data reduction pipeline and on-sky performance.

We observed Gl~486\,b on seven visits between 05 April 2021 and 26 June 2021.
Individual exposure times were set to the maximum possible value, 60\,s, and the duration of each observation averaged about 7.7\,h, with maximum and minimum durations of 8.34\,h and 7.45\,h, respectively.
We did not coadd or stack frames (imagettes).
The typical visit duration, of over seven times longer than the transit time duration of about 1.025\,h \citep{Trifonov2021}, allowed us to sample the pre- and post-transit phases and correct from systematics in the {\em CHEOPS} light curves.
Due to the increasing impact of the South Atlantic Anomaly and, especially, the longer occultations of the target by the Earth (due to the low-altitude orbit of the spacecraft) as the observing season progressed, the number of raw observations per visit decreased from 470 in the first visit to 284 in the last one.

We used the {\em CHEOPS} high-level products (level-2 output of the Data Reduction Pipeline, i.e. the light curve extracted for several aperture sizes and associated metadata) processed by the Science Operations Centre in Geneva, Switzerland, and available via the {\em CHEOPS} archive browser\footnote{\url{https://cheops.unige.ch/archive\_browser}}.
The \textit{CHEOPS} data are affected by systematics and instrumental artifacts that are associated with the spacecraft roll angle, flux ramps due to small scale changes in the shape of the point spread function, and internal reflections, among others.
Before proceeding with the joint RV+transit analysis, we corrected all our \textit{CHEOPS} light curves from these effects with the {\tt PyCheops}\footnote{\url{https://github.com/pmaxted/pycheops}} Python package. 
{\tt PyCheops} contains tools for downloading, visualising, and de-correlating \textit{CHEOPS} data, fitting transits and eclipsing exoplanets, and calculating light curve noise. 
We extensively used the {\tt diagnostic\_plot} function, which produces a series of plots of flux as a function of time, spacecraft roll angle, background noise, and $x$ and $y$ centroids, and the {\tt planet\_check} package, which allows to locate Solar System objects near the field of view of any observation.
Finally, we cleaned our light curves and freed them from instrumental artifacts and extra flux with the {\tt add\_glint} function, which removed periodic flux trends and ``spikes'' at certain spacecraft roll angles and contamination by the Moon, which introduced stray light.
We did not correct from ``glints'' from bright nearby stars. 
The post-processed light curves at the seven {\em CHEOPS} visits are shown in Fig.~\ref{fig:CHEOPS+TESS}.

\subsubsection{\em TESS}

{\em TESS} is a space telescope within NASA's Explorer program, designed to search for exoplanets using the transit method \citep{Ricker2015}. 
Since its launch in April 2018, it has unveiled a number of interesting planetary systems in the immediate vicinity of the Sun \citep[e.g.,][]{Gandolfi2018, Luque2019, Vanderspek2019, Nowak2020, Bluhm2021}, apart from shedding light on other astrophysical processes, such as stellar flares \citep{Guenther2020} or low-frequency gravity waves in blue supergiants \citep{Bowman2019}.

During sector 23 in March--April 2020, {\em TESS} monitored Gl~486, among many other stars, in 2\,min short-cadence integrations for 24.7\,d in a row, with a $\sim$5\,d gap in the middle.
The {\em TESS} Gl~486 data set here is the same one as in \citet{Trifonov2021}, who used the pre-search data conditioning simple aperture photometry (PDCSAP) light curve. 
We refer to \citet{Trifonov2021} for more details. 
The 13 Gl~486\,b transits in {\em TESS} sector 23 are overlaid on top of each other at the bottom of Fig.~\ref{fig:CHEOPS+TESS}.
The larger collecting area of {\em CHEOPS} with respect to {\em TESS} (32\,cm vs. 10\,cm) compensates the shorter time baseline and, therefore, reduced number of data points.

\subsection{High-resolution spectroscopy}

\begin{figure*}
    \centering
    \includegraphics[width=0.99\textwidth]{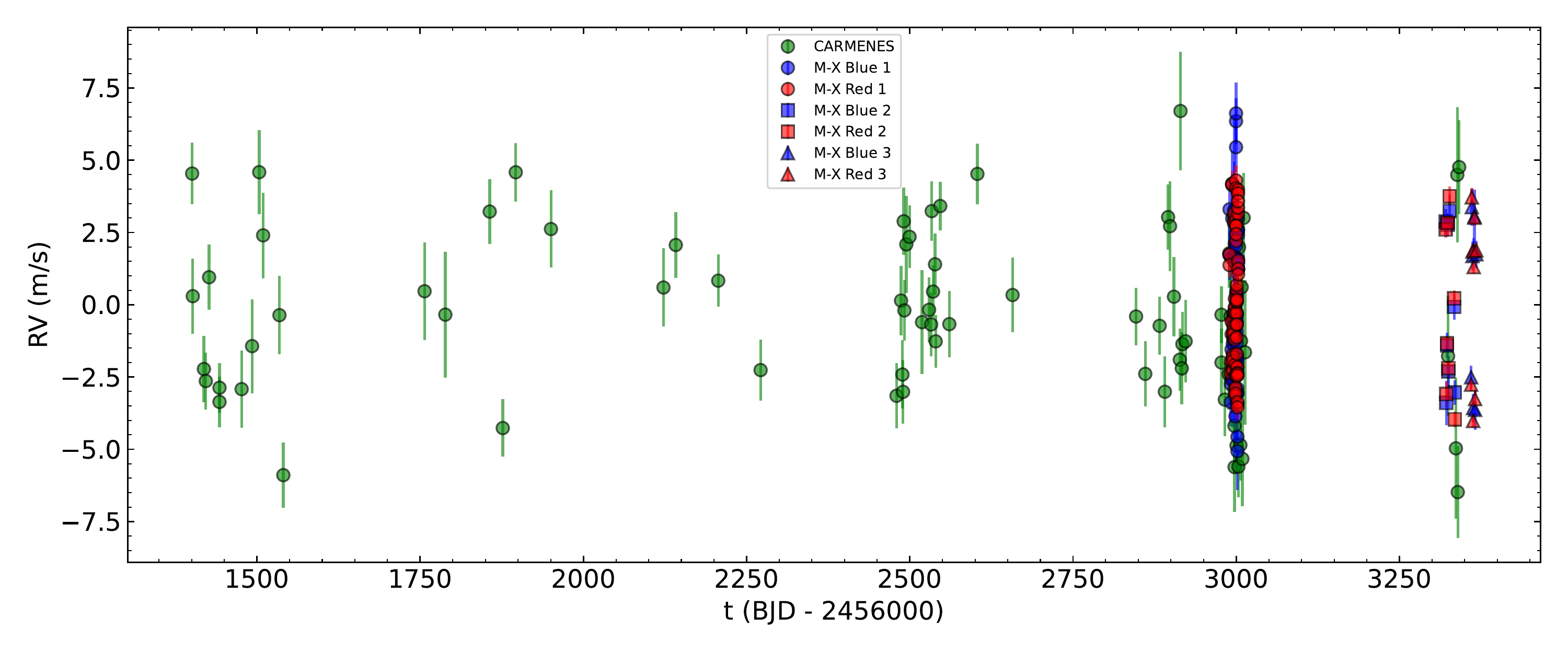}
        \caption{RV data from CARMENES (green circles), MAROON-X Red (red symbols), and MAROON-X Blue (blue symbols).
        MAROON-X data are splitted into runs 1 (circles), 2 (squares), and 3 (triangles).
        Compare with Fig.~S.1 in \citet{Trifonov2021}.}
        \label{fig:CARMENES+MAROON-X_gap}
\end{figure*} 

\subsubsection{MAROON-X}

MAROON-X\footnote{\url{https://www.gemini.edu/instrumentation/current-instruments/maroon-x}} is a red-optical (Blue arm: 5000--6700\,\AA, Red arm: 6500--9200\,\AA), high-resolution ($\mathcal{R} \approx$ 85,000) spectrograph on the 8.1\,m Gemini North telescope designed for high-precision RVs of M dwarfs \citep{Seifahrt2016, Seifahrt2018, Seifahrt2020}. 
In spite of having started its regular operations only in May 2020, MAROON-X has already contributed to a few publications on exoplanets \citep{Trifonov2021, Kasper2021, Winters2022, Reefe2022}.

We observed Gl~486 a total of 81 times during three runs in May-June 2020 (13\,d, run 1), April 2021 (14\,d, run 2), and May-June 2021 (8\,d, run 3).
The bulk of the observations were collected in run 1, which was used by \citet{Trifonov2021}.
Exposure times ranged from 300\,s to 600\,s depending on seeing conditions and cloud coverage, and the spectra were taken with simultaneous Fabry-P\'erot etalon wavelength calibration using a dedicated fibre.
The raw data were reduced using a custom Python 3 pipeline based on tools previously developed for the CRyogenic high-resolution InfraRed Echelle Spectrograph \citep[CRIRES;][]{Kaeufl2004, Bean2010}, while the RV and several spectral indices were computed with the SpEctrum Radial Velocity AnaLyser \citep[{\tt serval};][]{Zechmeister2018}.
In particular, we computed RV, dLW, CRX, H$\alpha$, and the three Ca~{\sc ii}~IRT indices in the Red channel, and 
RV, dLW, CRX, and the two Na~{\sc i}~D indices in the Blue channel (dLW and CRX stand for `differential line width' and `chromatic RV index', respectively; \citealt{Zechmeister2018}). 

There was an improvement in the S/N of the Blue channel spectra between the 2020 run 1 and the 2021 runs 2 and 3 due to an increase of the brightness of the Blue channel etalon in early 2021.
However, due to a systemic cooling pump failure on Gemini North in early May 2021, we found that there was also a large instrumental profile shift between our runs 2 and 3.
For the sake of caution, we built {\tt serval} spectral templates for the three runs separately, instead of reducing all data together.
While the template in each individual run is composed of fewer individual observations, especially in runs 2 and 3, there do not seem to be any dramatic RV shifts and there does not appear to be a meaningful increase in RV error.

\subsubsection{CARMENES}

CARMENES\footnote{\url{http://carmenes.caha.es}} (Calar Alto high-Resolution search for M dwarfs with Exo-earths with Near-infrared and optical Echelle Spectrographs) is a double-channel, high-resolution spectrograph at the 3.5\,m Calar Alto telescope that covers from 5200\,{\AA} to 17\,100\,{\AA} in one shot.
There is a beam splitter at 9600\,{\AA} that divides the stellar light between the optical (VIS, $\mathcal{R} \approx 94\,600$) and near-infrared (NIR, $\mathcal{R} \approx 80\,400$) channels.
Detailed descriptions of the CARMENES instrument at the 3.5\,m Calar Alto telescope and the exoplanet survey can be found in \citet{Quirrenbach2010, Quirrenbach2014} and \citet{Reiners2018}.

Gl~486 was one of the over 300 M-dwarf targets regularly monitored in the CARMENES guaranteed time observation program.
An updated list of past guaranteed time observation and new legacy project targets is included in \citet{Marfil2021}.
For Gl~486 we initially obtained 80 pairs of VIS and NIR spectra between January 2016 and June 2020 with a time baseline of about 4.5\,a.
This was the original data set that \citet{Trifonov2021} used in their analysis.
To these data, we added five additional visits in early May 2021 for anchoring CARMENES and new MAROON-X RVs.
The typical exposure time in all cases was about 20\,min, with the goal of achieving a signal-to-noise ratio of 150 in the $J$ band. 
A series of short-exposure spectra collected on 02 April 2021 within CARMENES legacy time for another science case was discarded from the analysis because of its low S/N.

All spectra were processed according to the standard CARMENES data flow \citep{Caballero2016b}. 
We used the latest version of the {\tt serval} data reduction pipeline (v2.10), re-computed the small nightly zero-point systematics of the CARMENES data, and corrected for them to achieve a metre-per-second precision \citep[e.g.,][]{TalOr2018, Trifonov2018, Trifonov2020}.
Because of the wider wavelength coverage, we were able to measure more indices with CARMENES than with MAROON-X.
New indices apart from dLW, CRX, H$\alpha$, Ca~{\sc ii}~IRT, and Na~{\sc i}~D were the atomic lines of He~{\sc i}~D$_3$, He~{\sc i}~$\lambda$10830\,\AA, and Pa$\beta$ and the molecular bands of TiO\,7050, VO\,7436, VO\,7942, TiO\,8430, and TiO\,8860 \citep{Schoefer2019}.
Besides, we also measured cross-correlation function (CCF) indicators as defined by \citet{Lafarga2020}: 
full-width-at-half-maximum (CCF FWHM), contrast (CCF CON), and bisector
inverse slope (CCF BIS).
Running {\tt serval} again and, therefore, creating a new template spectrum implies computing new RV velocities and indices.
Although very similar to those tabulated by \citet{Trifonov2021}, the CARMENES run-1 RVs and indices used here are not identical to what was already published.
The MAROON-X and CARMENES RVs are displayed in Fig.~\ref{fig:CARMENES+MAROON-X_gap}.

\subsection{Interferometry}
\label{sec:chara}

To extract the planetary parameters mass and radius from the combined RV and transit data we require the knowledge of the host star's radius, $R_\star$, and mass, $M_\star$. 
These are typically obtained from empirical relations or by using theoretical models to fit other observations of the host star \citep{Mann2015, Boyajian2012, Schweitzer2019}. 
In the case of Gl~486, \citet{Trifonov2021} obtained a precision of $\sim$3\,\% in stellar radius and $\sim$5\,\% in mass, not accounting for systematic errors. 
Compared to our latest {\em CHEOPS} and MAROON-X data, this precision turns out to be the limiting factor in measuring the planetary parameters. 
As described in Sect.~\ref{sec:star+planet}, \citet{Trifonov2021} determined $R_\star$ from $L_\star$ \citep{Cifuentes2020} and $T_{\rm eff}$ \citep{Passegger2019}, and $M_\star$ with the empirical mass-radius relation of \citet{Schweitzer2019}.
In the present work, however, we directly measured the angular size of Gl~486, from which we determined an $R_\star$ nearly independent of models or spectral-synthesis-based $T_{\rm eff}$.

\begin{table}
\centering
\small
\caption{Interferometric calibrator stars observed with CHARA.} 
\label{tab:chara_calibrators}
\begin{tabular}{ll ccc}
\hline
\hline
\noalign{\smallskip}
{\tt Cal\#} & Star & Sp. type & $H$ [mag] & $\theta_{H}^a$ [mas] \\ 
\noalign{\smallskip}
\hline
\noalign{\smallskip}
{\tt Cal1} & HD 120541 & A2\,V & $6.247 \pm 0.020$ & $0.1943\pm0.0053$ \\ %
{\tt Cal2} & HD 109860 & A0\,V & $6.301 \pm 0.031$ & $0.1771\pm0.0051$ \\ %
{\tt Cal3} & HD 111133 & A0\,V & $6.338 \pm 0.047$ & $0.1714\pm0.0053$ \\ 
{\tt Cal4} & HD 118245 & F2\,V & $6.496 \pm 0.017$ & $0.2037\pm0.0049$ \\ %
\noalign{\smallskip}
\hline
\end{tabular}
\tablefoot{
\tablefoottext{a}{Uniform disc diameter in band $H$ ({\tt UDDH}).}
}
\end{table}

We used the CHARA Array, a long-baseline optical-infrared interferometer located at Mount Wilson Observatory \citep{tenBrummelaar2005}. 
Observations of Gl~486 were taken on two nights (24 and 27 May 2021) with the MIRC-X beam combiner \citep{Anugu2020} in $H$ band using a five-telescope configuration (S1-S2-E1-E2-W1). 
In interferometry, frequent observations of calibrator stars are needed  
to measure visibility losses due to non-perfect atmospheric coherence and instrumental effects such as vibration, dispersion, and birefringence. 
Hence, we used an observing sequence alternating between our target and calibrator stars.
We selected calibrator stars from the second version of the Jean-Marie Mariotti Center Stellar Diameter Catalog\footnote{\url{http://www.jmmc.fr/jmdc}, VizieR {\tt II/346/jsdc\_v2}} \citep{Bourges2014, Duvert2016, Chelli2016}. 
The observed calibrator stars were chosen to be bright point sources within 15\,deg of the science star on the sky, and are presented together with their basic properties in Table~\ref{tab:chara_calibrators}.  
The data acquisition consisted of taking short 5\,min data sets plus 5\,min ``shutters'' of the science target ({\tt Obj}) and several calibrator stars ({\tt Cal\#}).
In the first run on 24 May 2021, we obtained two sets on the science target in an {\tt Obj - Cal1 - Obj - Cal1} sequence, while in the second run on 27 May 2021, we obtained five sets on the science target in a {\tt Cal2 - Obj - Obj - Cal3 - Obj - Cal4 - Obj - Cal1 - Obj - Cal3} sequence.  
Data were reduced and calibrated using the version 1.3.5 of the MIRC-X pipeline\footnote{\url{https://gitlab.chara.gsu.edu/lebouquj/mircx_pipeline}} 
to produce squared visibilities, $V^2$, and closure phases of the science target. 
During the reduction we used five coherent coadds, 150\,s maximum integration time (each 5\,min set was divided into two OIFITS\footnote{\url{https://www.chara.gsu.edu/analysis-software/oifits-data-format}} files), S/N threshold of 3, flux threshold of 5, and applied the bispectrum bias correction.

\subsection{Space spectroscopy}
\label{sec:spacespectroscopy}

\begin{table}
\centering
\tiny
\caption{{\em Hubble}/STIS line fluxes of Gl~486.} 
\label{tab:uvfluxes}
\begin{tabular}{l ccccc}
\hline
\hline
\noalign{\smallskip}
Ion & $\lambda_{\rm model}$ & $\log T_{\rm max}^a$ & $F_{\rm obs}$ & S/N & $\log{F_{\rm obs}/F_{\rm pred}}$ \\ 
    & (\AA) & ~ & (10$^{-17}$\,erg\,cm$^{-2}$\,s$^{-1}$) & ~ & ~ \\
\noalign{\smallskip}
\hline
\noalign{\smallskip}
\ion{Si}{iii} & 1206.5019 & 4.9 & 94.9 & 3.5 & --0.16 \\
\ion{N}{v} & 1238.8218 & 5.4 & 17.6 & 2.2 & +0.09 \\
\ion{N}{v} & 1242.8042 & 5.4 & 6.09 & 2.9 & --0.07 \\
\ion{C}{ii}$^b$ & 1335.7100 & 4.7 & 43.5 & 2.3 & +0.01 \\
\ion{Si}{iv} & 1393.7552 & 5.0 & 14,2 & 1.6 & --0.03 \\
\ion{Si}{iv} & 1402.7704 & 5.0 & 7.79 & 1.3 & +0.01 \\
\ion{Si}{ii} & 1526.7090 & 4.5 & 5.03 & 1.2 & +0.49 \\
\ion{C}{iv} & 1548.1871 & 5.1 & 89.0 & 2.8 & +0.06 \\
\ion{C}{iv} & 1550.7723 & 5.1 & 36.2 & 1.8 & --0.03 \\
\ion{Al}{ii} & 1670.7870 & 4.6 & 32.9 & 1.1 & +0.04 \\
\noalign{\smallskip}
\hline
\end{tabular}
\tablefoot{
\tablefoottext{a}{$T_{\rm max}$ (K) is the maximum temperature of formation of the line, unweighted by the emission measure distribution.}
\tablefoottext{b}{Blend with \ion{C}{ii} $\lambda$1335.6650\,{\AA} amounting more than 5\,\% of the line flux.}
}
\end{table}

For improving the coronal model and better constraining the transition region of Gl~486, we used {\em Hubble} low-resolution spectroscopic observations in the ultraviolet.
Since there are no public X-ray observations available after the {\em ROSAT} observations described by \citet{Stelzer2013}, we analysed instead two spectra collected on 15 March 2022 (P.I. Youngblood) with the {\em Hubble} Space Telescope Imaging Spectrograph (STIS), the FUV-MAMA detector, and the G140M (1140--1740\,\AA) and G140L (1150--1730\,\AA) filters.
The spectra were recently made available through the Milkuski Archive for Space Telescopes\footnote{\url{https://archive.stsci.edu/}} (MAST).
On those flux-calibrated spectra, we measured individual atomic lines useful for our purpose as \citet{SanzForcada2003}. 
The line fluxes, $F_{\rm obs}$, together with their S/Ns are displayed in Table~\ref{tab:uvfluxes}. 
The remaining tabulated parameters are discussed in Sect.~\ref{sec:stellarcoronalemission}.
The identified species are C~{\sc ii} and~{\sc iv}, N~{\sc v}, Al~{\sc iii}, and Si~{\sc ii},~{\sc iii}, and~{\sc iv}.

\subsection{Ground photometry}
\label{sec:groundphotometry}

For the seeing-limited optical photometric monitoring of Gl~486 we collected data with
ten different units of Las Cumbres Observatory Global Telescope\footnote{We use the acronym LCOGT instead of LCO (Las Cumbres Observatory).} \citep[LCOGT;][]{Brown2013}, 
the 0.9\,m T90 telescope of the Observatorio de Sierra Nevada \citep[OSN;][]{Amado2021} in Granada, Spain, 
the 0.8\,m Telescopi Joan Or\'o \citep[TJO;][]{Colome2010} at the Observatori Astron\`omic del Montsec in Lleida, Spain, 
and the 10\,cm Adonis refractor telescope of the Volkssterrenwacht AstroLAB IRIS\footnote{\url{https://astrolab.be/}} public observatory in Langemark, Belgium.
Besides, we also added public data of the the All-Sky Automated Survey for Supernovae \citep[ASAS-SN;][]{Shappee2014} and the Wide Angle Search for Planets \citep[SuperWASP;][]{Pollacco2006}. 
For a summary of the ground photometry data, see the bottom part of Table~\ref{tab:data}.

\begin{figure}[]
 	\centering
 	\includegraphics[width=0.45\textwidth]{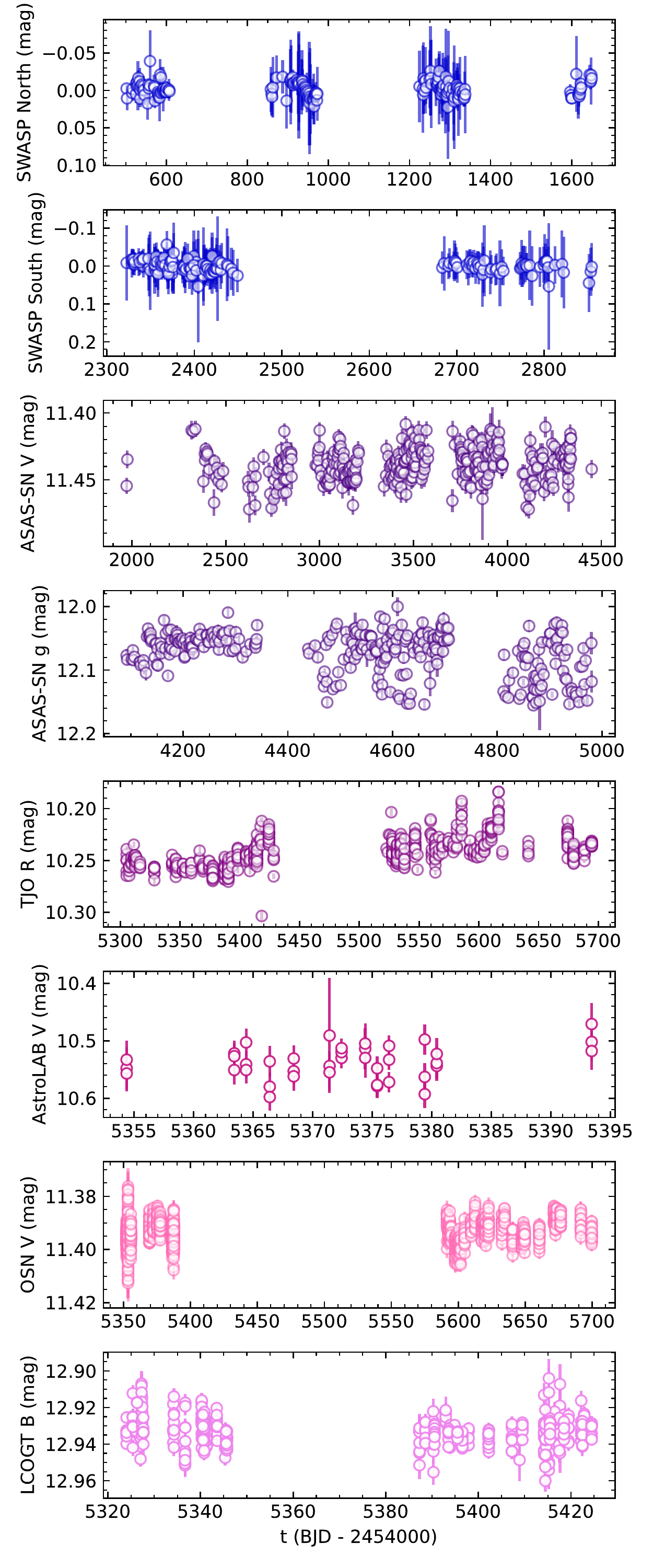}
 	\caption{Used photometric variability sets (Section~\ref{sec:groundphotometry}). 
    From top to bottom:
    SuperWASP (North and South), 
    ASAS-SN ($V$ and $g'$), 
    TJO ($R$),
    AstroLAB ($V$), 
    OSN ($V$), 
    and LCOGT ($B$). 
    For homogeneity, we transformed TJO, OSN, and LCOGT (normalised) fluxes to differential magnitudes.}
     \label{fig:Phot-all}
 \end{figure}

We did not use other Gl~486 photometry previously compiled by \citet{Trifonov2021}.
The long-term monitoring data of All-Sky Automated Survey \citep[ASAS;][]{Pojmanski1997} and Northern Sky Variability Survey \citep[NSVS;][]{Wozniak2004} have poor sampling and large scatter.
Because of their short duration, we did not use either observations during and immediately before and after planet transits with the Multicolor Simultaneous Camera for studying Atmospheres of Transiting exoplanets-2 \citep[MuSCAT2;][]{Narita2019} in May-June 2020, 
the Perth Exoplanet Survey Telescope (PEST\footnote{\url{http://pestobservatory.com/}}) in June 2020, 
LCOGT in May-June 2020, 
and TJO in April-May 2020.

Mostly because of its relative brightness, there are no useful data of Gl~486 in a number of long-time baseline, automated, wide surveys such as the Automated Patrol Telescope (APT\footnote{\url{https://rsaa.anu.edu.au/observatories/telescopes/unsw-automated-patrol-telescope-apt}}; C.\,G.~Tinney, priv. comm.),
Hungarian-made Automated Telescope Network (HATNet\footnote{\url{https://hatnet.org/}}; J.~Hartman, priv. comm.),
MEarth\footnote{\url{https://lweb.cfa.harvard.edu/MEarth/Welcome.html}} (J.~Irwin, priv. comm.),
Tennessee State University Automated Astronomy Group (TSU\footnote{\url{http://schwab.tsuniv.edu/}}; G.\,W.~Henry, priv. comm.), and Zwicky Transient Facility \citep{Bellm2019}.

The eight light curves used by us are displayed in Fig.~\ref{fig:Phot-all}.
The ASAS-SN and SuperWASP North and South data sets were used and described already by \citet{Trifonov2021}.
These authors also used TJO to cover the $\pm 3 \sigma$ phase window around the conjunction time predicted by the RV solution at the time of observations, but they performed an intensive monitoring only during four nights.
Here we present a completely new TJO data set that extends for about 11 months, ideal for a long-period determination.
We describe below the observations and preliminary data analysis with AstroLAB, LCOGT, and OSN, which were not used by \citet{Trifonov2021}, as well as with TJO.

\paragraph{AstroLAB.}
The Adonis telescope, a commercial 10\,cm Explore Scientific ED102 $f/7$ APO refractor, together with a G2-1600 Moravian CCD camera, provides a field of view of $66 \times 44$\,arcmin$^2$ and pixel size of 1.34\,arcsec, which matches the median natural seeing at Langemark, a village of Ieper (Ypres), at a height above sea level of only 15\,m.
We used the configuration above and an Astrodon Photometrics $V$ filter for monitoring Gl~486 for over six weeks between May and June 2021.
Because of the low declination of our target (+09:45) 
and the high latitude of AstroLAB (+50:51), 
we always observed it near culmination and at large airmass (1.6--2.4). 
The 39 collected images were processed with the {\tt LesvePhotometry}\footnote{\url{http://www.dppobservatory.net/}} reduction package. 
For the extraction of the light curve, we used differential photometry relative to one non-variable comparison star of similar brightness and colour within the field of view.

\paragraph{LCOGT.}
This network of astronomical observatories has been used for investigating a number of variable astrophysical processes, from supernovae \citep{Valenti2016}, through eclipsing binaries \citep{Steinfadt2010} and debris discs around white dwarfs \citep{Vanderbosch2020}, to transiting exoplanets \citep{Newton2019}. 
We refer to \citet{Brown2013} for the technical description of the network telescopes and basic data analysis, and the LCOGT website\footnote{\url{https://lco.global/}} for the latest updates.
Our $B$-band photometric observations of Gl~486 with ten 1\,m LCOGT robotic telescopes lasted from 22 April 2021 to 27 July 2021 and resulted in 521 images.
The data were first divided into ten subsets, one per telescope.
Upon visual inspection of each subset, we kept 440 images with S/N $>$ 8 and not affected by cosmic rays.
Aperture photometry on our target and eight reference stars was performed for each data set separately with {\tt AstroImageJ} \citep{Collins2017}.
The median of each data set was then subtracted to create the combined light curve, which has an rms of about 0.010\,mag and an approximate Nyquist frequency of 1.5\,d$^{-1}$.

\paragraph{OSN.}
We also monitored Gl~486 with the T90 telescope at the Observatorio de Sierra Nevada (2896\,m).
The Ritchey-Chr\'etien telescope is equipped with a 2k\,$\times$\,2k pixel VersArray CCD camera with a field of view of 13.2\,$\times$\,13.2 arcmin$^2$ and a pixel size of 0.387\,arcsec.
The $V$-band observations were carried out on nine nights in late Spring 2021, with typically 130 exposures per night, and 30 nights during early 2022, with typically 20 exposures per night, each with an integration time of 40\,s. 
We obtained synthetic aperture photometry from the unbinned frames, which were bias subtracted and properly flat-fielded with {\tt IRAF} beforehand, and selected the best aperture sizes and five reference stars for the differential photometry. 
In particular, we used the same T90 instrumental configuration and methodology as in previous works involving photometric monitoring of nearby M dwarfs with exoplanets \citep[e.g.,][]{Perger2019, Stock2020, Amado2021}.

\paragraph{TJO.}
For the March 2021-April 2022 run of TJO, we collected at least five exposures per observing night with the Large Area Imager for Astronomy (LAIA) imager and the Johnson $R$ filter.
LAIA is a 4\,k\,$\times$\,4\,k CCD camera with a field of view of 30\,arcmin and a scale of 0.4\,arcsec\,pixel$^{-1}$. 
The images were calibrated with dark, bias, and flat fields frames using the observatory pipeline. 
Differential photometry was extracted with {\tt AstroImageJ} using the aperture size and a set of 10 comparison stars selected to minimise the rms of the photometry.
We refer to \citet{GonzalezAlvarez2022} for a recent example of TJO being used for studying the rotation period of an M dwarf with a transiting planet and RV follow-up.

\section{Analysis and results}
\label{sec:AandR}

\subsection{Stellar radius}
\label{sec:Rstar}

\begin{figure*}
    \centering
    \includegraphics[width=0.49\textwidth]{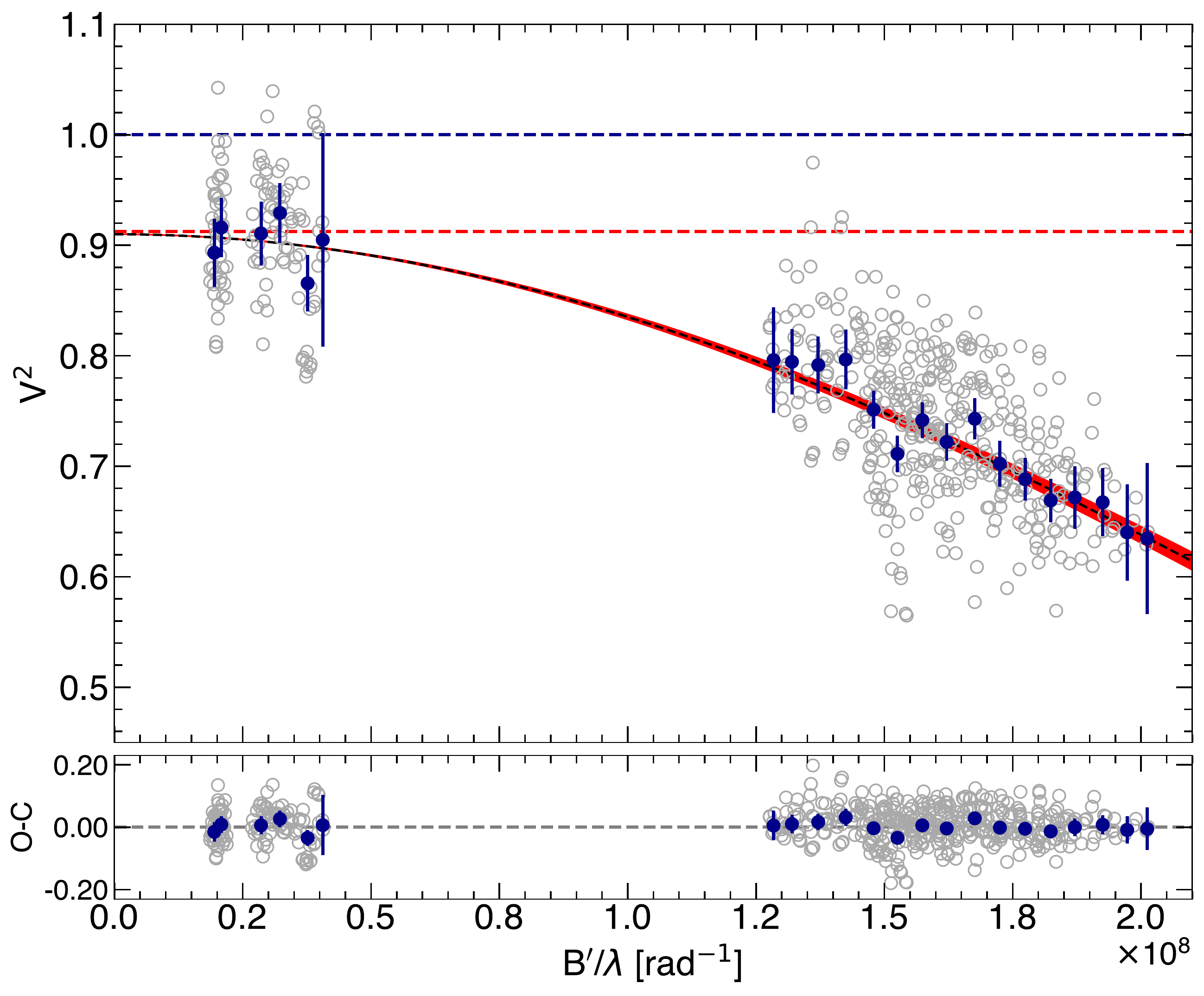}
    \includegraphics[width=0.49\textwidth]{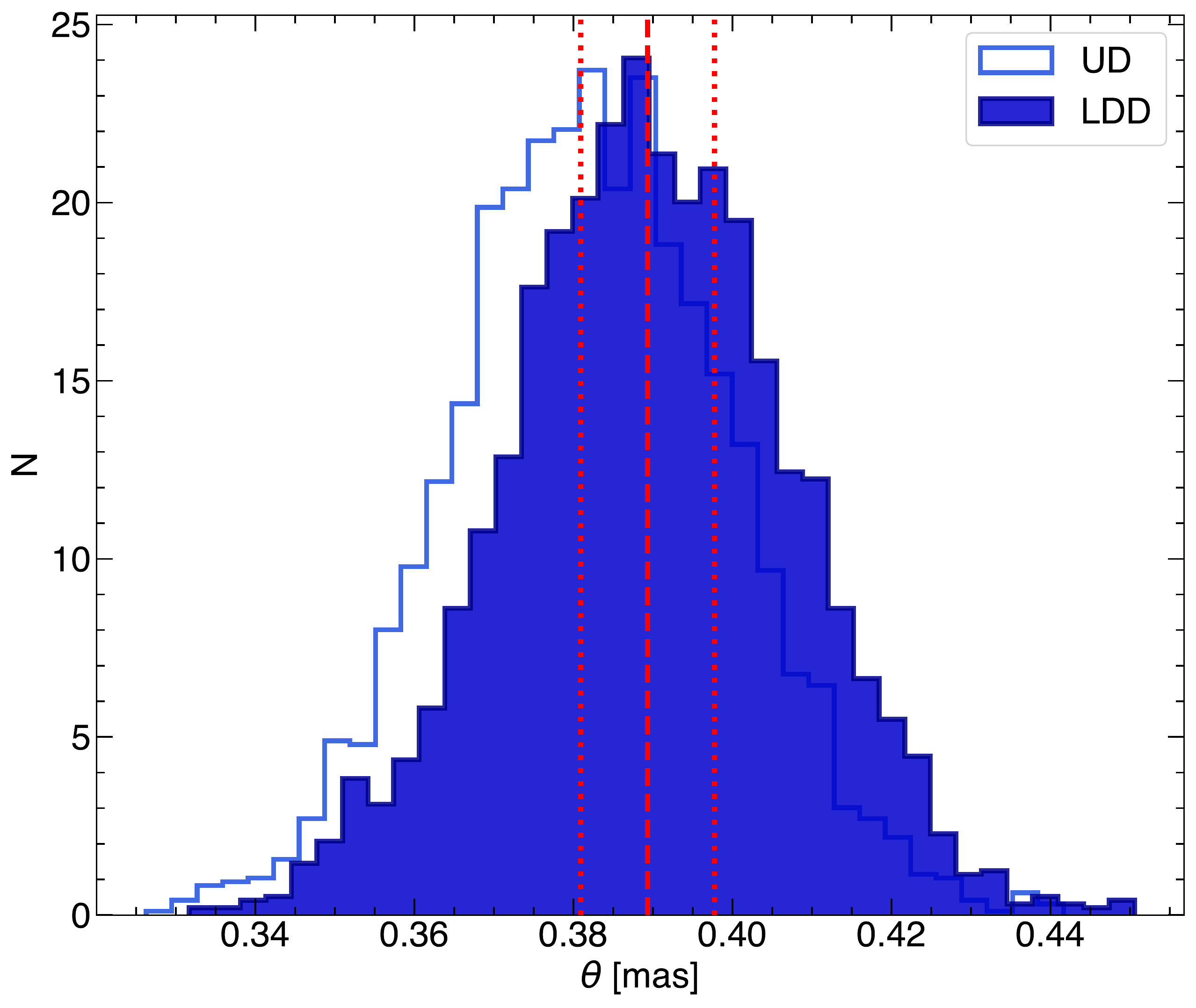}
        \caption{{\em Left:} CHARA MIRC-X squared visibility as a function of spatial frequency ($B'/\lambda$, baseline over wavelength).
        Open grey circles indicate actual measurements, dark blue filled circles with error bars are binned data with standard deviation, and dashed curve and red shadow are our angular diameter fit and its uncertainty.
        The horizontal dashed lines indicate the visibility at unity (blue) and at unity minus the correction $1 - V_{0,{\rm LDD}}^2 = 0.0891 \pm 0.0051$ (red).
        {\em Right:} Histograms of uniform (open, light blue) and limb-darkened disc (filled, dark blue) angular diameters.
        Red vertical lines indicate 15.8\,\% (dotted), 50.0\,\% (median, dashed), and 84.5\,\% (dotted) confidence intervals for the limb-darkened disc angular diameter.}
        \label{fig:CHARAvisibilitybinned}
\end{figure*} 

\citet{HanburyBrown1974} derived the relationship between the distribution of light on the sky, the uniform and limb-darkened disc (UD, LDD), and the squared visibilities, $V^2$, in interferometric observations for measuring apparent angular diameters of stars. 
The corresponding visibilities for a disc depends on the projected baseline $B^\prime$, the linear limb-darkening parameter, $\mu_\lambda$, the angular size of the object at a certain wavelength, $\theta_\lambda$, and the wavelength of observations, $\lambda$, as shown in Eqs.~\ref{eq:vis_squared}--\ref{eq:xlambda}: 

\begin{equation}
  \label{eq:vis_squared}
  V_\lambda^2 = V_{\lambda,0}^2 ~ \frac{ \mathcal{F}^2(\mu_\lambda, x_\lambda) }{ \mathcal{G}^2(\mu) },
\end{equation}

\begin{equation}
  \label{eq:Fmux}
    \mathcal{F}(\mu_\lambda, x_\lambda) = (1-\mu_\lambda) \frac{J_1(x_\lambda)}{x_\lambda}+\mu_\lambda \sqrt{\frac{\pi}{2}} \frac{J_{3/2}(x_\lambda)}{x_\lambda^{3/2}},
\end{equation}

\begin{equation}
  \label{eq:Gmu}
  \mathcal{G}(\mu_\lambda) = \frac{1-\mu_\lambda}{2}+\frac{\mu_\lambda}{3}, 
\end{equation}

\begin{equation}
  \label{eq:xlambda}
  x_\lambda = \frac{ \pi B^\prime \theta_\lambda }{ \lambda}, 
\end{equation}

\noindent where $J_\alpha(x)$ are the Bessel functions of the first kind ($\alpha$ = 1, 3/2).  
As explained below, the first term in Eq.~\ref{eq:vis_squared}, $V_0^2$, corrects for unknown systematic offsets \citep{diFolco2007, Woodruff2008}.

In order to determine the stellar angular diameter, we began by creating a large number ($N$ = 3000) of boot-strapped realisations of the calibrated $V^2$ data sets. 
The uniform disc model was then fit with the realisations of the data using the {\tt Scipy} non-linear least-squares minimisation routine \citep{Jones2001}. 
As part of the fitting process, we also allowed variations in the dependent parameter by sampling the uncertainty in the instrument's wavelength solution. 
We fit each night's data separately and then combined. 
We added the standard deviation of the night to night results in quadrature with the uncertainty from fitting the entire data set to better capture the true uncertainty in the fit. 
Since the distribution fits of the determined uniform disc, $\theta_{\rm UD}$, and its corresponding floating offset, $V_{0,{\rm UD}}^2$, are Gaussian, we tabulate the mean and standard deviations instead of median and 15.8\,\% and 84.1\,\% confidence intervals in Table~\ref{tab:chara_results}.

We then repeated this process for the limb-darkened disc model using the same technique. 
We estimated the $H$-band limb-darkening parameter with the Limb Darkening Toolkit, {\tt LDTK} \citep{ParviainenAigrain2015}, which uses the library of PHOENIX-generated specific intensity spectra by \citet[][hence, the stellar radius determination is not fully model-independent]{Husser2013}.
We provided the {\tt LDTK} module with first estimates of $T_{\rm{eff}}^{\tt LDTK}$, $\log{g}$, and metallicity, $Z$. 
For $\log{g}$ and $Z$ we used the values in Table~\ref{tab:star} and the approximate relation $Z = Z_\odot ~ 10^\text{[Fe/H]}$, with $Z_\odot \approx$ 0.013, while for $T_{\rm eff}$ we used the measured $\theta_{\rm{UD}}$ in combination with the stellar bolometric luminosity and the Stefan-Boltzmann law ($L_\star \propto \theta^2 T_{\rm eff}^4$).
We then iterated fitting the limb-darkened diameter until the final $T_{\rm{eff}}^{\tt LDTK}$ remained unchanged. 
This $T_{\rm{eff}}^{\tt LDTK}$, as listed in Table~\ref{tab:chara_results}, is not identical to the $T_{\rm{eff}}$ in Table~\ref{tab:star}, but equal within uncertainties, which supports our determinations.
We scaled the errors in $\mu_H$ by five to reflect more realistic values as compared to other limb-darkening grids \citep[e.g.,][]{ClaretBloemen2011}, though this has little impact on the resulting angular diameter, as the uncertainties in the squared visibilities dominate the error in diameter.

\begin{table}
\centering
\small
\caption{CHARA model input and interferometric results$^a$.} 
\label{tab:chara_results}
\begin{tabular}{lcc}
\hline
\hline
\noalign{\smallskip}
Parameter & \multicolumn{2}{c}{Value} \\ 
        & UD & LDD \\
\noalign{\smallskip}
\hline
\noalign{\smallskip}
$\theta$ [mas] & $0.382 \pm 0.017$ & $0.390 \pm 0.018$ \\
\noalign{\smallskip}
$T_{\rm eff}$ [K] & ... & $3283 \pm 78$ \\
\noalign{\smallskip}
$V_0^2$ & $0.9107 \pm 0.0051$ & $0.9109 \pm 0.0051$ \\
\noalign{\smallskip}
$\mu_H$ & 0.0 (fixed) & $0.2417 \pm 0.0014$ \\ 
\noalign{\smallskip}
\hline
\end{tabular}
\tablefoot{
\tablefoottext{a}{Mean and standard deviation for uniform (UD) and limb-darkened (LDD) fits.
Other derived parameters ($R_\star$, $M_\star$) are provided in Table~\ref{tab:star}. 
There is no $T_{\rm eff}$ in the UD fit.}
}
\end{table}

The mean and standard deviation for the limb-darkened angular diameter fit, $\theta_{\rm LDD}$, with the corresponding $\mu_{\rm H}$ and $V^2_{0,{\rm LDD}}$ values are again listed in Table~\ref{tab:chara_results}, while the model for both the limb-darkened and uniform-disc fits and the posterior distributions are shown in Fig~\ref{fig:CHARAvisibilitybinned}. 
Finally, our measured angular diameter coupled with the corrected {\em Gaia} star's parallax yields a stellar radius of $R_{\star} = 0.339 \pm 0.015\,R_\odot$, which is consistent within less than $1 \sigma$ with the value used by \citet{Trifonov2021} of $R_{\star} = 0.328 \pm 0.011\,R_\odot$. 

As noted above, we multiplied by an extra term, $V_0^2$, in Eq.~\ref{eq:vis_squared} to account for unknown systematic offsets.
Our results indeed showed that $V_0^2$ is non-unity for both the limb-darkened- and uniform-disc fits.  
In order to determine whether this offset was due to a bad calibrator (for example, an unknown binary or rotationally oblate object), we did a series of tests on the data sets. 
First, we calibrated each science target data set independently using the calibrator nearest in time to the science target, and found that all calibrators gave consistent results to each other.  
We then calibrated each calibrator against one another to ensure that their response was consistent with point sources, and found no evidence to reject any calibrator due to that. 
Lastly, we confirmed that the closure phases of the calibrators were consistent with zero, ensuring us that they are point-symmetric and should not produce spurious results. 
We suspect that the non-unity $V_0^2$ is a result of the colour mismatch between calibrators (four early A and one early F dwarfs) and science target (M3.5\,V) that is not completely treated in the standard reduction and calibration routines, although it may also be related to some systematic in the limb-darkening determination of M dwarfs.
While the origin of this floating offset is not well understood, we remain confident of the interferometric results as the stellar diameter is 
in agreement with past estimates based on independent techniques \citep[e.g.,][among many others]{Newton2015,Newton2017,Schweitzer2019,Trifonov2021}. 

Finally, as it was put forward in Sect.~\ref{sec:star+planet}, we determined the stellar mass, $M_\star = 0.333 \pm 0.019\,M_\oplus$, from $R_\star$ and the mass-radius linear relation of \citet{Schweitzer2019}.
The authors derived this relation from 55 detached, double-lined, double eclipsing, main-sequence M dwarf binaries from the literature.
The $M_\star$-$R_\star$ relation of \citet{Schweitzer2019} agrees (and may surpass) previous relations in the M-dwarf domain by \citet{Andersen1991}, \citet{Torres2010}, \citet{Torres2013}, \citet{Eker2018}, and references therein.

\subsection{Stellar rotation period}
\label{sec:Prot}

\begin{figure}
    \centering
    \includegraphics[width=0.49\textwidth]{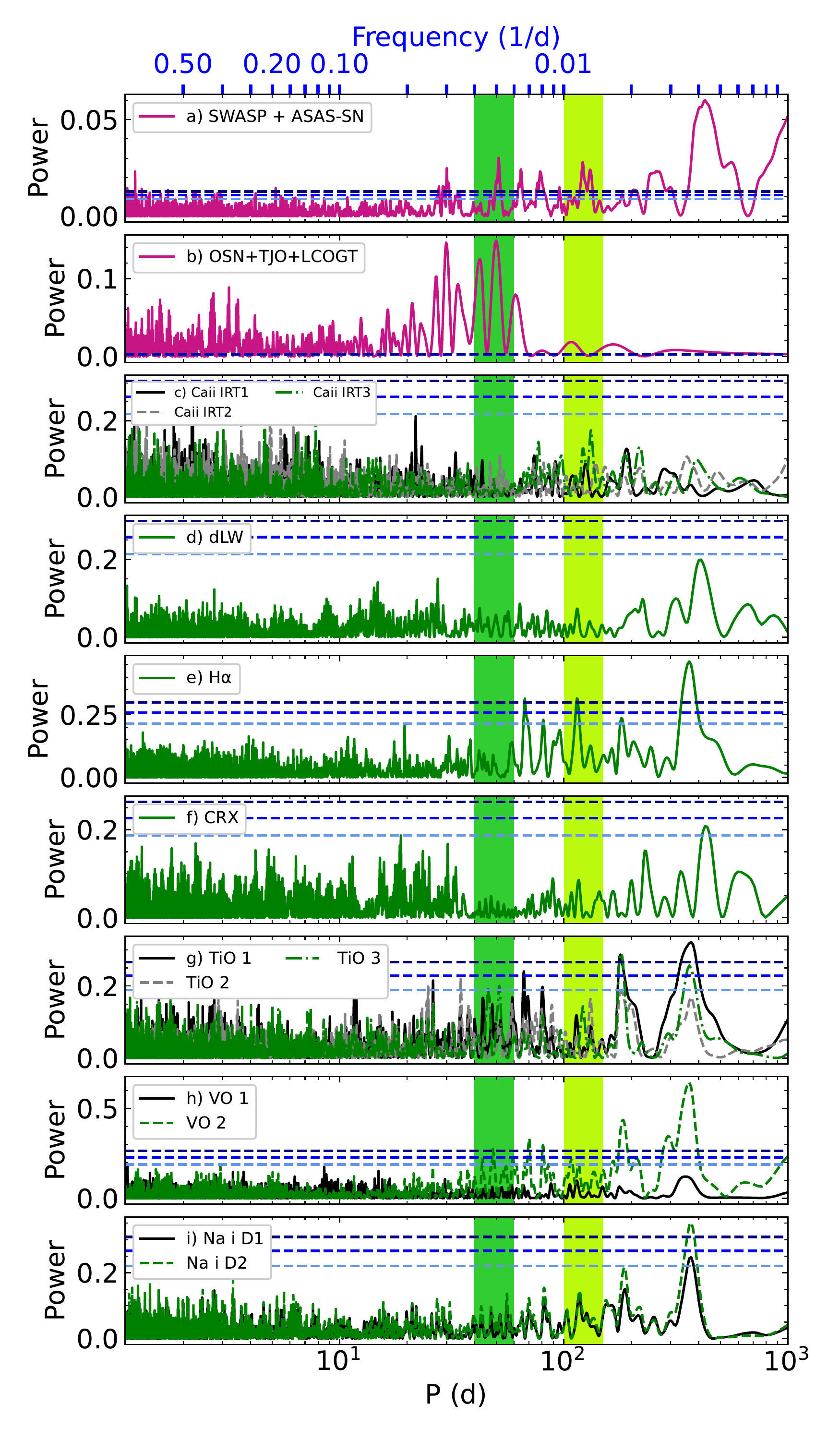}
        \caption{GLS periodograms of ground photometry (purple) and CARMENES spectral activity indicator (green) time series.
        From top to bottom:
        ($a$) SuperWASP and ASAS-SN, 
        ($b$) $B$~LCOGT, $V$~OSN, and $R$~TJO, 
        ($c$) CARMENES Ca~{\sc ii}~IRT$_{[1,2,3]}$, 
        ($d$) dLW, 
        ($e$) H$\alpha$, 
        ($f$) CRX,
        ($g$) TiO [1,2,3] (TiO~7050, 8430, and 8860),
        ($h$) VO [1,2] (VO~7436 and 7942),
        and ($i$) Na~{\sc i}~D$_{[1,2]}$.
        Dark and light green areas at 40--60\,d and 100--150\,d indicate the $P_{\rm rot}$ and cycle intervals, respectively.
        Blue dashed horizontal lines mark the 0.1\,\%, 1\,\%, and 10\,\% false alarm probabilities.}
        \label{fig:Phot}
\end{figure} 

The main difficulty in the determination of a robust $P_{\rm rot}$ of Gl~486 lies on its small amplitude of photometric variability.
For example, \cite{Clements2017} obtained 69 frames of the star during 5.02\,a with the 0.9\,m SMARTS telescope at the Cerro Tololo Inter-American Observatory and measured a $V$-band light curve standard deviation of 11.6\,mmag. 
\citet{DiezAlonso2019} measured greater standard deviations of 32\,mmag and 66\,mmag with poorer (but public) NSVS and ASAS data, respectively.
None of them could conclude anything on the star's photometric variability.
From the photometric data available to \citet{Trifonov2021}, they used only the ASAS-SN and SuperWASP (North and South combined) light curves for measuring a stellar rotation period (actually, ``a proxy obtained from a quasi-periodic representation of the photometric variability'') of Gl~486.
However, the signal corresponding to the $P_{\rm rot}$ tabulated by \citet{Trifonov2021}, of $P_{\rm rot} = 130.1^{+1.6}_{-1.2}$\,d, did not appear in the periodograms of all their data sets, which casted doubts on the $P_{\rm rot}$ determination.

The standard deviations of our eight ground-based light curves (Table~\ref{tab:data}, Fig.~\ref{fig:Phot-all}) range from 4--10\,mmag (LCOGT, OSN, TJO -- after subtracting a linear trend to the latter data set) to 28--34\,mmag (AstroLAB, ASAS-SN $g'$) after applying $N \sigma$-clipping filters for outliers ($N$ = 2.5--4.5).
We run generalised Lomb-Scargle (GLS) periodograms \citep{ZechmeisterKuerster2009} on the joint data sets of SuperWASP and ASAS-SN, with the longest time baseline, and LCOGT, OSN, and TJO, with the smallest scatter.
The first joint data set, although it is noisier, allows to investigate activity cycles much longer than the stellar $P_{\rm rot}$, while the second one allows to investigate a new range of low-amplitude signals.
In the two top panels of Fig.~\ref{fig:Phot}, we display the GLS periodograms of the two joint photometric data sets after considering different offsets between the light curves.
In the SuperWASP + ASAS-SN periodogram, there is power beyond 300\,d, apart from significant signals at $\sim$130\,d (as reported by \citealt{Trifonov2021}) and $\sim$50\,d, while the LCOGT + OSN + TJO periodogram shows several significant peaks in the 30--70\,d range.
A rotation period shorter than 100\,d matches better the $\log{R'_{\rm HK}}$-$P_{\rm rot}$ relations in the literature \citep[e.g.,][]{SuarezMascareno2016, AstudilloDefru2017a, Boudreaux2022} than the period reported by \citet{Trifonov2021}.

In Fig.~\ref{fig:Phot}, we also display the GLS periodograms of 13 representative activity indicators associated with H$\alpha$, Ca~{\sc ii}, Na~{\sc i}, TiO, VO, dLW, and CRX measured on the CARMENES spectra. 
The TiO indices are the only spectral activity indices that show a forest of (non-significant) peaks around the power maximum of the LCOGT + OSN + TJO periodogram at $\sim$50\,d.
This is in agreement with the weak activity of Gl~486\,b and the results of \citet{Schoefer2019}, who found that the TiO indices usually are the most sensitive ones to variable activity in weak M dwarfs.
However, most of the power of the periodograms falls beyond 100\,d. 
Because of the MAROON-X time sampling, the visits of which concentrated around a few nights on three short runs much shorter than $P_{\rm rot}$, a new joint periodogram analysis of the CARMENES and MAROON-X spectral activity indices does not improve the results of \citet{Trifonov2021}.

We did a joint fit of our three best light curves, LCOGT ($B$), OSN ($V$), and TJO ($R$), to a double-sinusoidal model with characteristic periods $P_{\rm rot}$ and $P_{\rm rot}/2$ \citep{Boisse2011, GonzalezAlvarez2022} and a wide uniform $P_{\rm rot}$ prior between 30\,d and 60\,d.
A pair of the most significant peaks in the LCOGT + OSN + TJO periodogram are also 1:2 alias of each other. 
Two adjusted sinusoids fixed at the fundamental period and its first
harmonic allow to remove a large fraction of the photometric jitter amplitude (\citealt{Boisse2011} did it for the two first harmonics).

\begin{figure}
    \centering
    \includegraphics[width=0.49\textwidth]{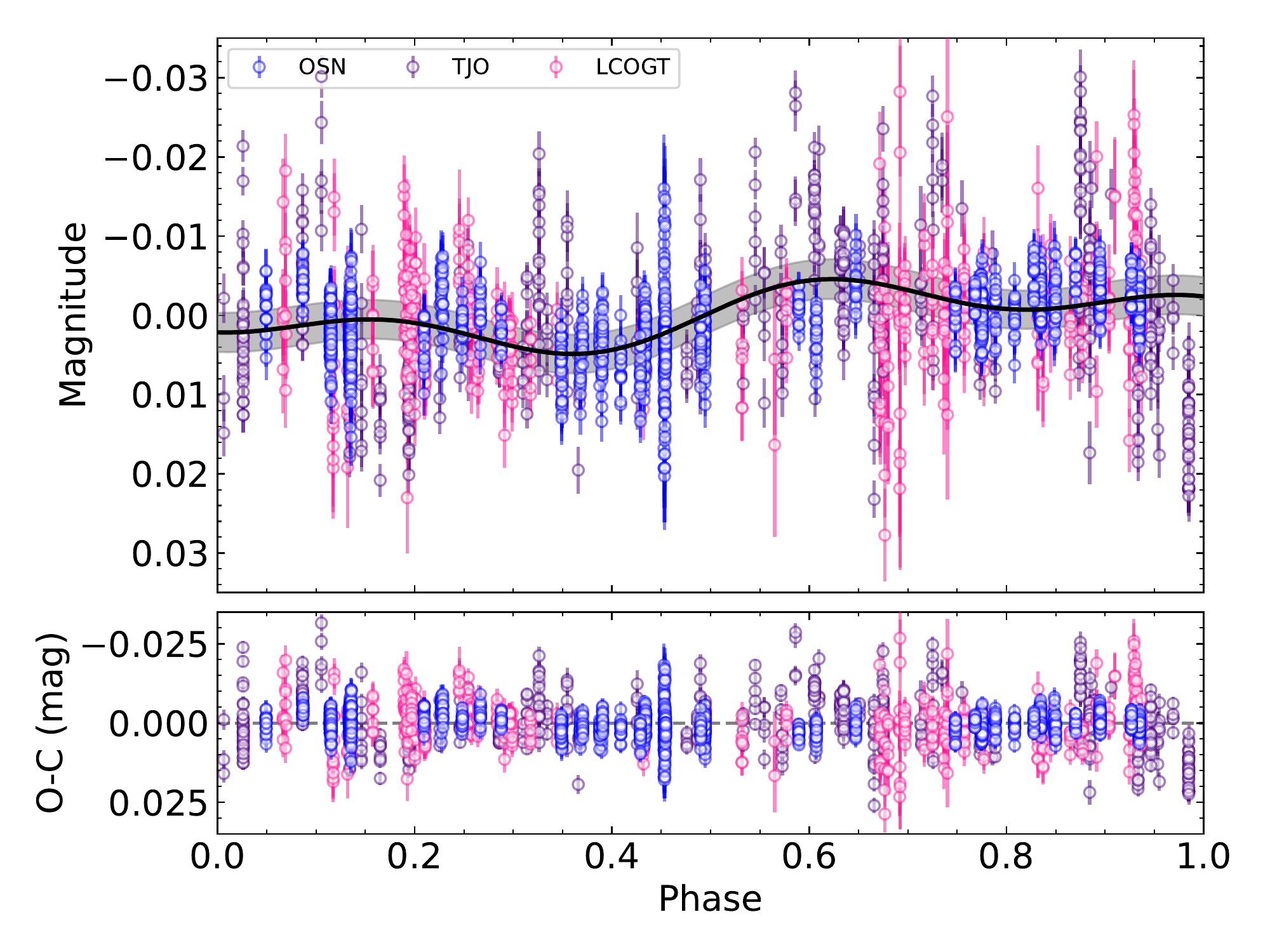}
        \caption{OSN (blue), TJO (violet), and LCOGT (magenta) light curves phase-folded to the double-sinusoidal rotation period ($P_{\rm rot}$ = 49.9 $\pm$ 5.5\,d).
        The grey dashed area indicates the fit uncertainty.
        }
        \label{fig:Phot-phase}
\end{figure}

The fit was performed using the {\tt juliet}\footnote{\url{https://github.com/nespinoza/juliet}} python package \citep{Espinoza2019}, which uses nested samplers, a numerical method for Bayesian computation that simultaneously provides both posterior samples and Bayesian evidence estimates. 
{\tt juliet} is mostly used for RV and transit best-fit optimisation (Sect.~\ref{sec:planetradiusandmass}), but can also determine rotation periods from light curves.
We determined a photometric rotation period $P_{\rm rot}$ = 49.9 $\pm$ 5.5\,d and an amplitude of just 3.4 $\pm$ 2.4\,mmag, which explains the $P_{\rm rot}$ non-detection by \citet[][SMARTS]{Clements2017} and \citet[][ASAS, NSVS]{DiezAlonso2019}, and the different $P_{\rm rot}$ determination by \citet[][ASAS-SN, SuperWASP]{Trifonov2021}.
It is likely that the period measured by them, about three times ours, is associated with a stellar activity cycle.
Fig.~\ref{fig:Phot-phase} displays the phase-folded LCOGT, TJO, and OSN data fitted to the the two sinusoids.

\subsection{Stellar abundances}
\label{sec:abundances}

\begin{figure*}
    \centering
    \includegraphics[width=0.99\textwidth]{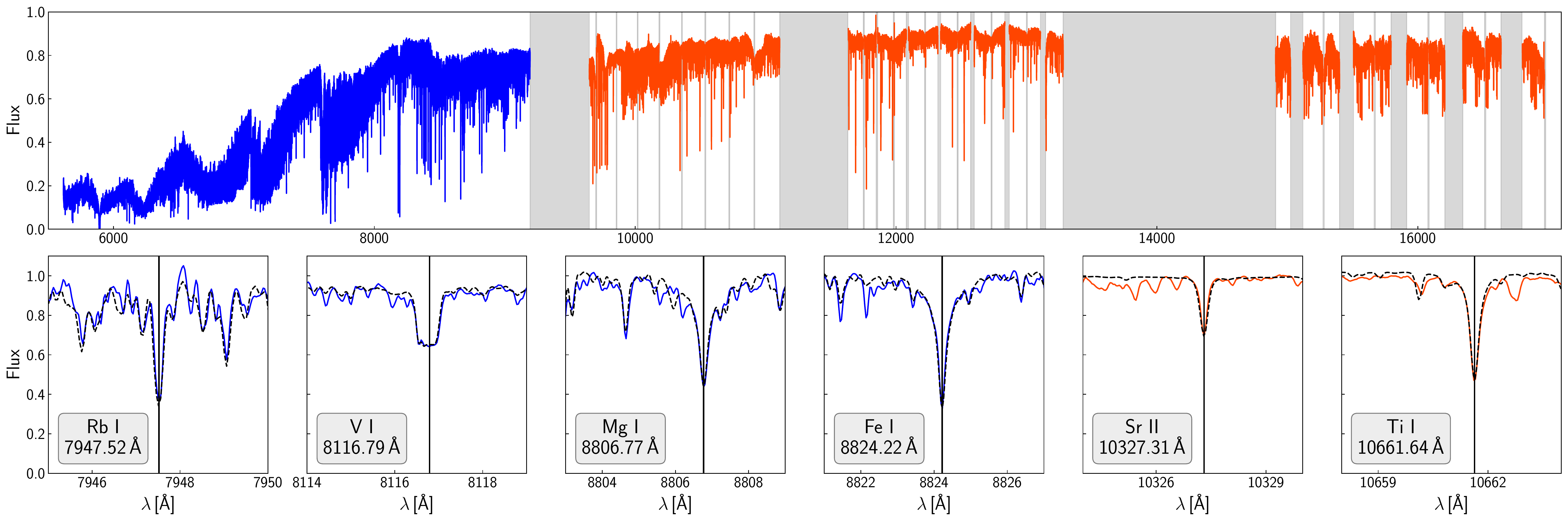}
         \caption{Element abundance determination of Gl~486\,b.
         {\em Top panel:} 
         co-added, order-merged, channel-merged CARMENES VIS (blue) and NIR (red) template spectrum.
         Interruptions (grey areas) are due to strong telluric contamination and inter-order and NIR detector array gaps.
         {\em Bottom panels:} zoom around six representative magnetic weakly-sensitive atomic lines from 
         \citet[][Rb~{\sc i}, Sr~{\sc ii}]{Abia2020},
         \citet[][V~{\sc i}]{Shan2021},
         \citet[][Mg~{\sc i}]{Passegger2019},
         and \citet[][Fe~{\sc i}, Ti~{\sc i}]{Marfil2021}.
         Black dashed lines are the synthetic fits.
         }
        \label{fig:lines}
\end{figure*} 

F-, G-, and K-type stars with orbiting Jovian planets are preferentially metal rich \citep{Gonzalez1997, Santos2004, FischerValenti2005}.
However, the frequency of low-mass planets, including rocky planets, does not seem to depend on metallicity \citep[e.g.,][]{Ghezzi2010, Mayor2011, Sousa2011, Petigura2018}.
Likewise, there is no robust indication of a larger frequency of Jovian planets around more metallic M dwarfs \citep[][but see \citealt{Pinamonti2019}]{Bonfils2007, Johnson2010, RojasAyala2012, Neves2013, Courcol2016}.
Jovian planets around M dwarfs are rare \citep{Delfosse1998, Marcy1998, Endl2006, Forveille2011, Morales2019, Sabotta2021}, which does not help in settling the issue.
In contrast, small planets around M dwarfs, such as mini-Neptunes and super-Earths, are numerous (Sect.~\ref{sec:intro}).
However, the ability to measure a correlation between metallicity and small planet frequency is hampered by the absence of reliable M-dwarf metallicity determinations until very recently.
The origin of this lack resides in that M-dwarf atmospheres are more complicated to model than their warmer FGK-type counterparts, though this difficulty is starting to be overcome with better calibration samples and improved models \citep[][and references therein]{Maldonado2020, Passegger2022}.

\begin{table}
\centering
\small
\caption{Element abundances of Gl~486$^a$.} 
\label{tab:abundances}
\begin{tabular}{l cc}
\hline
\hline
\noalign{\smallskip}
Element         & $A$(X)    & [X/H] \\ 
                &           & [dex] \\
\noalign{\smallskip}
\hline
\noalign{\smallskip}
Mg       & $7.63 \pm 0.09$  & $+0.03 \pm 0.09$ \\ 
\noalign{\smallskip}
Si       &  $7.42 \pm 0.13$  & $-0.09 \pm 0.13$ \\ 
\noalign{\smallskip}
V        & $3.84 \pm 0.08$   & $-0.08 \pm 0.08$ \\ 
\noalign{\smallskip}
Fe       & $7.35 \pm 0.14$  & $-0.15 \pm 0.13$ \\ 
\noalign{\smallskip}
Rb$^b$   & $2.35 \pm 0.12$   & ${+0.00 \pm 0.12}$ \\ 
\noalign{\smallskip}
Sr$^b$   & $2.80 \pm 0.10$   & ${-0.12 \pm 0.12}$ \\ 
\noalign{\smallskip}
Zr       & $2.70 \pm 0.12$   & ${+0.12 \pm 0.10}$ \\ 
\noalign{\smallskip}
\hline
\end{tabular}
\tablefoot{
\tablefoottext{a}{All element abundances computed in this work except for [Fe/H], which was computed by \citet{Marfil2021}.
$A$(X) is also known as $\log{\epsilon}$(X).
We also computed the all-metals relative abundance, [M/H] = $-0.15 \pm 0.10$\,dex, and the carbon-to-oxygen ratio, C/O = $+0.54 \pm 0.05$\,dex.}
\tablefoottext{b}{NLTE abundances \citep{Abia2020}.}
}
\end{table}

Relevant for this work, there are planet-formation models that use different stellar element abundances and ratios as inputs and that predict different planet composition and structure on the assumption that the protoplanetary disc preserves the original stellar abundances \citep{IdaLin2004, KampDullemond2004, Chambers2010, Emsenhuber2021}.
Some of these element ratios are Mg/Fe, Si/Fe, C/O, or N/O and will play a role in the future of comparative astrochemistry exoplanetology \citep{Dawson2015, Gaidos2015, Thiabaud2015, Santos2017, Cridland2020}. 

Here, we applied state-of-the-art M-dwarf element abundance analysis to the high-S/N VIS+NIR CARMENES template spectrum of Gl~486 computed with {\tt serval}.
First, we took the iron abundance with $\alpha$-enhancement correction, [Fe/H] = $-0.15 \pm 0.13$, from \citet{Marfil2021}, which is $1.5\sigma$ lower than the mean of seven previously published [Fe/H] values (Table~\ref{tab:Fe/H}).
However, in contrast to the other works \citep[cf.][]{Passegger2022}, the [Fe/H] values from \citet{Marfil2021} in the range of $T_{\rm eff}$ of our target star are in agreement with the metallicity distribution of FGK-type stars in the solar neighbourhood and correlate well with the kinematic membership of the targets in the Galactic populations.
Next, we applied recent techniques for the determination of other element abundances.
For internal consistency, apart from the $T_{\rm eff}$ of \citet{Marfil2021} we also used their $\log{g}$ (Table~\ref{tab:star}) in the following analysis.
Our procedure is illustrated by Fig.~\ref{fig:lines}.

First, we used the methodology of \citet{Abia2020} for determining abundances of three neutron-capture elements, namely Rb, Sr, and Zr.
In a first step, we determined an average metallicity [M/H] = $-0.15 \pm 0.10$\,dex with a number of Fe~{\sc i}, Ti~{\sc i}, Ni~{\sc i}, and Ca~{\sc i} lines, which matches the [Fe/H] of \citet{Marfil2021} and substantiates our choice.
Next, we determined the carbon-to-oxygen ratio C/O of Gl~486 with an iterative method that started with a synthetic spectrum with the carbon and oxygen abundances scaled to the metallicity of the model atmosphere.
We paid special attention to the strength of CO, OH, CN, and TiO absorption bands.
The determined C/O ratio, $+0.54 \pm 0.05$\,dex, becomes exactly solar with the latest revision of solar oxygen abundance with respect to the standard value, of +0.10\,dex \citep{Bergemann2021}.
Determining this ratio is vital, as almost all the available carbon in the atmospheres of M dwarfs is locked into CO and, therefore, the abundance of the other O-bearing molecules (TiO, VO, OH, H$_2$O) mainly depends on the C/O ratio.
After following all the steps enumerated by \citet{Abia2020}, and assuming non-local thermodynamic equilibrium (NLTE) for Rb and Sr, we obtained
[Rb/M] = $+0.15 \pm 0.12$\,dex, [Sr/M] = $+0.03 \pm 0.14$\,dex, [Zr/M] = $+0.00 \pm 0.13$\,dex, from which we determined the A(X) and [X/H] values in Table~\ref{tab:abundances} with solar values from \citet{Lodders2009}.
The derived [Rb,Sr,Zr/M] ratios in Gl~486 match well the corresponding [Rb,Sr,Zr,M] vs. [M/H] relationships of unevolved field stars obtained by \citet{Abia2020, Abia2021}.

Next, we used the methodology of \citet{Shan2021} for determining abundance of V.
In M dwarfs, including Gl~486\,b, many V~{\sc i} lines exhibit a distinctive broad and flat-bottom shape, which is a result of hyperfine structure.
We used four prominent V~{\sc i} lines at 8093\,\AA, 8117\,\AA, 8161\,\AA, and 8920\,\AA\ ($\lambda$ in air) for the fit, and got $A$(V) = $3.84 \pm 0.08$.
The line-to-line scatter and the errors from the input stellar parameters were added quadratically for determining the 
abundance uncertainty.
With the solar abundances of \cite{Asplund2009}, we arrived at [V/H] = $-0.08 \pm 0.08$.
The corresponding [V/Fe] = [V/H] -- [Fe/H] = $+0.07 \pm 0.15$ is a typical value for stars in the solar neighbourhood \citep{BattistiniBensby2015}.

Finally, we employed the spectral synthesis method, together with the PHOENIX BT-Settl atmospheric models \citep{Allard2012} and the radiative transfer code {\tt Turbospectrum} \citep{Plez2012} for determining Mg and Si abundances of Gl~486.  
We measured [Mg/H] = $+0.03 \pm 0.09$\,dex and [Si/H] = $-0.09 \pm 0.13$\,dex.  
Further details on the followed steps will be provided by Tabernero et~al. (in~prep.). 
To sum up, the composition of Gl~486 seems to be similar to the Sun, but slightly lower metallic, although consistent within the error bars.

\subsection{Stellar coronal emission}
\label{sec:stellarcoronalemission}

\begin{figure}
    \centering
    \includegraphics[width=0.49\textwidth]{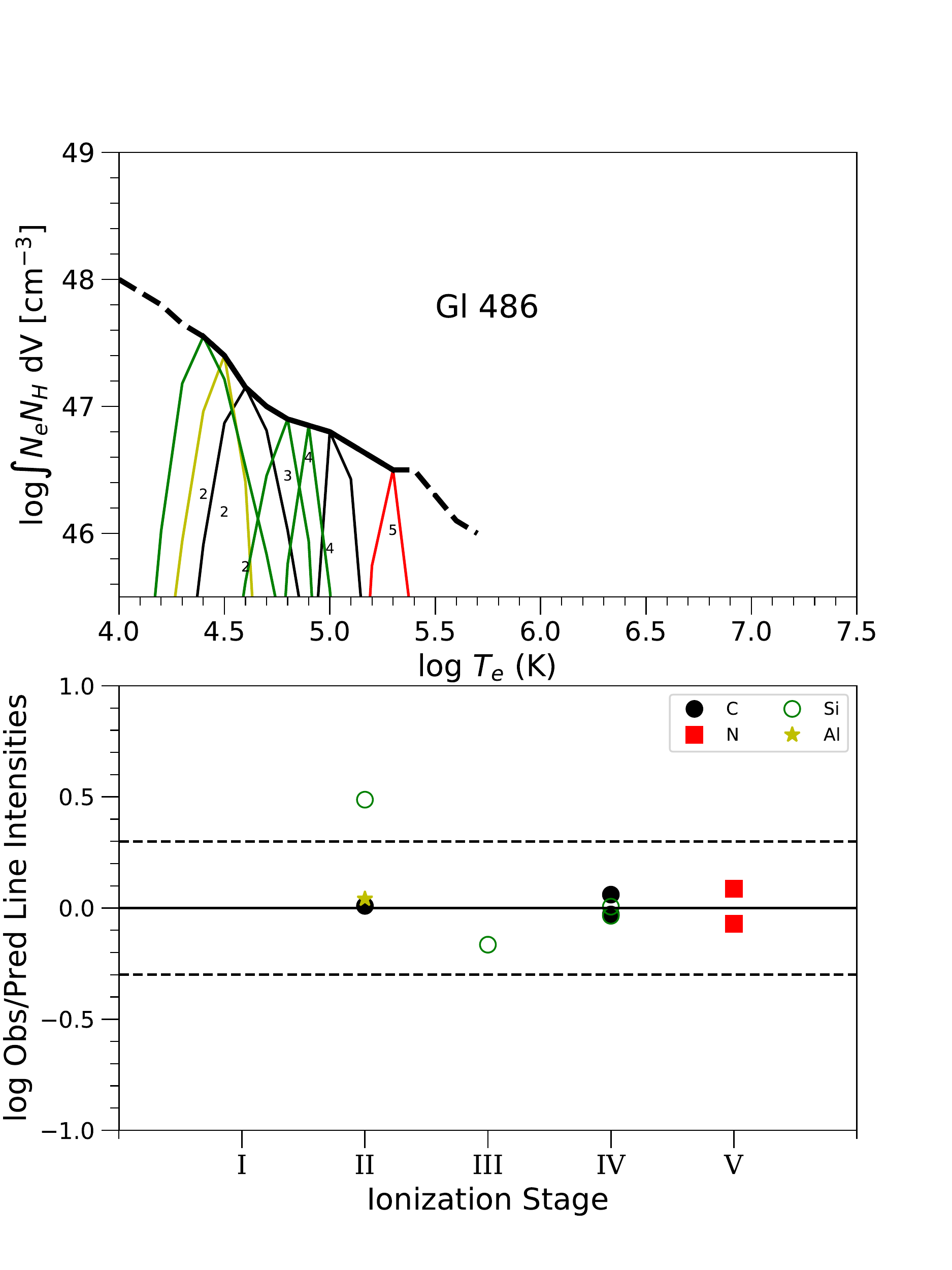}
    \caption{{\em Top:} Emission measure distribution from {\em Hubble}/STIS data.
    Thin coloured lines represent the relative contribution function for each ion (the emissivity function multiplied by the emission measure distribution at each point). 
    Small numbers indicate the ionisation stages of the species. 
    {\em Bottom:} Observed-to-predicted line flux ratios for the ion stages in the upper figure. The dashed lines denote a factor of two.}
    \label{fig:coronalemission}
\end{figure} 

For quantifying the coronal activity of Gl~486, we searched through public archives of space-borne high-energy observatories 
({\em Extreme Ultraviolet Explorer}, 
{\em Chandra}, 
{\em XMM-Newton}, 
{\em Neil Gehrels Swift}, 
eROSITA/{\em Spektr-RG}), 
and found a {\em ROSAT} X-ray (5--100\,\AA) upper limit by \citet[][see also a non-detection reported by \citealt{Wood1994}]{Stelzer2013}.
We converted it into an upper limit of the X-ray luminosity with the corrected {\em Gaia} distance.
The expected X-ray luminosity considering the stellar rotation period of {$\sim$ 50}\,d, together with the $V-K_s$ colour and $L_\star$ of Gl~486 and the $L_{\rm X}$-$P_{\rm rot}$ relation of \citet{Wright2011}, is $\log L_{\rm X} = 27.44$\,erg\,s$^{-1}$.
The value is higher than the upper limit calculated by \citet{Stelzer2013}, but still consistent given the spread of X-ray luminosity observed in the rotation-activity diagram, of up to one order of magnitude (Table~\ref{tab:star}).

We also computed our own upper limit for the extreme ultraviolet (EUV, 100--920\,\AA) luminosity after analysing the {\em Hubble}/STIS data presented in Sect.~\ref{sec:spacespectroscopy}.
On the G140L and G140M spectra, we measured the emission measure, $EM$, defined as:
\begin{equation}
    EM = \log \int N_{\rm e} \, N_{\rm H} \, {\rm d}V,  
\end{equation}
\noindent where $N_{\rm e}$ and $N_{\rm H}$ are the electron and hydrogen densities (in cm$^{-3}$), respectively, and $V$ is the volume. 
Although most measured C, N, Al, and Si lines do not reach the usually required $3\sigma$ detection (Table~\ref{tab:uvfluxes}), the combined fluxes give a consistent emission measure distribution in the range $\log T$~(K) = 4.1--5.5 following the techniques described by \citet{SanzForcada2003}. 
The resulting emission measure distribution is illustrated by Fig.~\ref{fig:coronalemission} and tabulated with their uncertainties in Table~\ref{tab:emd}.
To evaluate the coronal part of the model, we tried different values of $T$ and $EM$ consistent with the upper limit of $L_{\rm X} \approx 4.17 \cdot 10^{26}$\,erg\,s$^{-1}$ reported by \citet{Stelzer2013}. 
Since the low level of stellar activity indicates a low coronal temperature, we fixed it at a typical quiet solar value of $\log T$~(K) = 6.2, which implies $\log EM$(cm$^{-3}$) = 49.0 with a solar photospheric abundance.
Calculated coronal abundances are [C/H] = $+0.0\pm0.3$, [N/H] = $+0.0\pm0.3$, [Si/H] = $+0.2\pm0.4$, and [Al/H] = $+0.6\pm0.9$.
A more realistic value of coronal abundances would probably imply a Fe enhancement, similar to the solar corona, which would imply in turn an $EM$ value about one order of magnitude lower, but with little impact on the overall X-ray luminosity. 
With this coronal model, we predicted a more realistic upper limit of the EUV luminosity of $1.45 \cdot 10^{27}$\,erg\,s$^{-1}$.
The results are similar to applying the $L_{\rm X}$-$L_{\rm EUV}$ relation of \citet{SanzForcada2011}.
In the 100--504\,{\AA} spectral range, which is involved in the formation of the \ion{He}{i}~$\lambda$10830\,{\AA} triplet in planet atmospheres \citep{Nortmann2018}, the upper limit of the luminosity amounts to $1.27 \cdot 10^{27}$\,erg\,s$^{-1}$.

\subsection{Planet radius and mass}
\label{sec:planetradiusandmass}

\begin{table}
\centering
\small
\caption{Log-evidence and number of parameters of RV+transit {\tt juliet} models$^a$.}  
\label{tab:log-e}
\begin{tabular}{l c cc}
\hline
\hline
\noalign{\smallskip}
Model          & $N_{\rm par}$      & $\ln{\mathcal{Z}}$ & $| \Delta \ln{\mathcal{Z}} |$ \\
\noalign{\smallskip}
\hline
\noalign{\smallskip}
1pl            & 40    & 91\,725.250   & 14.302 \\
1pl+$e$        & 42    & 91\,724.236   & 15.316 \\
1pl+GP         & 43    & 91\,739.552   & 0 \\
1pl+$e$+GP     & 45    & 91\,739.043   & 0.506 \\
\noalign{\smallskip}
\hline
\end{tabular}
\tablefoot{
\tablefoottext{a}{
Models -- 
1pl: one planet.
$e$: non-circular orbit.
GP: Gaussian process ($P_{\rm rot,GP}$).
}
}
\end{table}

\begin{table}
    \centering
    \caption{Fitted and derived planet parameters of Gl~486\,b$^a$.}
    \label{tab:planet}
    \begin{tabular}{lc} 
        \hline
        \hline
        \noalign{\smallskip}
        Parameter$^a$ & Gl~486\,b  \\
        \noalign{\smallskip} 
        \hline
        \noalign{\smallskip}
        \multicolumn{2}{c}{\it Fitted parameters} \\ 
        \noalign{\smallskip}
        $P$ [d]                                     & $1.4671205 ^{+0.0000012} _{-0.0000011}$ \\
            \noalign{\smallskip} 
        $t_{0}$ (BJD)                               & $2459309.676506 ^{+0.000102} _{-0.000099}$ \\
            \noalign{\smallskip}
        $K$ [m\,s$^{-1}$]                          observations & $3.495 ^{+0.064} _{-0.066}$ \\
            \noalign{\smallskip}
        $e$                                         & 0.0 (fixed) \\ 
            \noalign{\smallskip} 
        $\omega$ [deg]                              & 90 (fixed) \\ 
            \noalign{\smallskip} 
       \multicolumn{2}{c}{\it Derived parameters} \\
            \noalign{\smallskip}
        $a_{\rm b}/R_\star$                         & $10.96^{+0.21}_{-0.44}$ \\ 
            \noalign{\smallskip} 
        $a_{\rm b}$ [au]                            & $0.01713^{+0.00091}_{-0.00098}$ \\ 
            \noalign{\smallskip} 
        $p = R_{\rm b}/R_\star$                     & $0.03635^{+0.00046}_{-0.00039}$ \\ 
            \noalign{\smallskip} 
        $b = (a_{\rm b}/R_\star)\cos i_{\rm b}$     & $0.21^{+0.14}_{-0.13}$ \\ 
            \noalign{\smallskip} 
        $i_{\rm b}$ [deg]                           & $88.90^{+0.69}_{-0.84}$ \\ 
            \noalign{\smallskip} 
        $T_{14}$ [h]                                & $1.083^{+0.086}_{-0.038}$ \\ 
            \noalign{\smallskip} 
        $T_{12} = T_{34}$ [h]                       &  $0.0383^{+0.0062}_{-0.0044}$ \\ 
            \noalign{\smallskip} 
        $\mathcal{P}_{\rm transit}$                  & $0.0912^{+0.0038}_{-0.0017}$ \\ 
            \noalign{\smallskip} 
        $\Delta_{\rm transit} = (R_{\rm b}/R_\star)^2$ [ppm] & $1321^{+34}_{-28}$ \\ 
            \noalign{\smallskip}
        $R_{\rm b}$ [$R_\oplus$]                    & $1.343^{+0.063}_{-0.062}$ \\ 
            \noalign{\smallskip} 
        $M_{\rm b}$ [$M_\oplus$]                    & $3.00^{+0.13}_{-0.13}$ \\ 
            \noalign{\smallskip} 
        $\rho_{\rm b}$ [10$^{3}$\,kg\,m$^{-3}$]     & $6.79^{+1.08}_{-0.91}$ \\ 
            \noalign{\smallskip} 
        $g_{\rm b}$ [m\,s$^{-2}$]                   & $16.3^{+1.7}_{-1.7}$ \\
            \noalign{\smallskip} 
        $v_{\rm e,b}$ [km\,s$^{-1}$]                & $16.71^{+0.53}_{-0.52}$ \\
            \noalign{\smallskip} 
        $S_{\rm b}$ [$S_\oplus$]                    & $41.3^{+4.4}_{-4.8}$ \\
            \noalign{\smallskip} 
        $T_\textnormal{eq,b} (1-A_{\rm Bond})^{-1/4}$ [K] & $706^{+19}_{-20}$ \\
            \noalign{\smallskip} 
       \multicolumn{2}{c}{\it Additional parameters} \\
            \noalign{\smallskip}
        $F_{\it XUV}$ [erg\,s$^{-1}$\,cm$^{-2}$]    & $<$ 6800  \\
            \noalign{\smallskip} 
        $\dot{M_{\rm b}}$ [10$^{7}$\,kg\,s$^{-1}$]  & $<$ 1.3 \\ 
        \noalign{\smallskip}
        \hline
    \end{tabular}
    \tablefoot{
      \tablefoottext{a}{Planet parameters from the 1pl+GP fit with the clipped data set.
      $T_{14}$ is the transit duration from first to last contact; $T_{12}$ and $T_{34}$ are ingress and egress durations.
      $\mathcal{P}_{\rm transit}$ is probability of transit;
      $\Delta_{\rm transit}$ is transit depth.}
      Remaining fitted parameters are in Table~{\ref{tab:priors+posteriors}.}
      }
\end{table}

\begin{figure*}
    \centering
    \includegraphics[width=0.49\textwidth]{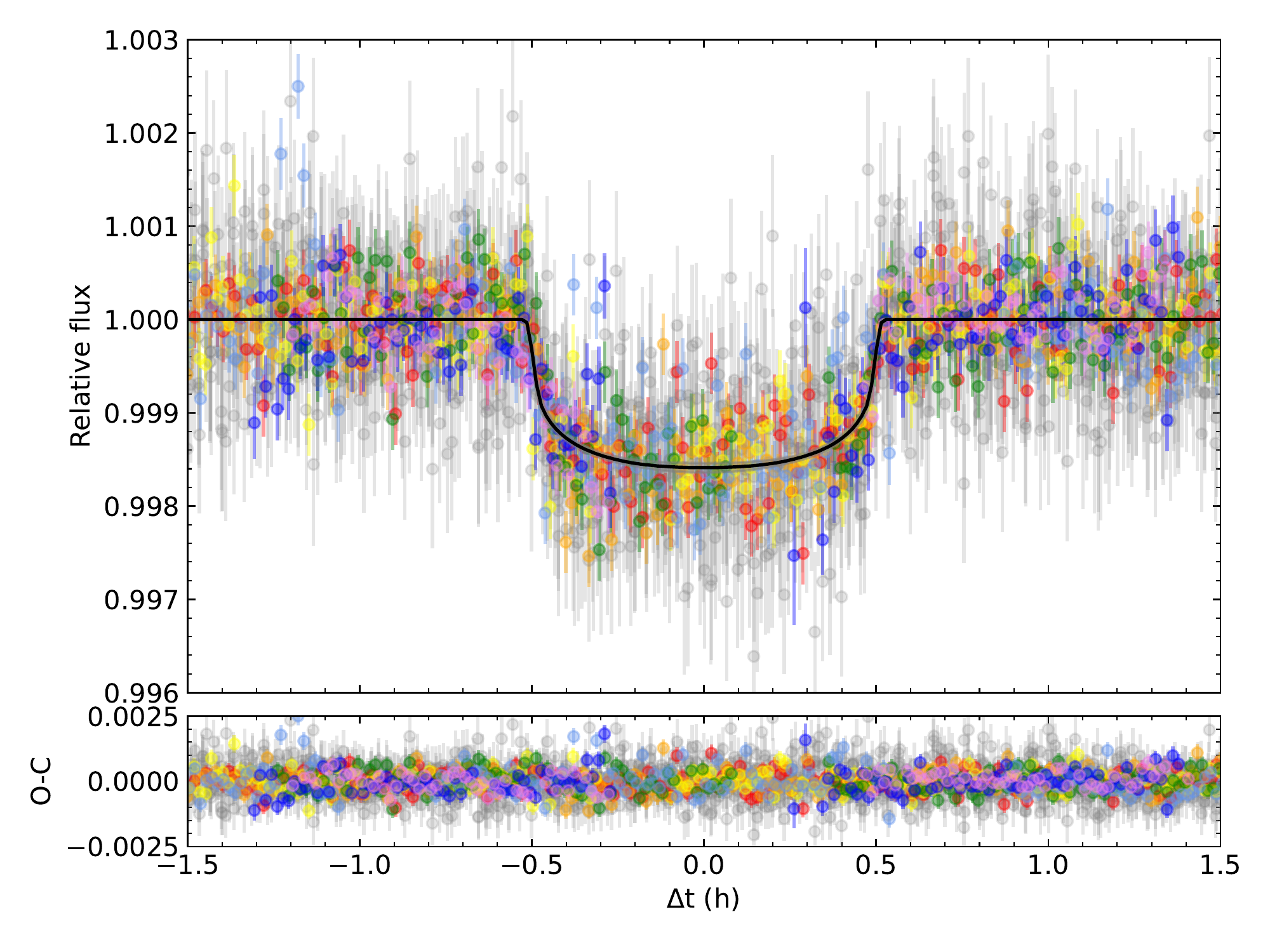}
    \includegraphics[width=0.49\textwidth]{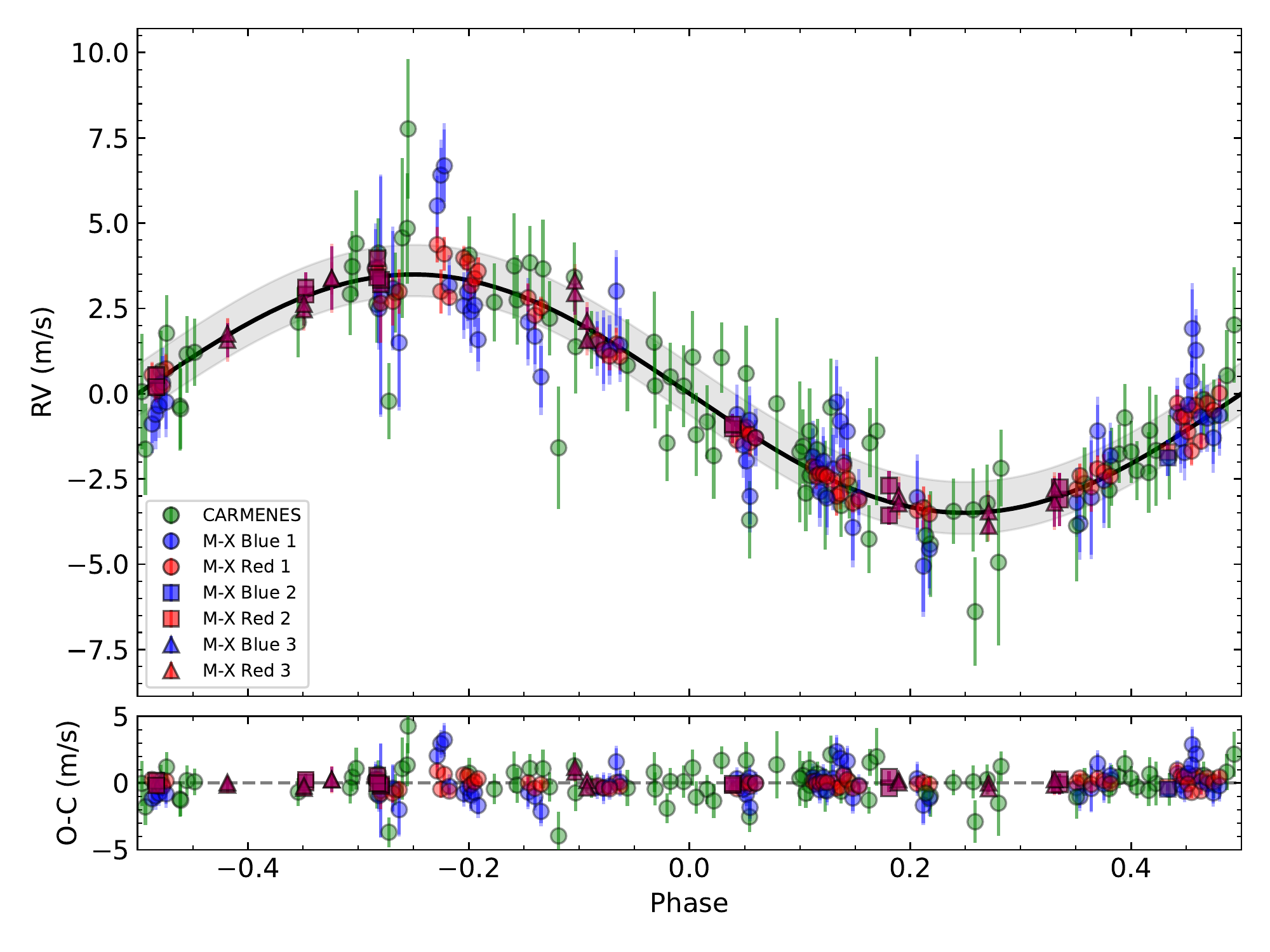}
        \caption{Phase-folded transit and RV data and 1pl+GP model fits.
        {\em Left:}
        light curve model fit (black line) and {\em CHEOPS}+{\em TESS} data with the same symbol colours as in Fig.~\ref{fig:CHEOPS+TESS} (grey: {\em TESS}, rainbow colours: {\em CHEOPS}).
        {\em Right:}
        RV curve model fit (black line with $\pm 1 \sigma$ uncertainty marked with a grey shadowed area) and CARMENES+MAROON-X data with the same colours as in Fig.~\ref{fig:CARMENES+MAROON-X_gap}.
        Error bars include original RV uncertainties (opaque) and jitter added in quadrature (semi-transparent).
        }
        \label{fig:RV+transit_phase}
\end{figure*} 

We combined the CARMENES and MAROON-X RV data with the {\em CHEOPS} and {\em TESS} photometric transit data. 
As in \citet{Trifonov2021}, we did not use HARPS or HIRES RV data, nor any of the plethora of noisier photometric data sets collected for the transit analysis (LCOGT, MuSCAT2, etc.) or by us for the stellar $P_{\rm rot}$ determination.
As already noticed by \citet{Trifonov2021}, the GLS periodograms of the CARMENES and MAROON-X data are dominated by the planet signal at $\sim$1.47\,d and its 1\,d alias at $\sim$3.14\,d (see their Fig.~S4).
Once the planet signal is subtracted, two signals with low significance of 1--10\,\% remain.
One of them is the yearly alias at $\sim$360\,d, which is visible in the RV window, while the other signal at $\sim$100\,d may correspond to a stellar activity cycle or twice $P_{\rm rot}$ (Sect.~\ref{sec:Prot} and below).

We implemented four different models, which are sorted by increasing complexity in Table~\ref{tab:log-e}, from one planet in circular orbit to one planet in eccentric orbit and a GP associated with the stellar photometric variability and applied to the RVs.
As for the stellar $P_{\rm rot}$ analysis (Sect.~\ref{sec:Prot}), the fits of the four models were performed using {\tt juliet}.
For the dynamic nested sampling, we used {\tt dynesty}, which is a generalisation of the nested sampling algorithm in which the number of ``live points'' varies to allocate samples more efficiently \citep{Higson2019}. 
For double checking the results obtained with {\tt juliet}, we also used the {\tt Exo-Striker}\footnote{\url{https://github.com/3fon3fonov/ exostriker}} exoplanet toolbox \citep{Trifonov2019, Trifonov2021}, a free Python tool with a fast graphical user interface for maximum productivity.

We modelled the planetary transits from the flattened photometric data to measure the orbital period ($P$) and relative planet-to-star size ($p \equiv R_{\rm b}/R_\star$), the time of transit centre of the planet ($t_{0, {\rm b}}$), the inclination of the planetary orbital plane ($i_{\rm b}$), and the star-planet separation-to-radius ratio ($a_{\rm b}/R_\star)$. 
The {\tt juliet} and {\tt Exo-Striker} tools use the {\tt batman} package \citep{Kreidberg2015} to this end. 
The stellar limb-darkening coefficients (quadratic law), $q_1$ and $q_2$, were parameterised following \cite{Kipping2013}.
We tested using as $q_1$ and $q_2$ priors the output of the Limb Darkening Calculator\footnote{\url{https://exoctk.stsci.edu/limb\_darkening}} with the star's $T_{\rm eff}$, $\log{g}$, and [Fe/H] from Table~\ref{tab:star}, Kurucz ATLAS9 models, and quadratic limb darkening profiles.
For the test, we also used the SDSS $i'$ bandpass response function as a proxy for those of {\em TESS} ($T$) and {\em CHEOPS}.
However, probably due to width of the space mission  response functions encompassing ranges of 4000--5000\,{\AA}\footnote{{\em TESS}: circa $R$, $I$, and $z'$, \url{https://heasarc.gsfc.nasa.gov/docs/tess/the-tess-space-telescope.html\#bandpass};
{\em CHEOPS}: circa $B$, $V$, $R$, and $I$, \url{https://www.cosmos.esa.int/web/cheops/performances-bandpass}}, the stellar density posterior, $\rho_\star$, did not match its prior from $R_\star$ and $M_\star$ in Table~\ref{tab:star} (see below).
Therefore, we eventually used uniform priors between 0 and 1 for the quadratic limb-darkening parameters.
As a sanity check, we compared our $q_1$ and $q_2$ parameters with those calculated with {\tt limb-darkening}\footnote{\url{https://github.com/nespinoza/limb-darkening}}, a Python code developed by \citet{EspinozaJordan2015}. 
We used the stellar parameters in Table~\ref{tab:star}, the {\em TESS} and {\em CHEOPS} response functions from the Filter Profile Service of the Spanish Virtual Observatory\footnote{\url{http://svo2.cab.inta-csic.es/theory/fps/}} \citep{Rodrigo2012,RodrigoSolano2020}, and the {\em A100} fitting technique (i.e., limb-darkening coefficients from ATLAS models and interpolating 100 $\mu$-points with a cubic spline as in \citealt{ClaretBloemen2011}).
The quadratic parameters computed with {\tt limb-darkening} ($q_{1,TESS}$ = 0.37, $q_{1,CHEOPS}$ = 0.46, $q_{2,TESS}$ = 0.16, $q_{2,CHEOPS}$ = 0.21) are identical within uncertainties to the ones determined by us with uniform priors between 0 and~1, which a posteriori justifies our approach.

In the fits with {\tt juliet}, instead of determining the planet's relative radius and impact parameter ($b \equiv (a_{\rm b}/R_\star)\cos i_{\rm b}$), we used the parameterisation of \citet{Espinoza2018} and \citet{Espinoza2019}, and determined $r_1$ and $r_2$, which vary between 0 and 1 and are defined to explore the physically meaningful ranges for $R_{\rm b}/R_\star$ and $b$. 
As \citet{Trifonov2021}, we adopted dilution factors, $D_{\it TESS}$ and $D_{\it CHEOPS}$, of 1.0, which translates into no contamination in the (relatively large) \textit{TESS} and \textit{CHEOPS} photometric apertures that may mimic a possible planetary transit (see Sect.~\ref{sec:companions} for a companion search). 
We added a photometric jitter to the nominal flux error bars, $\sigma_{\it TESS}$ and $\sigma_{\it CHEOPS}$, for symmetry with the RV jitter parameters, $\sigma_{\rm CARMENES}$ and $\sigma_{\rm MAROON-X}$, although the inclusion of the eight additional parameters (one from {\em TESS}, seven from {\em CHEOPS}) barely modifies the results of the fits.
We defined a prior on the stellar density, $\rho_{\star}$, instead of the scaled semi-major axis of the planets, $a_{\rm b} / R_\star$. 
For the periodicity associated with the 1pl+GP and 1pl+$e$+GP models, we used a Gaussian $P_{\rm rot,GP}$ prior centred on 49.9\,d and width 10.0\,d from the photometric analysis in Sect.~\ref{sec:Prot}.
We used a quasi-periodic kernel introduced by \cite{Foreman-Mackey2017} of the form:

\begin{equation}
\label{eq:GP_BCLProt}
k_{i,j}(\tau) = \frac{B_{\rm GP}}{2+C_{\rm GP}} e^{-\tau/L_{\rm GP}} \left[1 + C_{\rm GP} + \cos{ \frac{2\pi \tau}{P_{\rm rot}} } \right],
\end{equation} 

\noindent where $\tau= |t_{i}-t_{j}|$ is the time lag, $B_{\rm GP}$ and $C_{\rm GP}$ define the amplitude of the GP, $L_{\rm GP}$ is a timescale of the amplitude modulation of the GP, and $P_{\rm rot}$ is the period of the quasi-periodic modulations. 
In order to simplify the GP equation, we fixed the parameter $C_{\rm GP}$ to 0 and, therefore:

\begin{equation}
\label{eq:GP_BLProt}
k_{i,j}(\tau) = \frac{B_{\rm GP}}{2} e^{-\tau/L_{\rm GP}} \left[ 1+ \cos{ \frac{2\pi \tau}{P_{\rm rot,GP}} }  \right].
\end{equation} 

\noindent Finally, for the non-circular models, we set uniform priors on eccentricity, $e$, from 0.00 to 0.15 (three times the upper limit of \citealt{Trifonov2021}) and on the argument of periapsis, $\omega$ (periastron angle), from 0 to $2\pi$.

Table~\ref{tab:log-e} summarises the four different models with the corresponding values of Bayesian log-evidence, $\ln{\mathcal{Z}}$ \citep{Jeffries1961, Trotta2008} and number of parameters, $N_{\rm par}$.
As in \citet{Buchner2014}, we adopted the scale of \citet{Jeffries1961}: a Bayesian log-evidence difference, $\Delta \ln{\mathcal{Z}}$, above 2.0 is ``decisive'', above 1.5 ``very strong evidence'', between 1.0 and 1.5 ``strong evidence'', and between 0.5 and 1.0 ``substantial evidence''.
In case of $\Delta \ln{\mathcal{Z}} < 1.0$, one must keep the simplest model with the smallest $N_{\rm par}$.
In previous works on CARMENES RV follow-up of transiting planets detected by {\em TESS}, even more conservative criteria of $\Delta \ln{\mathcal{Z}} >$ 2.5 or 5.0 for decisive difference between models has been used \citep{Kossakowski2021, GonzalezAlvarez2022, Kemmer2022}.
Such conservative criteria in Bayesian analyses applied to exoplanets can be traced back to previous works \citep[e.g.,][]{Gregory2005, Feroz2011}.

The $\ln{\mathcal{Z}}$ of the four models vary within a small range, but the models with the highest $\ln{\mathcal{Z}}$ are 1pl+GP and 1pl+$e$+GP, which are decidedly better than 1pl and 1pl+$e$ ($|\Delta \ln{\mathcal{Z}}| \sim 15$).
Since the Bayesian log-evidences of the two best models differentiate by just $|\Delta \ln{\mathcal{Z}}| \sim 0.5$, we selected 1pl+GP, with a lower number of parameters than 1pl+$e$+GP, as our working model.

The rms of the 1pl+GP observed-minus-calculated (O-C) residuals are 693\,ppm and 411\,ppm for {\em TESS} and {\em CHEOPS}, and 1.22\,m\,s$^{-1}$ and 0.74\,m\,s$^{-1}$ for CARMENES and MAROON-X, respectively.
The phase-folded {\em TESS} + {\em CHEOPS} and CARMENES + MAROON-X data with the corresponding best model fit are shown in Fig.~\ref{fig:RV+transit_phase}, while the corner plot depicting the most relevant posterior distributions and derived parameters is shown in Fig.~\ref{fig:cornerplot}.
In total, the 1pl+GP model has one stellar, 22 photometry instrumental, 14 RV instrumental, four GP, and seven planet parameters, of which five were fixed ($e$, $\omega$, $D_{\it TESS}$, $D_{\it CHEOPS}$, and $C_{\rm GP}$), which makes 43 free parameters.
The priors, posteriors, units, and description of each parameter are given in Table~\ref{tab:priors+posteriors}.
See \citet{Espinoza2022} for a recent application of {\tt juliet} and an explanation of the {\em CHEOPS} and {\em TESS} $\mu$ and $\sigma$ units ($m_{\rm flux}$ and $\sigma_{\rm w}$ in their nomenclature).
According to \citet{Espinoza2018}, the extra jitter ($\sigma_{\rm w}$) is the term added in quadrature to each of the error bars of data points for speeding up the computation of the model log-evidence (and, therefore, the covariance matrix is diagonal and the noise is white).
The extra RV jitter terms added when no GP is included get significantly higher than the values in Table~\ref{tab:priors+posteriors}.
For example, $\sigma_{\rm CARMENES}$ gets as high as $1.41 \pm 0.19$\,m\,s$^{-1}$ in the 1pl (no GP) model, which is in line with the typical {\rm scatter} of CARMENES VIS RV measurements \citep{Reiners2018} and the actual rms measurement at 1.22\,m\,s$^{-1}$.

The most relevant fitted parameters for the 1pl+GP model are tabulated again, together with derived and additional planet parameters, in Table~\ref{tab:planet}.
The planet radius and mass are $R_{\rm b} = 1.343^{+0.063}_{-0.062}\,R_\oplus$ and $M_{\rm b} = 3.00^{+0.13}_{-0.13}\,M_\oplus$, respectively, from which we derive values of bulk density, surface gravity, and escape velocity that are only slightly larger than Earth ($\rho_{\rm b} \sim 1.2\,\rho_\oplus$, $g_{\rm b} \sim 1.7\,g_{\oplus}$, $v_{\rm e,b} \sim 1.5\,v_{{\rm e,}\oplus}$).
However, the instellation is over 40 times higher, which translates into a Bond albedo-corrected $T_{\rm eq}$ higher than on Earth (Sect.~\ref{sec:atmosphere}).
Apart from the parameters enumerated above, we also tabulate the transit probability, $\mathbb{P}_{\rm transit}$, transit depth, $\Delta_{\rm transit}$, transit duration from the first to the last contact, $T_{14}$, and the ingress and egress duration, $T_{12} = T_{34}$, computed as \citet{Winn2010}. 

Only displayed in Table~\ref{tab:priors+posteriors}, the value of $P_{\rm rot,GP}$ from the applied GP to the RVs, of $53.5 ^{+6.5} _{-4.2}$\,d, is consistent within $1\sigma$ with $P_{\rm rot,phot}$ from LCOGT+OSN+TJO optical photometry, of $49.9 \pm 5.5$\,d, as expected from the normal prior for $P_{\rm rot,GP}$ based on $P_{\rm rot,phot}$.
The GP parameter that accounts for the time scale of the cyclic variations, $L_{\rm GP} = 90 ^{+126} _{-46}$\,d, is in agreement with the period of 131\,d reported by \citet{Trifonov2021} and the RV signal at $\sim 100$\,d.
This long periodicity may actually be real and correspond to an activity cycle or the time scale of photospheric spots. 
On the other hand, the GP parameter that accounts for the activity semi-amplitude, $B_{\rm GP} = 2.21 ^{+1.26} _{-0.70}$\,m\,s$^{-1}$, is also in agreement with the expected RV semi-amplitude from the $\log{R'_{\rm HK}}$ relation of \citet{SuarezMascareno2018}, in the 1.0--2.5\,m\,s$^{-1}$ range,
and with the low amplitude of photometric variability \citep{Saar1998, Jeffers2022}.
As a sanity check, we also modelled our data under the 1pl+GP scenario using different $P_{\rm rot,GP}$ priors: normal around 130\,d and uniform from 50\,d to 130\,d.
In all cases, the fitted and derived parameters were identical within $1\sigma$ uncertainties and, therefore, independent from the stellar cyclic variation.

Finally, we compared the 43 parameters in common of the two models that are decisively best, namely 1pl+GP (our working model) and 1pl+$e$+GP.
In all cases, the differences between the parameters of the two models are well within $1 \sigma$.
These differences also apply to the derived parameters in Table~\ref{tab:planet}.
In particular, the differences in $R_{\rm p}$ and $M_{\rm p}$ of the two models are in the fourth significant figure, about three orders of magnitude smaller than the uncertainties.
The only two parameters that significantly differ between the two models are, obviously, $e$ and $\omega$.
For the 1pl+$e$+GP model, we measured $\omega = 134 ^{+84} _{-49}$\,deg and $e = 0.0116 ^{+0.0102} _{-0.0072}$ (they were fixed to 0 in the 1pl+GP model).
From this eccentricity value, we can infer a $1 \sigma$ upper uncertainty $e < 0.022$ for Gl~486\,b. 
This upper limit is more constraining by a factor two than the value determined by \citet{Trifonov2021}, who imposed $e < 0.05$.

There might also be a concern about the pre-detrending of the {\em CHEOPS} transit photometry previously to the joint transit+RV fit.
In our case, we first detrended the {\em CHEOPS} light curves using the {\tt add\_glint} function of {\tt PyCheops} (basically a cubic spline against the spacecraft roll angle -- Sect.~\ref{sec:cheops}) and then used these detrended light curves in our joint fit with {\tt juliet}.
Although such a pre-detrending approach is common in the literature, it may result in underestimating the uncertainties on the transit parameters (and derived planet  parameters, including the radius) by not accounting for the covariances between the detrending and transit parameters.  
Therefore, we modelled the correlated noise in the {\em CHEOPS} light curves simultaneously with the joint transit+RV fit and compared with the results in Table~\ref{tab:planet}.
For that, we re-run a modified 1pl+GP model by using a second GP as in \citet{Leleu2021}.
They used a Mat\'ern-3/2 kernel against roll angle plus a polynomial in time and $x$-$y$ photocentroid position, while we used an exponential-squared kernel only in linear time and the seven {\em CHEOPS} light curves.
For each light curve, we added two additional parameters (i.e., 14 in total) with very wide log-uniform priors between $10^{-6}$ and $10^{6}$ ($\sigma_{CHEOPS\,i,{\rm GP}}$), and between $10^{-3}$ and $10^{3}$ ($\rho_{CHEOPS\,i,{\rm GP}}$).
We compared the posteriors of the rest of parameters with the results of the 1pl+GP fit in Table~\ref{tab:priors+posteriors} and, again, found no appreciable differences within $1\sigma$.
This result is in line with the findings of \citet{Lendl2020}, who also compared the results from applying different GP kernels and decorrelation techniques, as well as pre-detrending with {\tt PyCheops}, and ``obtained values that are fully compatible''.
All in all, Gl~486\,b is a well-characterised warm planet at the boundary between exo-Earths and super-Earths.

\subsection{Search for additional companions}
\label{sec:companions}

\begin{figure*}
    \centering
    \includegraphics[height=0.40\textwidth]{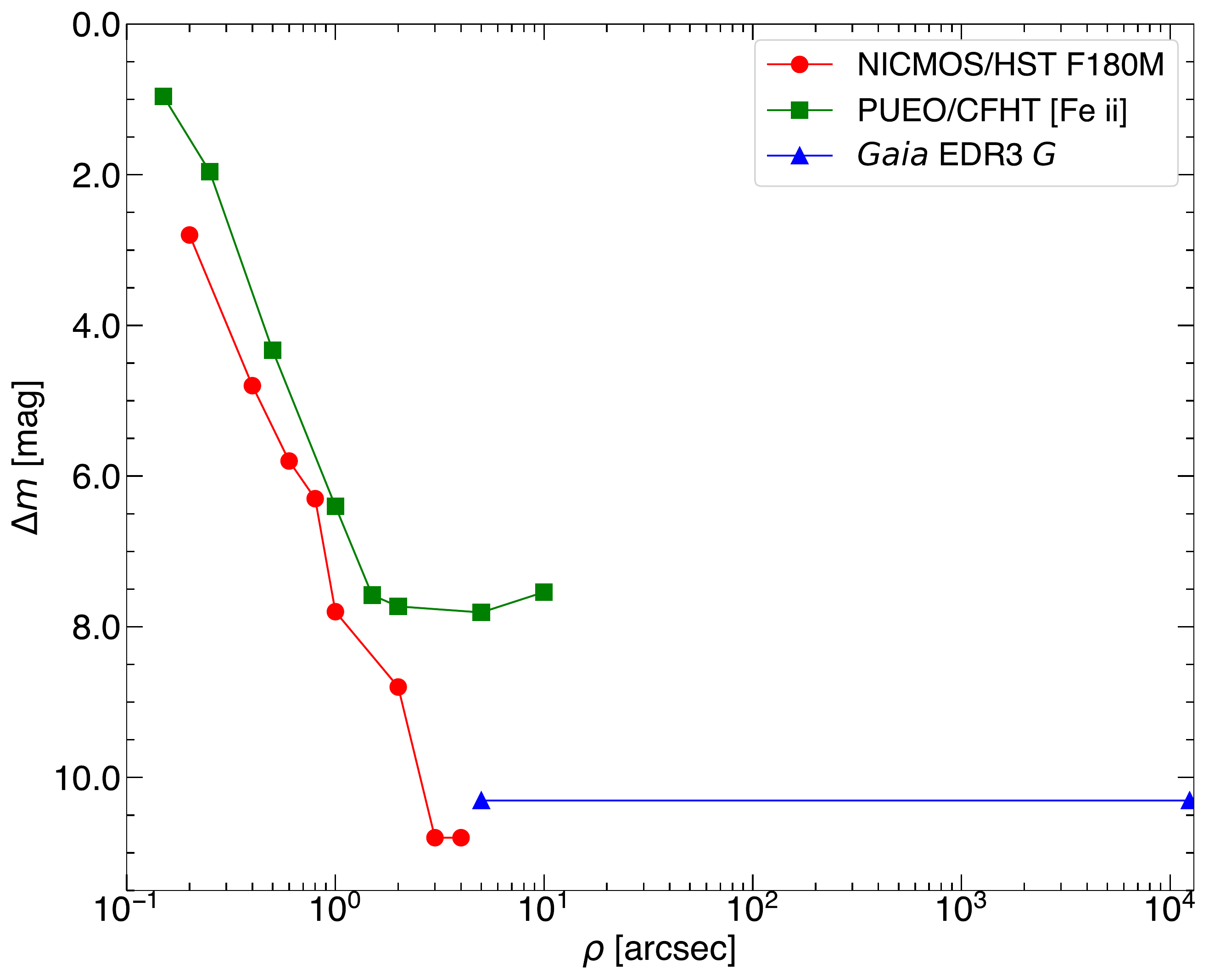} 
    \includegraphics[height=0.40\textwidth]{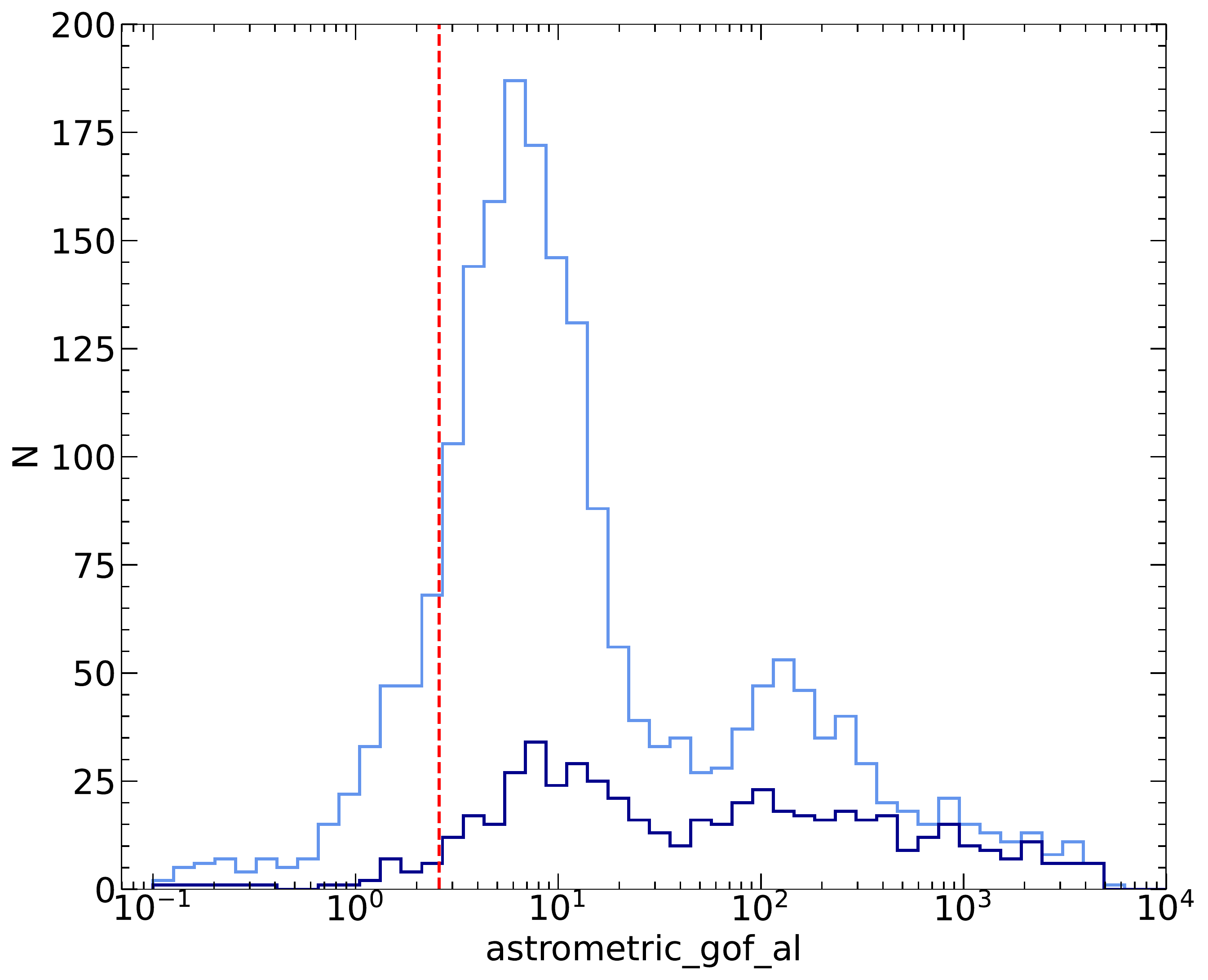} 
    \caption{{\em Left}: Contrast curves of Gl~486 from \citet[][NICMOS/{\em HST} F180M, red circles]{Dieterich2012}, \citet[][PUEO/CFHT {[Fe~{\sc ii}]}, green squares]{WardDuong2015}, and our common-parallax and proper-motion search at wide angular separations with {\em Gaia} EDR3 data (blue triangles).
    {\em Right}: Distribution of {\em Gaia} EDR3 goodness of fit statistic of model with respect to along-scan observations ({\tt astrometric\_gof\_al}) of over 2200 Carmencita M dwarfs.
    Light blue: all Carmencita stars; dark blue: Carmencita stars with companions at less than 5\,arcsec \citep{Caballero2016b,CortesContreras2016,Jeffers2018,Baroch2018,Baroch2021}.
    The vertical dashed line marks the value for Gl~486. 
    }
    \label{fig:contrast+Gaia} 
\end{figure*}

Due to its proximity, the star Gl~486 has been the subject of several searches for close companions, sensitive to very low-mass stars, brown dwarfs, or planets.
The results of the Doppler surveys by \citet{MarcyBenitz1989} and \citet{Davison2015} were not useful for this purpose because they collected only one RV point each.
\citet{Bonfils2013}, although they collected four RV points with HARPS, set a preliminary upper limit to the short-term RV scatter of a few m\,s$^{-1}$.
\citet{Jodar2013} and \citet{Rodriguez2015} also looked for companions (and debris discs) around Gl~486 with the FastCam lucky imager in the red optical and with the ESA {\em Herschel} space mission in the mid-infrared, both with null results.
However, the imaging observations by \citet{Dieterich2012} and \citet{WardDuong2015} were more restrictive for the presence of close companions, as illustrated by the left panel of Fig.~\ref{fig:contrast+Gaia}.
\citet{Dieterich2012} used NICMOS on the {\em Hubble} with the F180M near-infrared filter, and \citet{WardDuong2015} used archival adaptive optics imaging obtained with the KIR infrared imager of AOB/PUEO at the Canada-France-Hawai'i Telescope and the [Fe~{\sc ii}] narrow filter at 1.6\,$\mu$m.
Both of them established magnitude differences and angular separation limits for which hypothetical companions could be ruled out for Gl~486, which ranged from 0.96\,mag at 0.15\,arcsec (1.2\,au) to 10.8\,mag at 3.0--4.0\,arcsec (24--32\,au).
Besides, \citet{WardDuong2015} complemented their AOB/PUEO survey with a common proper motion search on SuperCOSMOS digitised photographic plates \citep{Hambly2001} to extend their limits at 7.5--7.8\,mag from 4.0\,arcsec to 19.9\,arcsec (161\,au).
According to D.~Ciardi (priv.~comm.), beyond a few arcseconds, it will be very hard to beat {\em Hubble} without spending a very large amount of time on Keck, Palomar, Lick, or Paranal observatories with adaptive optics facilities.  

\begin{figure*}
    \centering
    \includegraphics[height=0.40\textwidth]{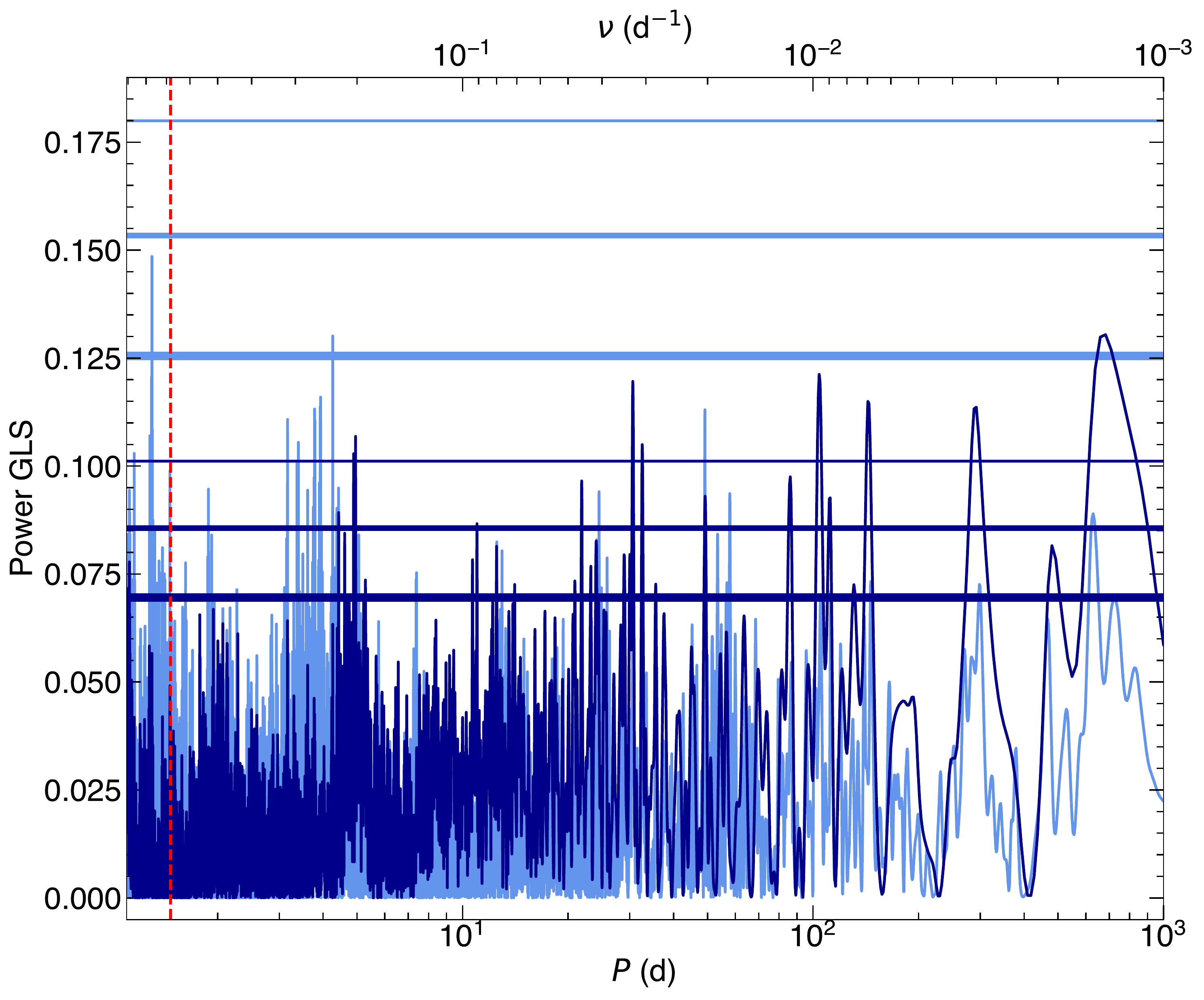} 
    \includegraphics[height=0.40\textwidth]{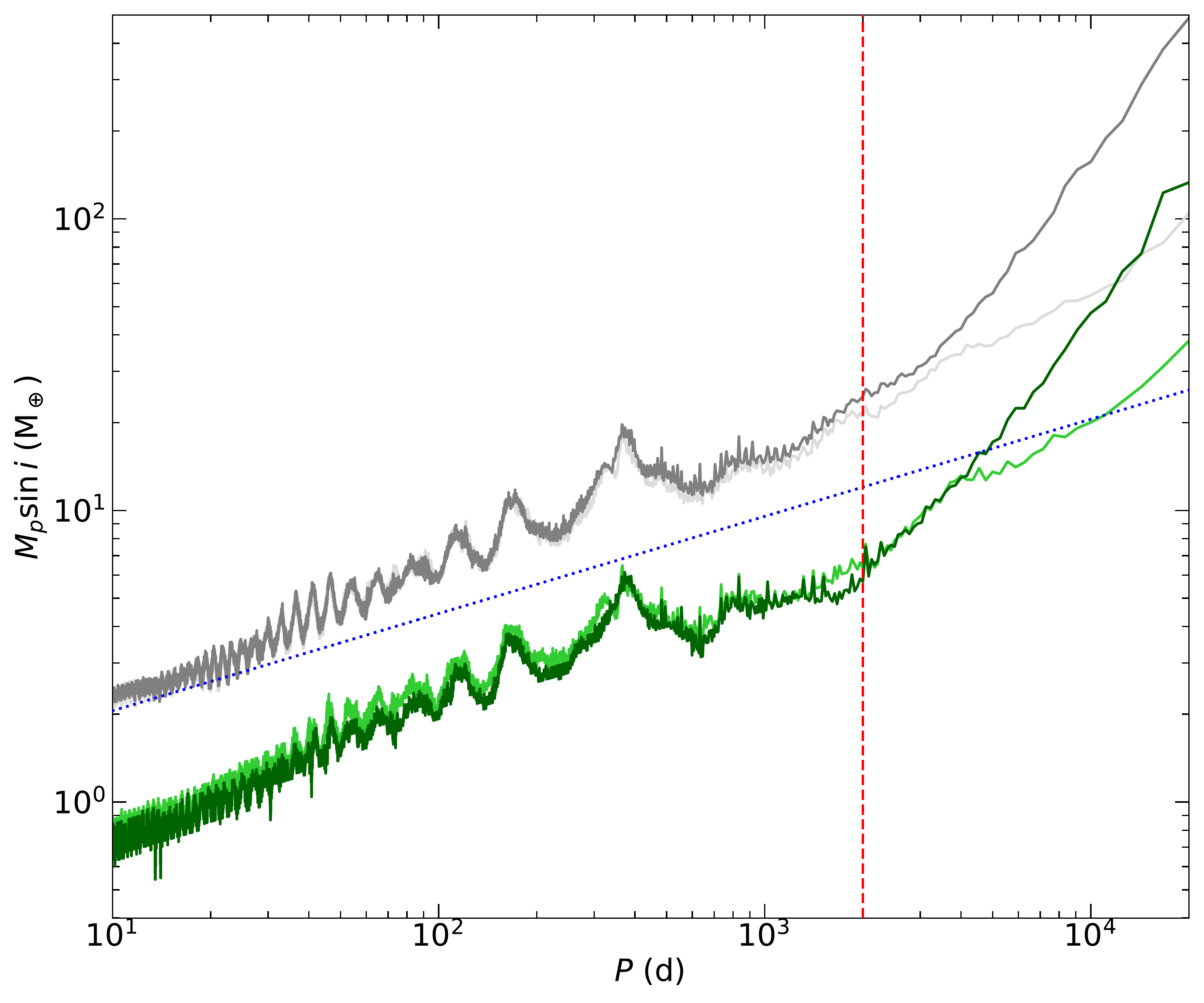}     
    \caption{Signal injection on RV data and planet retrieval.
    In both panels, the CARMENES+MAROON-X and CARMENES+MAROON-X+HARPS+HIRES data sets are plotted with dark and light colours, respectively.
    {\em Left:} GLS periodogram of the RV residuals of the 1pl model (blue).
    Horizontal lines of different thickness mark the 0.1\,\%, 1\,\%, and 10\,\% false alarm probabilities, from top to bottom, while the red vertical line marks the orbital period of Gl~486\,b.
    Compare with panels A--H of Fig.~S4 of \cite{Trifonov2021}.
    {\em Right:} Detection limits on minimum mass following \citet{Bonfils2013} for the RV data of Gl~486 before (grey) and after (green) subtracting out the 1pl model.
    Any planet above the green lines is excluded with 99.9\,\% confidence. 
    The blue dashed line is the minimum mass corresponding to a semi-amplitude of 1.22\,m\,s$^{-1}$ ($M_{\rm p} \sin{i} \propto M_*^{2/3} P^{1/3}$).
    The red vertical line indicates the approximate time baseline of the CARMENES+MAROON-X RV data set at about 2000\,d. }
    \label{fig:injection} 
\end{figure*}

We also performed our own multiplicity analysis with {\em Gaia} EDR3 data.
First, we searched for any previously undetected close multiplicity. 
We analysed the distribution of four representative {\em Gaia} EDR3 astrometric quality indicators of over 2200 Carmencita stars:
Goodness of fit statistic of model with respect to along-scan observations ({\tt astrometric\_gof\_al}),
along-scan chi-square value ({\tt astrometric\_chi2\_al}),
excess noise of the source ({\tt astrometric\_excess\_noise}), and
significance of excess noise ({\tt astrometric\_excess\_noise\_sig}).
Because of the similarity of the four {\em Gaia} EDR3 astrometric quality indicators, we plot in the right panel of Fig.~\ref{fig:contrast+Gaia} only the distribution of {\tt astrometric\_gof\_al}.

Carmencita is the CARMENES input catalogue of M dwarfs observable from Calar Alto, from which we selected the CARMENES survey targets, including Gl~486: the brightest stars in their spectral type bin and without physical or visual companions at less than 5\,arcsec \citep{Caballero2016b, AlonsoFloriano2015, Reiners2018}.
Close binaries with angular separations $\rho \sim$ 0.05--0.50\,arcsec unresolved by {\em Gaia}, but known to have companions (e.g., resolved instead with adaptive optics, lucky imaging, speckle, or with {\em Hubble}, or unresolved with imaging but known to be long-term spectroscopic binaries), have in general relatively large astrometric quality indicator values.
The low values of Gl~486 are typical of stars with no companions at close separations.
This is in line with its re-normalised a posteriori mean error of unit weight error ({\tt RUWE}) of 1.094, below the conservative value for stars with well-behaved astrometric solutions at $\sim$1.4 \citep{Arenou2018, Lindegren2018, Cifuentes2020, Penoyre2022}, and the previous non-detections by \citet{Dieterich2012}, \citet{Jodar2013}, \citet{Rodriguez2015}, and \citet{WardDuong2015}.

Next, we searched for {\em Gaia} common parallax and proper motion companions to Gl~486 at wide separations not explored before, as in \citet{Montes2018} and \citet{Cifuentes2021}.
In particular we searched between 5\,arcsec (there are no closer {\em Gaia} EDR3 sources) and the angular separation that corresponds to a projected physical separation of 100\,000\,au \citep{Caballero2009} with the following criteria: 
proper motion ratio $\Delta \mu / \mu < 0.15$, proper motion position angle difference $\Delta PA < 15$\,deg (see Eqs.~1 and~2 in \citealt{Montes2018}), and distance ratio, $\Delta d \equiv |d_{\rm A} - d_{\rm B}| / d_{\rm A} < 0.10$ (i.e., within 0.8\,pc for this target).
This search was complete down to the {\em Gaia} EDR3 $G$-band completeness limit at 20.41\,mag \citep{GaiaSmart2021}, which translates into an absolute magnitude $M_G \approx$ 20.87\,mag at the distance of Gl~486 \citep{PecautMamajek2013, Cifuentes2020}.
This absolute magnitude corresponds to substellar objects at the L-T spectral type boundary.
Eventually we ruled out the presence of stellar and high-mass brown dwarf companions to Gl~486 from the limit of the {\em Hubble} observations at 24--32\,au up to 100\,000\,au.
At 8\,pc, this projected physical separation implies a search radius of 3.5\,deg and, with a total proper motion of 1100\,mas\,a$^{-1}$, projection effects of about 8\,mas\,a$^{-1}$, much lower than $\Delta \mu \sim$ 160\,mas\,a$^{-1}$ from the $\Delta \mu / \mu$ criterion. 
Therefore, we are not missing any possible wide companion. 

\begin{figure*}
    \centering
    \includegraphics[width=0.99\textwidth]{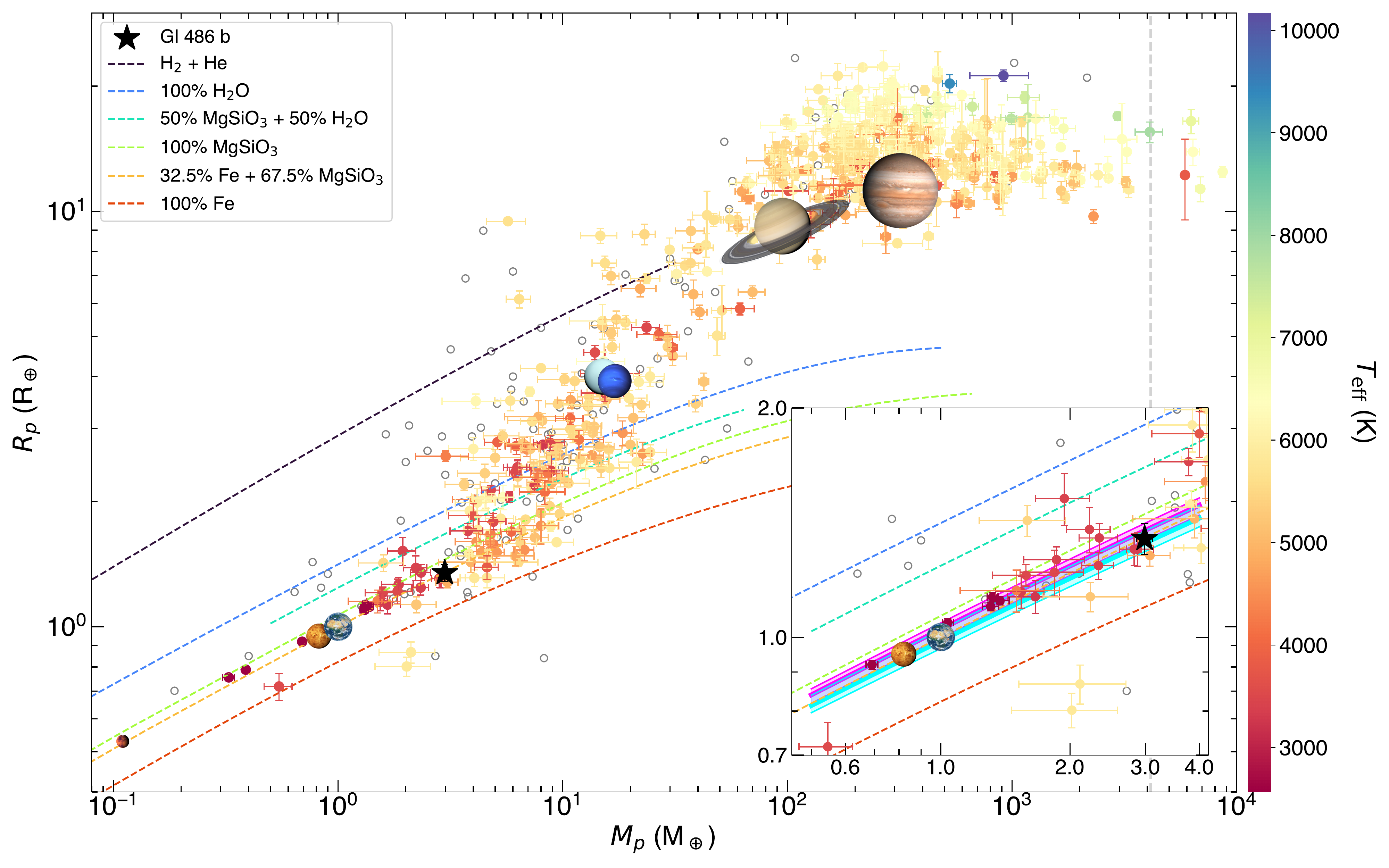}
        \caption{Mass-radius diagram of all currently known transiting exoplanets with mass determination (from RV or transit time variations), in comparison with the Solar System planets.
        Filled circles with error bars colour-coded by their host's $T_{\rm eff}$ are planets with mass and radius uncertainties less than 30\,\%, open grey circles are the others.
        The filled black star is {Gl~486\,b}.
        Dashed coloured curves are theoretical models of \citet{Zeng2019}, specified in the legend. 
        The Earth-like model is orange.
        Grey vertical dashed line: deuterium burning mass limit at 13\,$M_{\rm Jup}$ (``planet''-brown dwarf boundary; see \citealt{Caballero2018} and references therein).
        In the inset, we zoom around the smallest planets and add mass-radius relationships informed by stellar abundances (Sect.~\ref{sec:planetinteriorcharacterisation}). 
 	    We plot median and $1\sigma$ error regions following nominal relative abundances of Fe, Mg, and Si of the host star without (pink) and with (cyan) the empirical correction of \citet{Adibekyan2021} based on well-characterised super-Earths.
 	    The two outliers with very high densities and $M \sim$ 2.0\,$M_\oplus$ are \object{Kepler-1972}\,b and~c, which are two transiting planets with masses determined from transit time variations \citep{Leleu2022}.
 	    }
        \label{fig:MvsR}
\end{figure*}

Finally, we searched for additional planets in the system at much shorter separations with our RV data.
The GLS periodogram of the RV residuals of the 1pl+GP fit, displayed in the left panel of Fig.~\ref{fig:injection}, does not show any significant signal in the investigated frequency domain (which corresponds to periods in the 1.1--1000\,d range).
To estimate the sensitivity of our RV data to additional planets in the system, we computed detection limits following the injection and retrieval procedure presented by \citet{Bonfils2013} and used by \citet{Kossakowski2021}. 
In short, we injected signals to the actual CARMENES + MAROON-X RV data set corresponding to circular orbits ($e=0$) with increasing semi-amplitude $K$ until they were detected with a 99.9\,\% confidence. 
We performed the experiment both using the measured RVs and the residuals after subtracting a new 1pl model of Gl~486\,b with the 1pl+GP posteriors as narrow priors.
We did not use the 1pl+GP model in the experiment because the GP may erase any additional long-term signal besides the stellar rotation period. 
The test on residuals indicates that additional planets in the system above the mass range $\sim$2--5\,$M_{\oplus}$ can be excluded for periods within the time span of the data at $\sim$2000\,d, and above $\sim$10\,$M_{\oplus}$ for periods up to 10\,000\,d, although in this last case, eccentric orbits may impact the results.
These periods correspond, through Kepler's Third Law of planetary motion, to semi-major axes of $\sim$2.2\,au and 5.5\,au, respectively, which are 130--320 times greater than the orbital semi-major axis of Gl~486\,b.

Next, we extended the time baseline of the RV monitoring for a better investigation of periods longer than $\sim$2000\,d. 
For that, we added to our CARMENES and MAROON-X data sets the HIRES and HARPS RVs compiled by \citet{Trifonov2021}, which led us to extend the time span from 1967\,d (243 RVs) to 8529\,d (281 RVs).
Because of the greater noise of the HIRES and HARPS data with respect to CARMENES and MAROON-X, the improvement in the injection is appreciable only beyond 5000\,d.
With the additional data, we excluded planets with minimum masses $\sim$30\,$M_{\oplus}$ with periods up to 20\,000\,d ($\sim$11\,au).
We also considered adding to the new injection 12 additional RVs from the SOPHIE and ELODIE Archive\footnote{\url{http://atlas.obs-hp.fr/}}, but they are even noisier than those of HARPS and HIRES and overlap in time.
The
injection analysis is illustrated in the right panel of Fig.~\ref{fig:injection}.

As discussed below, Gl~486 will be observed again by {\em TESS} in Sector 50.
These new observations, together with the {\em CHEOPS} and {\em TESS} data used in this paper, photometric data from MuSCAT2/Telescopio Carlos S\'anchez and LCOGT presented by \citet{Trifonov2021}, and three new {\em CHEOPS} visits scheduled for 2022 will allow us to search for new low-mass planets in the system, especially through transit time variations.
This analysis will be part of a forthcoming publication.

\section{Discussion}
\label{sec:discussion}

\subsection{Comparison with other exoplanets}

For a proper comparison of Gl~486\,b with exoplanets from the literature, we compiled all the confirmed exoplanets with measured mass and radius.
For that, we started off with the complete list of exoplanets provided by the {Exoplanet Archive}\footnote{\url{https://exoplanetarchive.ipac.caltech.edu}}. 
Firstly, we discarded unconfirmed {\em Kepler} and {\em TESS} candidates by choosing {\tt soltype} == {\tt Published Confirmed}.
Secondly, we selected only the exoplanets discovered by transits, RV, or transit timing variations using the {\tt discoverymethod} column.
Next, we kept only those for which values of both mass ({\tt pl\_masse}) and radius ({\tt pl\_rade}) were tabulated.
Among them, we chose those whose masses were derived directly from observations and not inferred from their radii by selecting planets with tabulated densities {\tt pl\_denslim} == 0. 
Finally, we selected the default set of planetary parameters tabulated in the {Exoplanet Archive} by selecting {\tt default\_flag} == 1. 
Only for \object{Kepler-7}\,b, 
\object{Kepler~51}\,b, c, and~d, 
\object{Kepler~138}\,b, and \object{LTT~3780}\,b and~c 
we chose the parameters from a source different from the default choice.

Besides, we added by hand the uncertainties in stellar mass for the hosts of 23 planets, including the seven planets in the TRAPPIST-1 system, since they are incorrectly rounded when downloaded.
Additionally, we incorporated four planets not yet included in the {Exoplanet Archive}: 
\object{GJ~3929}\,b \citep{Kemmer2022}, \object{TOI-1759}\,b \citep{Espinoza2022}, and \object{TOI-1238}\,b and \object{TOI-1238}\,c \citep{GonzalezAlvarez2022}. 
In total, the sample contains {651} confirmed transiting exoplanets with mass and radius determination, displayed in Fig.~\ref{fig:MvsR}. 
Coloured symbols stand for planets with uncertainties in mass and radius less than 30\,\%.

As already mentioned by \citet{Trifonov2021}, Gl~486\,b falls in the expected region for rocky planets with an Earth-like composition, which is well traced towards Martian masses by Gl~637\,b and the seven planets in the TRAPPIST-1 system.
Gl~486\,b seems to be placed in the most massive extreme of the sequence of pure rocky planets.
Planets with greater mass are more scattered in the mass-radius diagram as they are probably made of varied mixtures of shallow-to-deep gas and ice envelopes, rocky mantles of various depths, and metallic cores of different sizes.

\begin{table*}
\centering
\small
\caption{Comparison of Gl~486\,b and Solar System telluric planets interior and atmospheric parameters$^a$.} 
\label{tab:solarsystem}
\begin{tabular}{l c ccc c cccc c}
\hline
\hline
\noalign{\smallskip}
Planet & Model$^b$ & $\log{m_{\rm upper}/M_p}$ & $m_{\rm mantle}/M_p$ & $m_{\rm core}/M_p$ & $R_{\rm core}/R_p$ & $A_{\rm Bond}$ & $T_{\rm eq}$ [K] & $T_{\rm surf}$ [K] & $\tau$ & $H$ [km] \\
\noalign{\smallskip}
\hline
\noalign{\smallskip}
            & MR-S & $-2.6 ^{+1.0}_{-2.5}$  & $0.76 ^{+0.15}_{-0.24}$   & $0.23 ^{+0.24}_{-0.14}$   & $0.46 ^{+0.14}_{-0.13}$  & \\
\noalign{\smallskip}
Gl~486\,b    & MRA-S & $-4.91 ^{+1.46}_{-0.98}$  & $0.83 ^{+0.10}_{-0.12}$   & $0.17 ^{+0.12}_{-0.10}$   & $0.399 ^{+0.082}_{-0.104}$  & \multicolumn{5}{c}{See text} \\
\noalign{\smallskip}
            & MRA-SH & $-4.5 ^{+1.5}_{-1.3}$  & $0.83 ^{+0.10}_{-0.11}$   & $0.17 ^{+0.12}_{-0.10}$   & $0.403 ^{+0.079}_{-0.102}$  & \\
\noalign{\smallskip}
\hline
\noalign{\smallskip}
Mercury     &   & ...                   & $\sim 0.15$               & $\sim 0.85$               & $0.828 \pm 0.012$  & 0.088     & 437     & ...       & ...   & ... \\
Venus       &   & $-4.00 \pm 0.01$      & $\sim 0.67$               & $\sim 0.33$               & $0.525 \pm 0.045$  & 0.76      & 229     & 737       & 160   & 15.9 \\
Earth       &   & $-3.62 \pm 0.03$      & $\sim 0.67$               & $\sim 0.33$               & $0.4536 \pm 0.016$  & 0.306     & 254     & 287       & 0.94  & 8.5 \\
Mars        &   & $-7.41 \pm 0.02$      & $\sim 0.85$               & $\sim 0.15$               & $0.539 \pm 0.012$ & 0.250     & 210     & 213       & 0.09  & 11.1 \\
\noalign{\smallskip}
\hline
\end{tabular}
\tablefoot{
\tablefoottext{a}{All parameters from the NASA Fact Sheets or computed by us except for the core-to-mantle mass ratios of Solar System planets, which actually are the iron mass fraction estimated by \citet{LoddersFegley1998}.
Actual Mercury $T_{\rm surf}$ ranges from about 100\,K on the dark side and at the bottom of deep craters to about 700\,K at the subsolar point during aphelion.}
\tablefoottext{b}{Models -- MR-S: baseline model with water steam upper layer (894\,000 samples); MRA-S: MR-S plus stellar abundance constraints (725\,000 samples); MRA-SH: MRA-S plus hydrated magma (224\,000 samples).
Nomenclature follows the ``data type-model setup'' structure.}
}
\end{table*}

While our relative radius error of Gl~486\,b matches the bulk for other well-characterised small transiting planets with $R_{\rm p} \lesssim 2\,R_\oplus$, it is one of the very few transiting planets at all sizes with an interferometric determination of its stellar host radius, namely 55~Cnc, HD~189733, HD~209458, HD~219134, HD~97658, and Gl~436 \citep{vonBraun2011, vonBraun2012, Boyajian2012, Boyajian2015, Ligi2016, Ellis2021}.
Due to the precision of the CARMENES and MAROON-X RV data, together with the relative brightness and very weak activity of Gl~486, our relative mass error is comparable to the most precise determinations available to date among low- and intermediate-mass planets with $M_{\rm p} \lesssim 100\,M_\oplus$, at about 4\,\%.
The sources and propagation of error of radius and mass for Gl~486\,b in particular and for transiting planets in general are detailed in Sect.~\ref{sec:errors}.

\subsection{Planet interior modelling}
\label{sec:planetinteriorcharacterisation}

Here we model the interior of Gl~486\,b with available data.
First, we compared the planet mass and radius from Sect.~\ref{sec:planetradiusandmass} to the stellar abundances from Sect.~\ref{sec:abundances}. 
If the planet reflected the relative abundances of refractory elements of its host star (i.e., Fe, Mg, Si), mass-radius curves informed by stellar abundances would fit the planet data \citep{Dorn2015}. 
As shown in the inset of Fig.~\ref{fig:MvsR}, this is the case of Gl~486\,b when the uncertainties of the planet parameters match the mass-radius relation informed by stellar proxies (pink).
Recently, empirical studies have demonstrated that planets are likely enriched in iron compared to their stellar abundance proxies \citep{Plotnykov2020, Adibekyan2021}. 
If we use the empirical correction suggested by \citet{Adibekyan2021}, the planetary data fit better the iron-enriched mass-radius relation (cyan). 

Second, we computed inference analysis of interior parameters.
Our 1D planetary interior model was based on the generalised Bayesian inference method of \citet{Dorn2015} and \cite{Dorn2017} with updates from \citet{DornLichtenberg2021}. 
We described the planet in hydrostatic equilibrium with three possible components: an upper layer, a mantle of variable composition, and an iron-dominated core.

In our model the upper layer is made of pure H$_2$O, for which we used the equation of state of \citet{Haldemann2020}. 
For Gl~486\,b, the $T_{\rm eq}$ of about 700\,K (see Table~\ref{tab:planet} and Sect.~\ref{sec:atmosphere}) forces any water to be present as a steam atmosphere. 
We assumed that the transit radius derived in Sect.~\ref{sec:planetradiusandmass} is at a pressure of $p_{\rm transit}=1$\,hPa (i.e. $10^{-3}$\,bar).
By assuming any volatile in the upper layer to be in the form of water, we provide upper limits on the possible amount of water.

Also in our model, the mantle is made of MgO, SiO$_2$, and FeO that form different minerals.
To compute stable mineralogy for a given composition, pressure, and temperature of the solid mantle, we used the {\tt Perple\_X} thermodynamical model of \citet{Connolly2009}, which employs equations of state from \citet{Stixrude2011}.
For the different components of the liquid mantle, we used a compilation of equations of state from \citet{Melosh2007}, \citet{Faik2018}, \citet{Ichikawa2020}, and \citet{Stewart2020} and the additive volume law to compute mixtures from \citet{DornLichtenberg2021}.
Finally, for the core we also used a compilation of different equations of state to account for different phases of iron \citep{Dorogokupets2017, Hakim2018, Ichikawa2020, Kuwayama2020, Miozzi2020}.
We allowed the mantle to vary in iron content, which is why an iron-rich mantle with no core was allowed in the posterior solution.
Whenever we allowed for water to be present in dissolved form in possible magma layers, we followed the approach of \citet{DornLichtenberg2021}, while the partitioning of water between the surface reservoir and the magma was determined by solubility laws \citep{Kessel2005, Lichtenberg2021, Bower2022}. 
We accounted for the fact that water increases melt fraction by lowering melting temperature \citep{Katz2003}. 
Also, the presence of water decreases melt density, which is nearly independent of pressure and temperature \citep{Bajgain2015}.

We tested three different planet interior scenarios in increasing order of complexity.
Scenario MR-S was the baseline model of a rocky interior with a thin volatile layer and no further restrictions.
In scenario MRA-S, we added the constraints from nominal relative stellar abundances (Fe, Mg, Si), which significantly reduced uncertainty on interior parameters and also made the mantle mass to significantly increase at the expense of the core mass.
The upper layer also became less massive.
However, this scenario did not take into account the abundance enrichment suggested by \citet{Adibekyan2021}, which is necessary to locate Gl~486\,b within a modelable $M_p$-$R_p$ trend. 
As a comparison, \citet{Demangeon2021} did not measure the LP~98--59 Mg and Si abundances directly, but instead adopted the mean abundances of Mg and Si of a large list of thin-disc stellar analogues as proxy.
Finally, in scenario MRA-SH, apart from adding the stellar abundance constraints, we allowed water to be not only present on the surface but to be dissolved in molten parts of the mantle magma ocean, if temperature and pressure conditions did not prevent to do so. 
This added complexity corrects possible water mass fractions by roughly an order of magnitude.  
Therefore, this addition, which was recently proposed by \citet{DornLichtenberg2021}, is critical for warm and hot planets such as Gl~486\,b: by neglecting deep water reservoirs in the mantle, bulk water estimates lead to underestimated values.
Unfortunately, the actual amount of water critically depends on the stellar proxy of refractory elements (and whether they are added as constraints), how water is modelled in the interior, and the atmosphere mass loss during the entire life of the system (as oxygen and hydrogen loss in the Venus atmosphere) and, therefore, on the incoming stellar flux, as well as the atmospheric XUV (5--920\,\AA) heating and eroding efficiency \citep{KastingPollack1983, Barabash2007, SanzForcada2011}.
While the mantle and core equations of state are based on the knowledge of planets of our Solar System, we must still learn more on M-dwarf stellar abundances and their impact on planet interior models, as well as on XUV radiation and its effect on planet atmospheres. 

\begin{figure}[]
 	\centering
 	\includegraphics[width=0.49\textwidth]{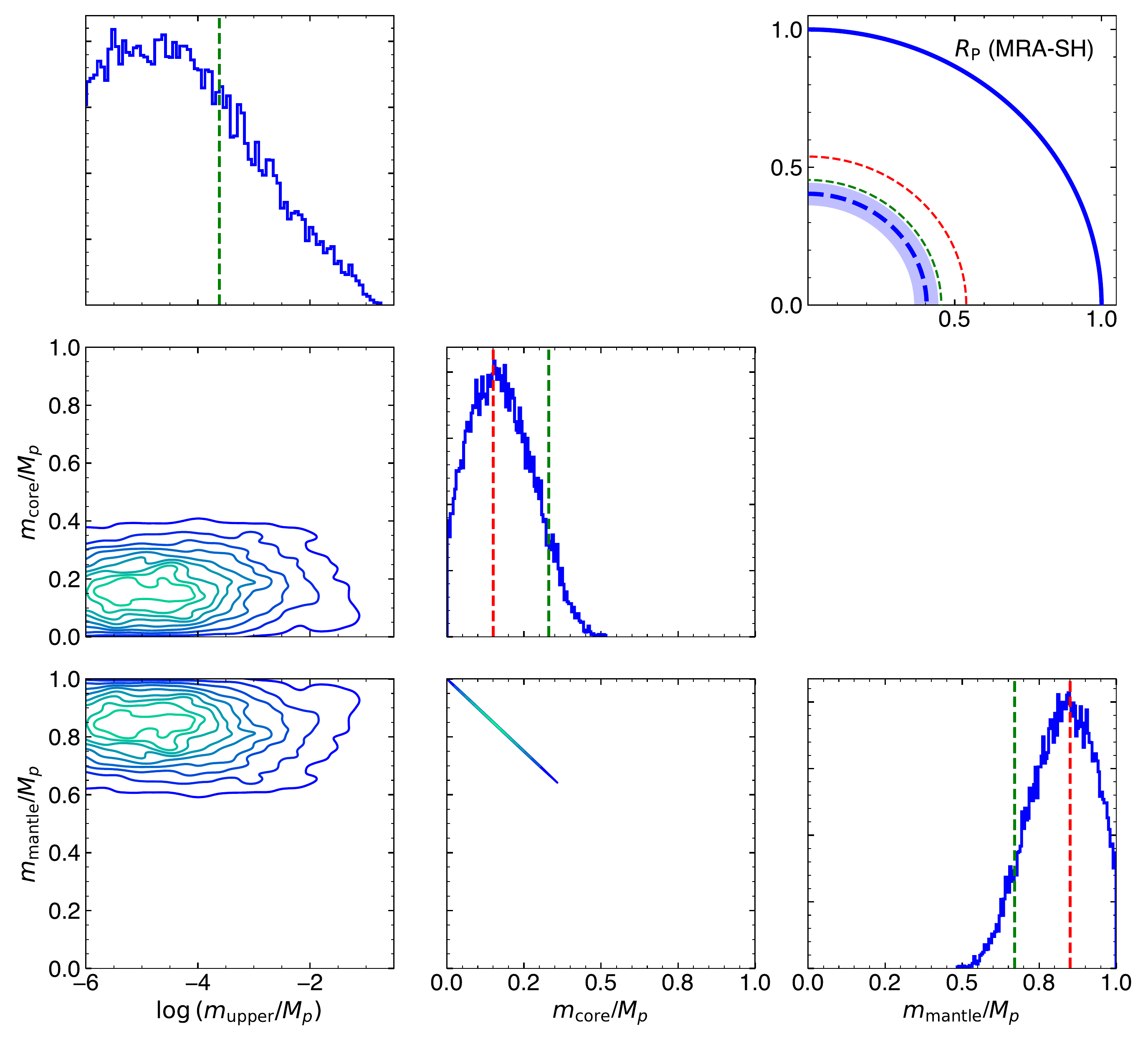}
 	\caption{One- and two-dimensional posterior distributions for inferred mass fractions of water $\log{m_{\rm upper}/M_p}$, mantle $m_{\rm mantle}/M_p$, and core $m_{\rm core}/M_p$ for the MRA-SH scenario.  
 	The top right panel shows the $R_{\rm core}/R_{\rm p}$ ratio of Gl~486\,b (dashed blue with $1\sigma$ dashed area) and of the four Solar System telluric planets. 
 	Green and red vertical lines denote the Earth and Mars values; comparisons to Mercury and Venus data must be done with caution.
 	Posterior distributions and radii for MR-S and MRA-S scenarios are displayed in Fig.~\ref{fig:interior-appendix}.}
    \label{fig:interior}
\end{figure}

Under the MRA-SH model, with the Gl~486\,b mass and radius, Fe, Mg, and Si relative stellar abundance constraints, and water allowed to dissolve in molten rock, we computed confidence regions for mass fractions of the water upper layer, $\log{m_{\rm upper}/M_p} = -4.47^{+0.70}_{-0.68}$, mantle $m_{\rm mantle}/M_p = 0.826 ^{+0.051}_{-0.054}$, and core, $m_{\rm core}/M_p = 0.174 ^{+0.054}_{-0.051}$.  
The core radius relative to the planet radius is $R_{\rm core} / R_{\rm p} = 0.404 ^{+0.040} _{-0.045}$.
The results for the MRA-SH model, together with the MR-S and MRA-S models, are shown in Table~\ref{tab:solarsystem}.
Our analysis is illustrated by Figs.~\ref{fig:interior} (MRA-SH) and~\ref{fig:interior-appendix} (MR-S and MRA-S).

In Table~\ref{tab:solarsystem}, we also compare Gl~486\,b to the telluric planets of the Solar System.
For the mass of the upper layer, $m_{\rm upper}$, we first estimated the mass of the atmospheres of Earth, Venus, and Mars from the definition of surface atmospheric pressure:

\begin{equation}
    p_{\rm surf} = \frac{F}{\mathcal{S}} = \frac{m_{\rm atm} g}{4 \pi R^2},
\end{equation}

\noindent where $g$ is the planet surface gravity and $R$ is the mean planet radius.
In the case of the Earth, our value of $m_{\rm atm}$ matches that measured by \citet{TrenberthSmith2005} at $5.1480 \cdot 10^{18}$\,kg.
However, this value is about 300 times lower than the total mass of the Earth hydrosphere, $m_{\rm hydro}$.
Therefore, for the Earth we instead estimated $m_{\rm upper}$ from the volume of Earth oceans from \citet{EakinsSharman2007} and the average sea water density (i.e. $m_{\rm upper} = m_{\rm hydro} + m_{\rm atm} \approx m_{\rm hydro}$).
We do not tabulate a $m_{\rm upper}$ for Mercury, as it has a very tenuous and highly variable atmosphere with a pressure level of just 1\,nPa ($10^{-14}$\,bar).

As in the case of the Earth, our models predict that Gl~486\,b has a solid inner core, a liquid outer core, a solid lower mantle, and a liquid upper mantle.
Since the most probable surface temperature is below the critical value for molten rocks (Sect.~\ref{sec:atmosphere}), the planet must also have a thin solid crust as an interface between the liquid upper mantle and the gaseous atmosphere\footnote{In the Earth, the solid lithosphere includes the crust and a thin part of the upper mantle above the asthenosphere.}.
However, because of the expected crust thinness, we did not include it in our analysis. 

\begin{figure*}
    \centering
    \includegraphics[width=0.49\textwidth]{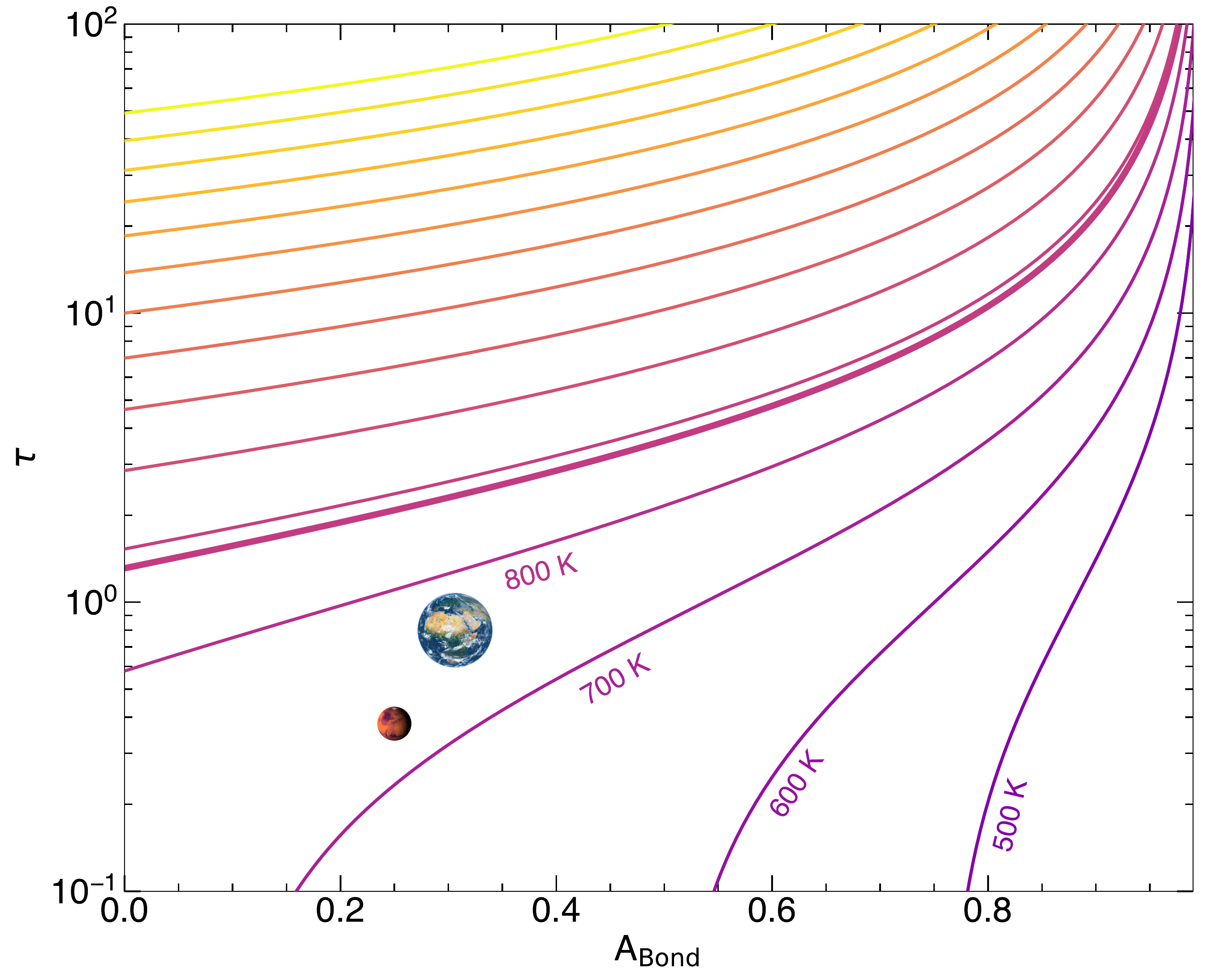} 
    \includegraphics[width=0.49\textwidth]{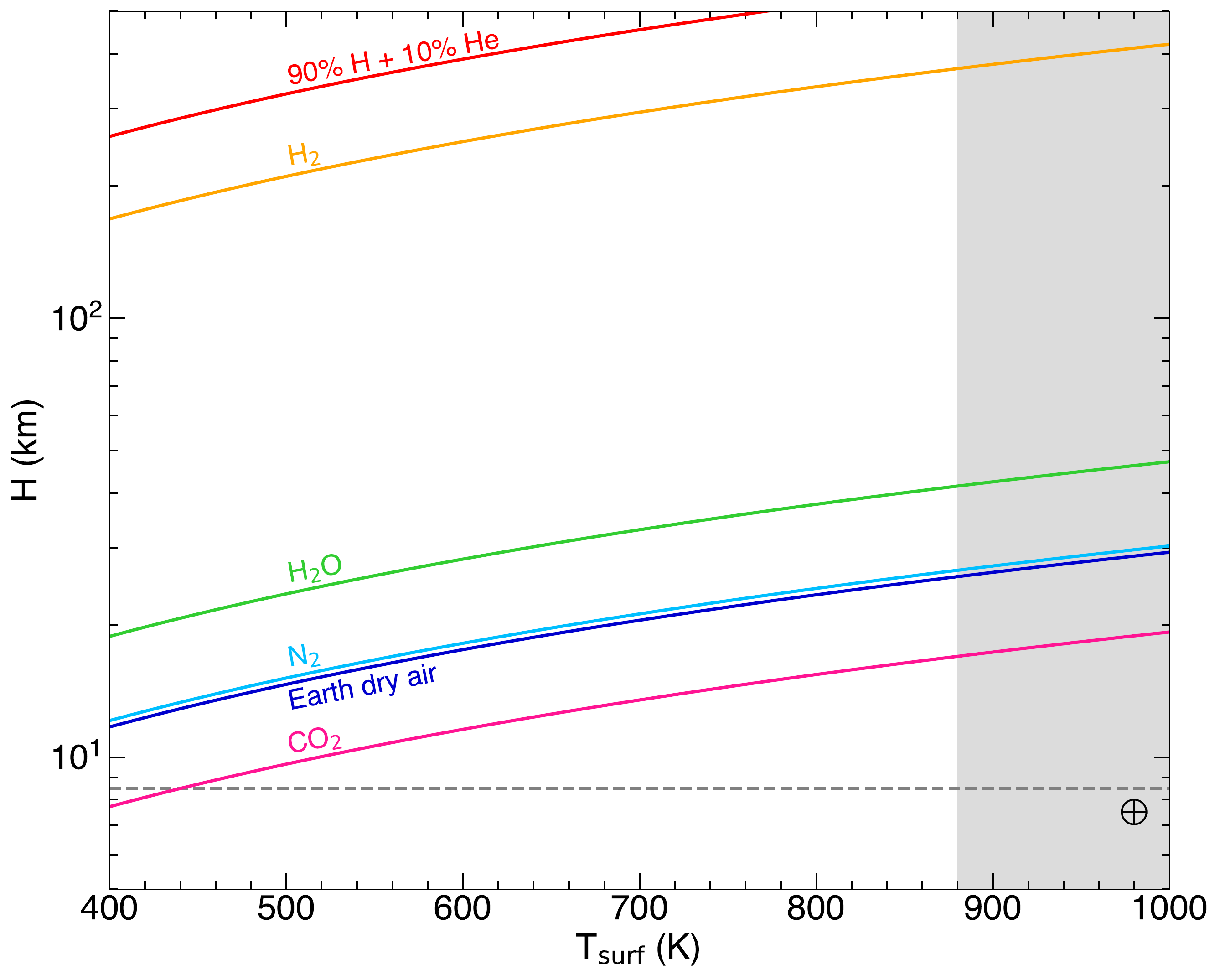} 
    \caption{{\em Left:}
    Surface temperature as a function of Bond albedo and atmospheric optical thickness, from 500\,K to 2000\,K in steps of 100\,K.
    The coolest isotherms are labelled.
    The isotherm of 880\,K, at which silicon-based surfaces melt, is drawn with a thick line.
    The Earth and Mars are indicated as a reference (but their $T_{\rm surf}$ do not correspond to those of Gl~486\,b because of different $L_\star$ and $a$; see Eq.~\ref{eq:Teq}). 
    Venus, with $\tau \approx$ 160, is out of limits.
    {\em Right:}
    Height scale as a function of surface temperature and mean molecular mass for different atmospheric compositions.
    From top to bottom: 
    90\,\% H + 10\,\% He (1.31\,u, red), 
    H$_2$ (2.02\,u, orange), 
    H$_2$O (18.0\,u, green), 
    N$_2$ (28.0\,u, cyan), 
    Earth dry air (29.0\,u, blue), 
    CO$_2$ (44.0\,u, magenta). 
    The dotted horizontal line indicates the Earth scale height ($H_\oplus$ = 8.5\,km), while the grey shaded area marks the surface temperature for rock volatilisation ($T_{\rm surf} \gtrsim$ 880\,K).
    }
    \label{fig:tsurf} 
\end{figure*}

The core-to-mantle mass ratio of Gl~486\,b, between 17:83 (MRA-S and MRA-SH scenarios) and 23:76 (MR-S), is similar to that of Mars, of about 15:85, and to the ratios measured for L~98--59\,b and c, and $\nu^{02}$~Lup~b, c, and d, at about 13:87 \citep{Delrez2021, Demangeon2021}.
The hypothetical iron core of Gl~486\,b could proportionally be as massive as those of the Earth and Venus, 33:67, only in the most extreme cases of our simulations, especially in the baseline MR-S scenario. 
However, the Gl~486\,b core relative size, between 0.40\,$R_{\rm p}$ (MRA-S and MRA-SH) and 0.46\,$R_{\rm p}$ (MR-S), resembles more to the Earth core relative size, of 0.454\,$R_{\rm p}$ but is far smaller than that of  Gl~367\,b, a well-characterised, dense, ultrashort-period sub-Earth around an M dwarf \citep{Lam2021}.
The Gl~486\,b upper layer mass strongly depends on the planet interior scenario, being more massive than Earth for the model MR-S and less massive than Venus for the models MRA-S and MRA-SH (Figs.~\ref{fig:interior} and~\ref{fig:interior-appendix}).
In any case, as mentioned before, the actual upper layer mass is also determined by the incoming XUV radiation.

\subsection{Prospects for atmospheric characterisation}
\label{sec:atmosphere}

Here we present possible atmospheric models of Gl~486\,b and discuss whether next observations, especially with {\em Webb}, will be able to differentiate them.
One of the key atmosphere parameters that is an input for atmosphere models is the scale height, $H$.
It is the increase in altitude for which the atmospheric pressure decreases by a factor of $e$ (Euler's number):

\begin{equation}
    H = \frac{k_{\rm B} T}{m g},
    \label{eq:H}
\end{equation}

\noindent where $T$ is the temperature, $m$ is the mean molecular mass, $g$ is the surface gravity, and $k_{\rm B}$ is the Boltzmann constant.
In the Solar System, $H$ of the three telluric planets with atmospheres and Titan are defined immediately above the surface, where the troposphere is located and, therefore, $T \approx T_{\rm surf}$.
Above the surface and below the tropopause, rotational turbulence mixes the layers of the atmosphere (the Earth's tropopause height varies from about 9\,km at the poles to 17\,km at the equator), and the mean molecular mass of the ``dry air'' (without water vapour content) keeps constant.
$H$ is also a key input parameter for the equivalent height of the absorbing atmosphere ($\delta \approx 2 H R_{\rm b} / R_\star^2$), which is widely used in exoplanet atmosphere studies \citep[e.g.,][]{Nortmann2018,OrellMiquel2022}.

The surface temperature, $T_{\rm surf}$, strongly depends on the equilibrium temperature, $T_{\rm eq}$.
In radiative equilibrium, a planet $T_{\rm eq}$ depends on the stellar bolometric luminosity, the planet Bond albedo, and the star-planet separation through:

\begin{equation}
    T_{\rm eq} ^4 = \frac{L_\star (1 - A_{\rm Bond})}{16 \pi \sigma \epsilon a^2},
    \label{eq:Teq}
\end{equation}

\noindent where $a$ is the semi-major axis (when $e = 0$), $\sigma$ is the Stefan-Boltzmann constant, and $\epsilon$ is the atmosphere emissivity. 
As it is customary in the literature, we assumed $\epsilon$ = 1.0, by which the whole surface of the planet emits as a black body (if the planet is in radiative equilibrium and there is no heat transfer between both hemispheres, $\epsilon$ = 0.25).
There are many formal ways of estimating a planet surface temperature, $T_{\rm surf}$, from $T_{\rm eq}$ and the properties of an Earth-like atmosphere.
Given the vast amount of uncertainties in the Gl~486\,b planet system parameters, here we used the simple approach of \citet{Houghton1977}, who quantified an atmosphere greenhouse effect with an effective optical thickness,~$\tau$:

\begin{equation}
    T_{\rm surf} ^4 = T_{\rm eq} ^4 \left(1 + \frac{2}{3} \tau \right).
\end{equation}

\noindent In the Solar System, $\tau \approx$ 0.09, 0.94, and 160 for Mars, the Earth, and Venus, respectively (Table~\ref{tab:solarsystem}).
As a result, the minimum $T_{\rm surf}$ of a rocky planet is the $T_{\rm eq}$ for $A_{\rm Bond} = 1$ (total reflectance) and $\tau = 0$ (total transparent atmosphere; i.e. no atmosphere), while the maximum $T_{\rm surf}$ is attained for Earth's open ocean or C-type asteroid-like albedoes ($A_{\rm Bond}$ = 0.03--0.10) and Venus-like optical thickness (the greatest known to date).
Since $A_{\rm Bond}$ is bolometric, it depends on the wavelength of the incident flux and, therefore, the spectral type of the stellar host.
In planets around M dwarfs, the spectral energy distributions of which peak at the red optical-near infrared boundary \citep[][and references therein]{Cifuentes2020}, large values of $A_{\rm Bond}$ are unlikely because of the expected strong atmospheric absorption at these wavelengths, such as the telluric absorption bands in Fig.~\ref{fig:lines} \citep[e.g.,][]{IrvinePollack1968, Kieffer1977, Sudarsky2000, Rogers2009}.
This unlikeliness of large values, together with a simpler computation and the smooth dependence $T_{\rm eq} \propto (1-A_{\rm Bond})^{1/4}$, explains why many works assume $A_{\rm Bond}$ = 0 for exoplanets around M dwarfs.

In the left panel of Fig.~\ref{fig:tsurf}, we plot the isotherms of $T_{\rm surf}$ for different values of $A_{\rm Bond}$ from 0.0 to 1.0 and $\tau$ from 0.1 to 100.
A priori and without any actual observation of the atmosphere of Gl~486\,b, we can only hypothesise its approximate location in the bottom left quadrant of the diagram, with most probable albedoes less than 0.4 ($\sim 1.3\,A_{{\rm Bond,}\oplus}$) and optical thickness values less that 3 ($\sim 3.2\,\tau_\oplus$).
With the determined $L_\star$ and $a$ and most $A_{\rm Bond}$-$\tau$ combinations, $T_{\rm surf}$ is below the critical value at 880\,K, above which surface rocks can be partially devolatilised \citep{Mansfield2019}.
Even with the highest $A_{\rm Bond}$ and lowest $\tau$, Gl~486\,b would never be habitable according to the standard definition \citep{Kasting1993,Kopparapu2013}.
However, a better $T_{\rm eq}$ and $T_{\rm surf}$ determination, perhaps from multi-dimensional climate models, would help in future atmospheric retrievals as the ones presented below \citep[e.g.,][and references therein]{Fauchez2020}.

In the right panel of Fig.~\ref{fig:tsurf}, we plot lines of constant mean molecular mass as a function of $T_{\rm surf}$ for different atmosphere compositions, with mean molecular masses from $\sim$1.3\,u (90\,\% H, 10\,\% He) to $\sim$44.0\,u (pure carbon dioxide). 
Depending on the actual $T_{\rm surf}$ and atmospheric composition, Gl~486\,b $H$ can vary by almost two orders of magnitude from Earth's $H_\oplus \approx$ 8.5\,km to up to 400\,km of a primordial hydrogen and helium atmosphere.

\begin{table*}
\centering
\small
\caption{Forthcoming space observations of Gl~486 and Gl~486\,b in the optical and infrared$^a$.} 
\label{tab:jwst}
\begin{tabular}{llllcccl}
\hline
\hline
\noalign{\smallskip}
Program     & Telescope & Instrument    & Mode  & $\Delta \lambda ^b$ & $\mathcal{R}$ & Visit         & Remark    \\ 
            &           &               &       &       & ($\lambda / \Delta \lambda$) & time [h]      &           \\ 
\noalign{\smallskip}
\hline
\noalign{\smallskip}
Sector 50 & {\em TESS} & Camera 1 & 2\,min cadence & 6000--10\,000\,\AA & $\sim$2 & $\sim$650 & Photometry \\
GO 1743 & {\em Webb} & MIRI & LRS slitless prism & 5--12\,$\mu$m & $\sim$100 & $2 \times 6.27$ & Secondary eclipse \\ 
GO 1981 & {\em Webb} & NIRSpec & BOTS G395H/F290LP & 2.87--5.18\,$\mu$m & 1900--3700 & $2 \times 5.14$ & Transmission spectra \\ 
GO 16722 & {\em Hubble} & COS & NUV/G230L & 1700--3200\,\AA & 2100--3200 & 0.15 \\ 
      &              & COS & FUV/G160M & 1405--1775\,\AA & 13\,000--24\,000 & 1.88 \\ 
      &              & COS & FUV/G130M & 900--1236\,\AA & $<$11\,500 & 2.11 \\ 
      &              & STIS & FUV/G140M & 1140--1740\,\AA & 11\,400--17\,400 & 1.36 & Ly$\alpha$ \\ 
\noalign{\smallskip}
\hline
\end{tabular}
\tablefoot{
\tablefoottext{a}{Scheduled {\em TESS} observations in Year 4+ from 25 March 2022 to 22 April 2022 accessed from \url{https://heasarc.gsfc.nasa.gov/cgi-bin/tess/webtess/wtv.py?Entry=gj+486}.
Accepted {\em Webb} general observer cycle 1 (2nd half 2022 -- 1st half 2023) programs accessed from \url{https://www.stsci.edu/jwst/science-execution/approved-programs/cycle-1-go}.
Accepted {\em HST} general observer cycle 29 (1 October 2021 -- 30 September 2022) programs accessed from \url{https://www.stsci.edu/hst/proposing/approved-programs}.
Joint {\em Hubble} and {\em XMM-Newton} program GO~16722 accessed from \url{https://archive.stsci.edu/hst/joint_programs.html}.}
\tablefoottext{b}{For the sake of readability, we mix wavelength units.}
}
\end{table*}

\begin{figure}
    \centering
    \includegraphics[width=0.49\textwidth]{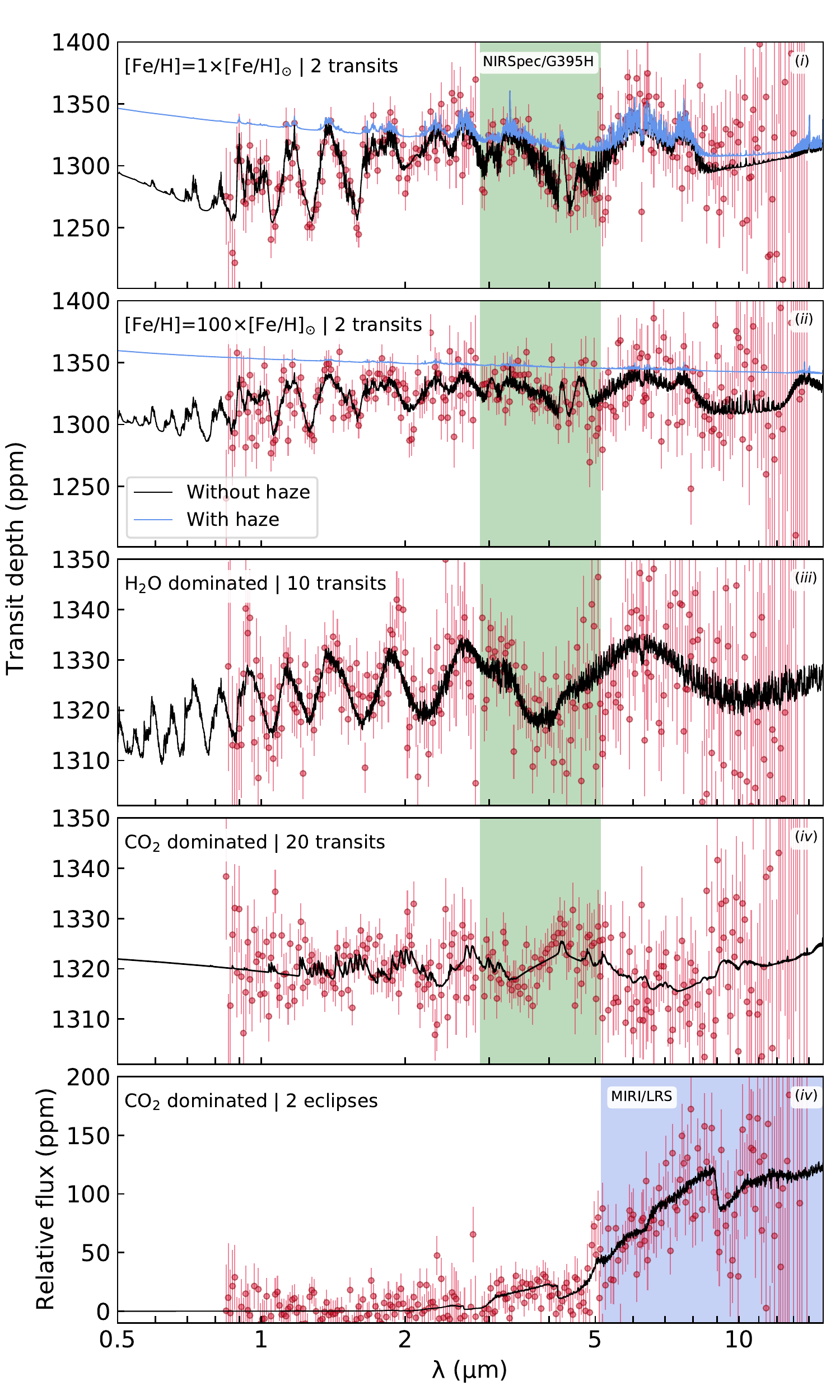} 
    \caption{Synthetic {\em Webb} atmospheric spectra of Gl~486\,b.
    From top to bottom: transmission spectra under optimistic ($i$), feasible ($ii$), realistic and interior-compatible ($iii$), and realistic ($iv$) scenarios, and emission spectrum for the realistic ($iv$) scenario.
    Red circles with estimated uncertainties are shown for different number of transits and eclipses as stated in the panels, with {\em Webb} NIRISS/SOSS, NIRSpec/G395H, and MIRI/LRS configurations.     
    Black and blue lines represent clear (i.e., no haze opacity contribution) and hazy atmospheres, respectively. 
    There is no hazy atmosphere in the realistic scenarios.
    NIRSpec/G395H and MIRI/LRS (Table~\ref{tab:jwst}) wavelength coverages are highlighted by green and blue shades, respectively.
    }
    \label{fig:atmJWST} 
\end{figure}

As summarised by Table~\ref{tab:jwst}, the Gl~486 system has been already scheduled for future observations with space missions such as {\em TESS}, {\em Webb}, and {\em Hubble} (besides {\em XMM-Newton} and {\em Chandra}).
Only the observations with {\em Webb}, namely low-resolution spectroscopy in the near- and mid-infrared with NIRSpec and MIRI, are aimed to investigate the atmosphere of the planet, while the other observations are equally necessary but focused on quantifying the stellar coronal activity of the host star.

We quantitatively assess the suitability of Gl~486\,b for atmospheric characterisation with both NIRSpec and MIRI.
However, before that, one must assess the actual presence of an atmosphere surrounding the planet.
In absence of high-quality X-ray and ultraviolet data, we used the $L_{\rm X}$ and $L_{\rm EUV}$ upper limits in Table~\ref{tab:star} for computing maximum values of the XUV flux that arrives to Gl~486\,b and the subsequent current atmosphere mass loss rate, given at the bottom of Table~\ref{tab:planet}.
For computing $F_{\rm XUV}$ and $\dot{M_{\rm b}}$, we used the procedures of \citet{SanzForcada2011}.
Although the mass loss rate during the first stages of evolution of the host star was necessarily greater \citep{Ribas2005, Kubyshkina2018}, the tabulated upper limit of present-day $\dot{M_{\rm b}}$ translates into 0.07\,$M_\oplus$\,Ga$^{-1}$, which is comparable or relatively low in comparison with many transiting exoplanets with atmospheres investigated to date 
(e.g., \object{K2-100\,b} with more than 0.5\,$M_\oplus$\,Ga$^{-1}$, \citealt{Barragan2019}; 
\object{TOI-849\,b} with 0.95\,$M_\oplus$\,Ga$^{-1}$, \citealt{Armstrong2020}; \object{Gl~436\,b} with 0.019\,$M_\oplus$\,Ga$^{-1}$, \citealt{SanzForcada2011, VillarrealDAngelo2021, Foster2022}; see Fig.~11 of the latter authors for a population study).
Therefore, available XUV data suggest that Gl~486\,b may still retain an atmosphere.  

For the quantitatively assessment of the Gl~486\,b suitability for atmospheric characterisation, we synthesised a suit of atmospheric spectra with five scenarios, all of them consistent with the planet interior characterisation in Section~\ref{sec:planetinteriorcharacterisation}: 
($i$) Optimistic -- a substantial H/He gaseous envelope with a relatively low mean molecular weight (i.e., solar abundances);
($ii$) Feasible -- a H/He gaseous envelope with enhanced metallicity (i.e., 100$\times$ solar abundances); 
($iii$) Realistic and compatible with our interior MRA-SH scenario model -- a H$_2$O-dominated atmosphere;
($iv$) Realistic -- a CO$_2$-dominated atmosphere; 
and 
($v$) Pessimistic -- bare rock with a tenuous atmosphere. 

The atmospheric composition, spectra, and uncertainties were calculated using the photo-chemical model {\tt ChemKM} \citep{Molaverdikhani2019a, Molaverdikhani2019b, Molaverdikhani2020}, the radiative transfer code {\tt petitRADTRANS} \citep{Molliere2019}, and the community tool {\tt Pandexo} for transiting exoplanet science with {\em Webb} and {\em Hubble} \citep{Batalha2017}. 
The system parameters, including planet mass, radius, equilibrium temperature, and transit duration, and stellar mass, radius, magnitude, and metallicity were taken from Tables~\ref{tab:star} and~\ref{tab:planet}. 

The resulting synthetic transmission spectra for the optimistic scenario ($i$) show strong absorption features due to H$_2$O, CH$_4$, and CO$_2$ over the wavelength range of 0.8--10\,$\mu$m, as shown in the top panel of Fig.~\ref{fig:atmJWST}. 
The spectral signature amplitudes are on the order of 50--100\,ppm when no haze is assumed. 
However, haze opacity contribution obscures the mentioned absorption features \citep[e.g.,][]{Fortney2005, Kreidberg2014, Nowak2020, Trifonov2021}, particularly at short wavelengths. 
Here, haze is defined the same way as in \citet{Nowak2020}, who included C$_8$H$_6$, C$_8$H$_7$, C$_{10}$H$_3$, C$_{12}$H$_3$, C$_{12}$H$_{10}$, C$_{14}$H$_3$, C$_2$H$_4$N, C$_2$H$_3$N$_2$, C$_3$H$_6$N, C$_4$H$_3$N$_2$, C$_4$H$_8$N, C$_5$HN, C$_5$H$_3$N, C$_5$H$_4$N, C$_5$H$_6$N, C$_9$H$_6$N, C$_3$H$_3$O, C$_3$H$_5$O, C$_3$H$_7$O, and C$_4$H$_6$O. 
All these precursor molecules that represent the haze particles are collectively called soot or haze \citep[e.g.,][]{LavvasKoskinen2017}. 

An increase in the atmospheric metallicity (i.e. scenario $ii$) enhances this obscuration effect both due to a higher mean molecular weight and a higher haze production rate, as illustrated by the dampened spectral features in the second panel of Fig.~\ref{fig:atmJWST}. 
Such a H/He gaseous envelope with enhanced metallicity might be a feasible scenario in a rocky planet, for instance, due to resupply of hydrogen from magma outgassing \citep{Chachan2018, Kite2019, Kite2020}. 
Both scenarios (i.e., optimistic, $i$, and feasible, $ii$) would be distinguishable with only two transits observed by {\em Webb} over a wide range of wavelengths if the atmosphere were haze-free. 
However, given that a hazy atmosphere for this class of planets has been proposed to be likely \citep[e.g.,][]{Gao2018, Yu2021}, using only NIRSpec/G395H 
may require more than two transits to differentiate between different (solar or enhanced metallicity) H/He-dominated scenarios.

Terrestrial planets in the Solar System and models of evolution of exoplanetary atmospheres suggest that small planets are less likely to maintain a H-dominated atmosphere \citep[e.g.,][]{Gao2015, Woitke2021}.
For example, \citet{Ortenzi2020} showed that, for a $\sim$\,3\,$M_\oplus$ planet such as Gl~486\,b, a very efficient H$_2$O outgassing, and hence an H$_2$O dominated atmosphere, is a more likely scenario. 
Therefore, in our realistic scenario $iii$, we simulated such an atmosphere based on the outcome of our planet interior MRA-SH model. 
We assumed an isothermal atmosphere with a temperature equal to the equilibrium temperature of the planet up to 100\,hPa (0.1\,bar), which roughly marks the tropopause temperature. 
Then the temperature profile follows an adiabat. 
The resulting spectrum is illustrated in the third panel of Fig.~\ref{fig:atmJWST}, with water features reduced in amplitude by a factor of a few compared with the enhanced-metallicity H/He-dominated scenario ($ii$), which is mainly due to a higher mean molecular weight. 
The uncertainties shown for ten transits suggest that a steam atmosphere would be detectable under these circumstances, although haze and cloud form might obscure water features significantly.

In spite of the fact that an H$_2$O-dominated atmosphere is a more likely scenario according to \citet{Ortenzi2020}, a CO$_2$-dominated atmosphere cannot be ruled out in case of very dry and oxidised mantle. 
Moreover, examples in the Solar System, like Venus and Mars, support such a possibility. 
Hence, we also assumed a CO$_2$-dominated atmosphere for Gl~486\,b to synthetise another realistic spectrum, namely scenario $iv$. 
Such an atmosphere could have evolved from an initial secondary atmospheric composition (e.g., H$_2$O, CO, and CO$_2$), as hinted by carbonaceous chondrite outgassing measurements \citep{Thompson2021}. 
In this scenario, a higher mean molecular weight results in a smaller atmospheric scale height and, therefore, smaller spectral features. 
This effect is shown in the fourth panel of Fig.~\ref{fig:atmJWST}, where a total of 20 transits were employed to achieve a reasonable S/N for the detection of CO$_2$ features on the order of 5--10\,ppm. 
The number of required transits would increase even more if haze contribution, like a Venusian atmosphere, were also considered. 
A workaround to this issue is to observe this planet in emission.

The day-side temperature of Gl~486\,b is expected to be cold enough to maintain an atmosphere, but hot enough to be suitable for emission spectroscopy. 
The bottom panel of Fig.~\ref{fig:atmJWST} illustrates a synthetic {\em Webb} emission spectrum of a realistic CO$_2$-dominated atmosphere. 
The relative flux remains below 20\,ppm with slight spectral modulations below 5\,$\mu$m. 
However, it increases to 100--150\,ppm at longer wavelengths, where MIRI/LRS spans its coverage. 
Uncertainty estimations suggest that the spectral features, particularly around 9\,$\mu$m, should be detectable with only two eclipses. 
Therefore, both transmission and emission spectroscopy are necessary to characterise the atmosphere of Gl~486\,b over different scenarios.
The two eclipses of GO~1743 (MIRI/LRS) and the two transits GO~1981 (NIRSpec/G395H) programmed for {\em Webb} cycle~1 (Table~\ref{tab:jwst}) will certainly shed light on the composition and structure of the atmosphere of the exoearth. 

\begin{figure}
    \centering
    \includegraphics[width=0.49\textwidth]{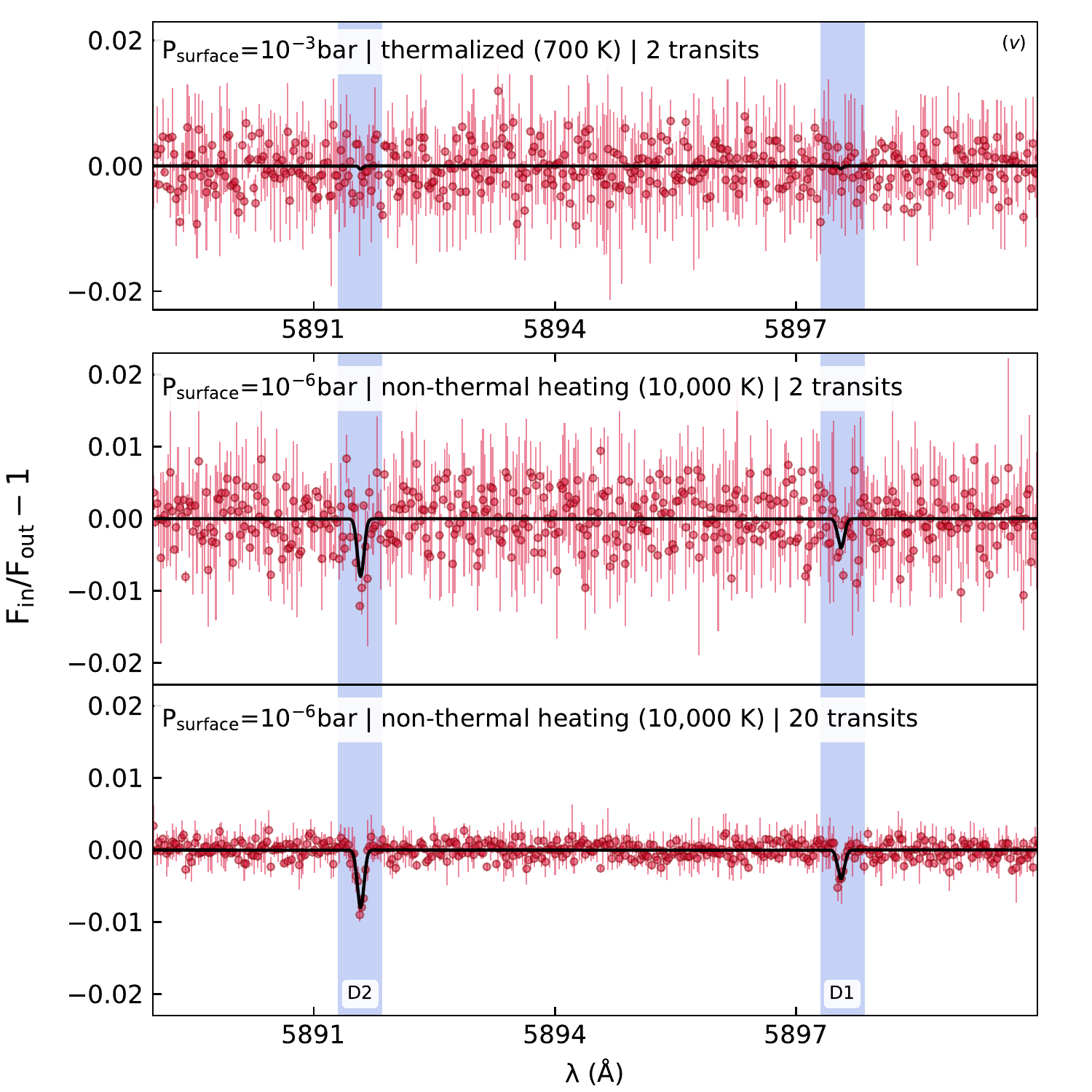} 
    \caption{Synthetic CARMENES atmospheric spectra of Gl~486\,b.
    Transmission spectra around the Na~{\sc i} D$_1$ and D$_2$ doublet under the pessimistic scenario ($iv$: a bare rock with a tenuous atmosphere) for a thermalised atmosphere and two transits ({\em top}), and for an exosphere with non-thermal heating and two transits ({\em middle}) and 20 transits ({\em bottom}).
    }
    \label{fig:atmCARMENES} 
\end{figure}

Still, the pessimistic scenario ($v$) of a bare rock with a tenuous atmosphere would be out of reach for {\em Webb} instruments. 
Therefore, we examined such a scenario for two cases by employing ground-based high-resolution spectrographs, such as CARMENES or MAROON-X. 
In the first case, we assumed a Mars-like surface pressure of 1\,hPa ($10^{-3}$\,bar). 
With such a pressure, the atmosphere is likely to be thermalised at the surface and, therefore, we used the $T_{\rm eq}$ for null albedo of about 700\,K.
\citet{Welbanks2019} conducted a homogeneous retrieval of sodium abundance for 19 exoplanets and found that about half of these planets show an abundance of 10$^{-9}$ or non-detection. 
Hence, the synthetic spectrum around the D$_2$ and D$_1$ sodium doublet, Na~{\sc i} $\lambda\lambda$5889.95,5895.92\,\AA, is shown in the top panel of Fig.~\ref{fig:atmCARMENES} for a Mars-like surface pressure and a Na abundance of 10$^{-9}$. 
The estimated uncertainties for CARMENES over two transits are much larger than the signals that, hence, remain undetected. 
The Na D$_2$ and D$_1$ doublet is one of the strongest planetary lines and, thus, other spectral lines are expected to face a similar fate with a tenuous, thermalised atmosphere.

The second case of a bare rock is by assuming an even more tenuous atmosphere. 
Although this may seem contradictory at a first glance, a lower surface pressure might help with heating up the surface species to more than 10\,000\,K through non-thermal processes, such as surface sputtering by ions \citep[e.g.,][]{Bida2000}. 
In this case, we assumed a surface pressure of around 0.001\,hPa (10$^{-6}$\,bar, still several magnitudes higher than that of Mercury), a temperature of 10\,000\,K, and the same Na abundance of 10$^{-9}$. 
We estimated the absorption excess of the Na D$_2$ line, illustrated by the middle panel of Fig.~\ref{fig:atmCARMENES}, at around 1\,\%, which is large enough for an instrument such as CARMENES to resolve the line. 
The bottom panel of Fig.~\ref{fig:atmCARMENES} illustrates the same atmospheric scenario but observed through 20 transits. 
Characterisation of such atmospheres would be a more suitable task for facilities with more light-gathering power, such as ESPRESSO at the 8.4\,m Very Large Telescope or ANDES at the 39\,m Extremely Large Telescope. 
Moreover, molecular formation under such conditions is unlikely, but atomic and ionic species that do not have strong resolved lines might be still detectable with the cross-correlation technique \citep[e.g.,][]{Snellen2010}.

We estimate that two transits would suffice to meet S/N = 5 for the realistic (CO$_2$-dominated) atmosphere model, while a single transit would do it for the other four models, namely optimistic (H/He and 1$\times$[Fe/H]$_\odot$), feasible (H/He and 100$\times$[Fe/H]$_\odot$), interior-compatible realistic (H$_2$O-dominated), and pessimistic (emission).
However, the definition of S/N becomes arbitrary and model dependent when dealing with low-resolution spectra over a wide wavelength range.
For example, the S/N calculated directly by dividing signals by their corresponding noises (uncertainties) for each spectral bin would not ensure differentiation of atmospheric scenarios.
Therefore, we defined back-of-the-envelope baseline models (a flat line for transmission, the ratio of two black body functions for emission), removed them from the signals, and used the residuals to estimate the sum of all S/Ns for all spectral bins and within each instrument range.
As a result, our estimations must be used with care.
Instead of reporting S/N, we suggest comparing the models and choosing the best one according to their log-evidence.

Altogether, Gl~486\,b provides an exceptional opportunity to characterise the atmosphere of an exoearth. 
However, our analysis shows synergistically planned observations between ground-based facilities and {\em Webb} instruments would be needed to achieve that goal.

\section{Conclusions}
\label{sec:conclusions}

At a distance of only 8.1\,pc, Gl~486\,b is the third closest transiting planet to the Sun and presents an important addition to the demographics of known transiting rocky planets. 
The relatively bright, very weakly active, M-dwarf host star, its visibility from both Earth hemispheres, and the short orbital period and warm expected surface temperature make this planet one of the best targets for planet atmosphere emission and transit spectroscopy with {\em Webb} and future ground-based extremely large telescopes.

In this work, we slightly improve the precision and accuracy of the planet mass and radius determination, with which we develop different planet interior and atmosphere scenarios.
Except for the eccentricity, which upper limit is set at 0.025, the improvement of most parameters with respect to what \citet{Trifonov2021} tabulated is slim.
However, there are a few differences with respect to previous work that make this analysis unique.
Instead of estimating the stellar radius from luminosity and model-dependent spectral synthesis, we directly measured the angular radius of the planet host star with MIRC-X at the CHARA Array.
We reduced the input data error contribution by gathering extremely precise RV data collected by CARMENES and MAROON-X and transit data obtained by {\em TESS} and, presented here for the first time, {\em CHEOPS}.
The selected joint RV and transit fit model, 1pl+GP, was supported by an independent photometric monitoring with small and medium-size telescopes for determining the stellar rotation period, which turned to be shorter than previously reported and of very low amplitude.
The additional errors in planet radius and mass introduced by the transit and RV data with respect to the uncertainties in star radius and mass were just 0.3\,\% and 0.4\,\%, respectively, possibly at the limit of what is technically possible nowadays.

As a novelty in M dwarfs, we determined Mg, Si, V, Fe, Rb, Sr, and Zr abundances of the stellar host, which constrained two of the three considered planet interior scenarios.
We also considered different planet atmosphere scenarios and their detectability with forthcoming {\em Webb} observations with NIRSpec and MIRI after taking into account different possibilities on composition and planet surface temperature and pressure.

In the most probable combination of scenarios, Gl~486\,b is a warm Earth-like planet of $R \sim 1.343\,R_\oplus$ and $M \sim 3.00\,M_\oplus$ with a relatively low-mass, metallic core surrounded by a silicate mantle with dissolved water, and an upper layer probably composed of a mixture of water steam and carbon dioxide.
Now it is time to do comparative planetology and investigate topics on solid grounds that were provocative until very recently.
For example, if there is a liquid outer core, and because of the fast tidally-locked planet rotation of 1.47\,d, there may be a strong magnetic field that protects the Gl~486\,b atmosphere from stellar erosion \citep{Scalo2007} or, if a thick atmosphere is indeed preserved, there may be jet streams that transport heat from the illuminated to the dark hemisphere as in hot Jupiters \citep{ShowmanPolvani2011}.
These and other novel studies based on the results presented here, such as deriving simultaneously stellar and planetary mass and radius using interferometry and probability density functions \citep{Crida2018} or constraining the oxygen fugacity of the planet \citep{Doyle2019}, will come soon.

\begin{acknowledgements}
We thank
  the reviewer for helpful comments,
  Kate Isaak for her support on {\em CHEOPS} observations,
  Mahmoudreza Oshagh for her preliminary transit-time variation analysis,
  David Ciardi for helpful comments on deep adaptive optics imaging,
  Vera~M. Passegger for her Mg~{\sc i} fit,
  {Sandra~V. Jeffers for comments on stellar activity},
  and Joel Hartman, Greg~W. Henry, Jonathan Irwin, and Chris~G. Tinney for their information on HATNet, TSU, MEarth, and APT photometric data.

CHEOPS is an ESA mission in partnership with Switzerland with important contributions to the payload and the ground segment from Austria, Belgium, France, Germany, Hungary, Italy, Portugal, Spain, Sweden and the United Kingdom.

The development of the MAROON-X spectrograph was funded by the David and Lucile Packard Foundation, the Heising-Simons Foundation, the Gemini Observatory, and the University of Chicago. This work was enabled by
observations made from the Gemini North telescope, located within the Maunakea Science Reserve and adjacent to the summit of Maunakea. We are grateful for the privilege of observing the Universe from a place that is unique in both its astronomical quality and its cultural significance.

CARMENES is an instrument at the Centro Astron\'omico Hispano en Andaluc\'ia (CAHA) at Calar Alto (Almer\'{\i}a, Spain), operated jointly by the Junta de Andaluc\'ia and the Ins\-ti\-tu\-to de Astrof\'isica de Andaluc\'ia (CSIC).
  
CARMENES was funded by the Max-Planck-Gesellschaft (MPG), 
the Consejo Superior de Investigaciones Cient\'{\i}ficas (CSIC),
the Ministerio de Econom\'ia y Competitividad (MINECO) and the European Regional Development Fund (ERDF) through projects FICTS-2011-02, ICTS-2017-07-CAHA-4, and CAHA16-CE-3978, 
and the members of the CARMENES Consortium 
  (Max-Planck-Institut f\"ur Astronomie,
  Ins\-ti\-tu\-to de Astrof\'{\i}sica de Andaluc\'{\i}a,
  Landessternwarte K\"onigstuhl,
  Institut de Ci\`encies de l'Espai,
  Institut f\"ur Astrophysik G\"ottingen,
  Universidad Complutense de Madrid,
  Th\"uringer Landessternwarte Tautenburg,
  Instituto de Astrof\'{\i}sica de Canarias,
  Hamburger Sternwarte,
  Centro de Astrobiolog\'{\i}a and
  Centro Astron\'omico Hispano-Alem\'an), 
with additional contributions by the MINECO, 
the Deutsche Forschungsgemeinschaft (DFG) through the Major Research Instrumentation Programme and Research Unit FOR2544 ``Blue Planets around Red Stars'', 
the Klaus Tschira Stiftung, 
the states of Baden-W\"urttemberg and Niedersachsen, 
and by the Junta de Andaluc\'{\i}a.

This work is based upon observations obtained with the Georgia State University (GSU) Center for High Angular Resolution Astronomy Array at Mount Wilson Observatory. The CHARA Array is supported by the National Science Foundation under Grant No. AST-1636624 and AST-2034336.  Institutional support has been provided from the GSU College of Arts and Sciences and the GSU Office of the Vice President for Research and Economic Development. 

We would like to recognise the observing team, scientists, and support staff at the CHARA Array. Observation time for this work was generously allocated via discretionary time from CHARA Array director Theo ten~Brumelaar and via NOIRLab community access program (proposals 2021A-0247 and 2021A-0141).

MIRC-X received funding from the European Research Council (ERC) under the European Union's Horizon 2020 research and innovation program (Grant No. 639889), as well as from NASA (XRP NNX16AD43G) and NSF (AST 1909165).

Data were partly collected with the 90\,cm telescope at the Observatorio de Sierra Nevada operated by the Instituto de Astrof\'\i fica de Andaluc\'\i a (IAA-CSIC). 

This work made use of observations from the Las Cumbres Observatory Global Telescope network.
LCOGT observations were partially acquired via program number TAU2021A-015 of the Wise Observatory, Tel-Aviv University, Israel.

We acknowledge financial support from the Agencia Estatal de Investigaci\'on of the Ministerio de Ciencia, Innovaci\'on y Universidades and the ERDF through projects 
  PID2019-109522GB-C5[1:4],	
  PID2019-107061GB-C64, 
  PID2019-110689RB-100, 
  PGC2018-095317-B-C21, 
  PGC2018-102108-B-I00, 
and the Centre of Excellence ``Severo Ochoa'' and ``Mar\'ia de Maeztu'' awards to the Instituto de Astrof\'isica de Canarias (CEX2019-000920-S), Instituto de Astrof\'isica de Andaluc\'ia (SEV-2017-0709), and Centro de Astrobiolog\'ia (MDM-2017-0737),
DFG through FOR2544 (KU 3625/2-1) and Germany's Excellence Strategy to the Excellence Cluster ORIGINS (EXC-2094 - 390783311),
European Research Council (Starting Grant 639889),
Bulgarian National Science Fund through VIHREN-2021 (KP-06-DB/5), 
Schweizerischer Nationalfonds zur F\"orderung der wissenschaftlichen Forschung / Fonds national suisse de la recherche scientifique (PZ00P2\_174028),
United Kingdom Science Technology and Facilities Council (630008203),
NASA (80NSSC22K0117),
National Science Foundation (2108465 and Graduate Research Fellowship DGE 1746045),
Universidad La Laguna through the Margarita Salas Fellowship from the Spanish Ministerio de Universidades and under the EU Next Generation funds (UNI/551/2021-May~26),
and the Generalitat de Catalunya (CERCA programme).

We used the NASA Exoplanet Archive, which is operated by the California Institute of Technology, under contract with the National Aeronautics and Space Administration under the Exoplanet Exploration Program, and
{\tt Uncertainties}, a Python package for calculations with uncertainties developed by E.\,O.~Lebigot (\url{https://pythonhosted.org/uncertainties}).

\end{acknowledgements}

\bibliographystyle{aa} 
\bibliography{biblio}

\appendix


\section{Short tables and diagrams}
\label{sec:appendix}

\begin{table}[!ht]
\centering
\small
\caption{Published heliocentric radial velocities of Gl\,486$^a$.} 
\label{tab:gamma}
\begin{tabular}{cr}
\hline
\hline
\noalign{\smallskip}
$\gamma$ & Reference \\
{[km\,s$^{-1}$]} & \\
\noalign{\smallskip}
\hline
\noalign{\smallskip}
$+19.1 \pm 0.8$         & Giz02 \\
$+19.090 \pm 0.010$     & Nid02 \\
$+19.180 \pm 0.001$     & Fou18$^b$ \\
$+19.50 \pm 0.12$       & Jef18 \\
$+18.970$               & Rei18$^c$ \\
$+19.106 \pm 0.013$     & Sou18 \\
$+19.395 \pm 0.042$     & Laf20$^d$ \\
\noalign{\smallskip}
\hline
\end{tabular}
\tablebib{
Giz02: \citet{Gizis2002}; 
Nid20: \citet{Nidever2002}; 
Fou18: \citet{Fouque2018};
Jef18: \citet{Jeffers2018}; 
Laf20: \citet{Lafarga2020}; 
Rei18: \citet{Reiners2018};
Sou18: \citet{Soubiran2018}. 
}
\tablefoot{
\tablefoottext{a}{The obsolete values of $+5.0 \pm 0.7$\,km\,s$^{-1}$ and $+11 \pm 5$\,km\,s$^{-1}$ of \citet{Joy1947} and \citet{Newton2014}, respectively, are not tabulated.}
\tablefoottext{b}{Although VizieR tabulates an uncertainty of 1\,m\,s$^{-1}$ for the $\gamma$ value of \citet{Fouque2018}, the actual accuracy of the heliocentric RVs measured by ESPaDOnS and reduced with {\tt LIBRE-ESPRIT} \citep{Donati1997} is about 20--30\,m\,s$^{-1}$ \citep{Moutou2017}.}
\tablefoottext{c}{\citet{Reiners2018} did not tabulate $\gamma$ uncertainties.}
\tablefoottext{d}{We tabulate their cross-correlation function RV weighted mean plus the gravitational redshift, with uncertainties summed quadratically.}}
\end{table}


\begin{table}[!ht]
\centering
\small
\caption{Multiband photometry of Gl\,486$^a$.} 
\label{tab:phot}
\begin{tabular}{lcr}
\hline
\hline
\noalign{\smallskip}
Band & Magnitude & Reference \\
    & [mag] & \\
\noalign{\smallskip}
\hline
\noalign{\smallskip}
$u'$    & $15.183\pm0.006$      & SDSS DR9 \\
$B$     & $12.933\pm0.020$      & UCAC4 \\
$g'$    & $12.099\pm0.020$      & UCAC4 \\
$G_{BP}$& $11.6426\pm0.0030$    & {\it Gaia} EDR3 \\
$V_T$   & $11.379\pm0.006$      & TYC \\
$V$     & $11.393\pm0.020$      & UCAC4 \\
$r'$    & $10.829\pm0.040$      & UCAC4 \\
$G$     & $10.1051\pm0.0028$    & {\it Gaia} EDR3 \\
$i'$    & $ 9.316\pm0.070$      & UCAC4 \\
$G_{RP}$& $ 8.8883\pm0.0038$    & {\it Gaia} EDR3 \\
$J$     & $7.195\pm0.026$       & 2MASS \\
$H$     & $6.666\pm0.046$       & 2MASS \\
$K_s$   & $6.362\pm0.018$       & 2MASS \\
$W1$    & $6.206\pm0.101$       & AllWISE \\
$W2$    & $5.955\pm0.044$       & AllWISE \\
$W3$    & $5.979\pm0.015$       & AllWISE \\
$W4$    & $5.810\pm0.041$       & AllWISE \\
\noalign{\smallskip}
\hline
\end{tabular}
\tablebib{
TYC: Tycho-2, \citet{Hog2000};
2MASS: Two Micron All-Sky Survey, \citet{Skrutskie2006};
SDSS DR9: Sloan Digital Sky Survey, \citet{Ahn2012}; 
UCAC4: The Fourth US Naval Observatory CCD Astrograph Catalog, \citet{Zacharias2013};
AllWISE: Wide-field Infrared Survey Explorer, \citet{Cutri2014};
{\em Gaia} EDR3: \citet{GaiaBrown2021}.
}
\tablefoot{
\tablefoottext{a}{{\em TESS} $T$ magnitude in Table~\ref{tab:star}.}
}
\end{table}


\begin{figure}[]
 	\centering
 	\includegraphics[width=0.49\textwidth]{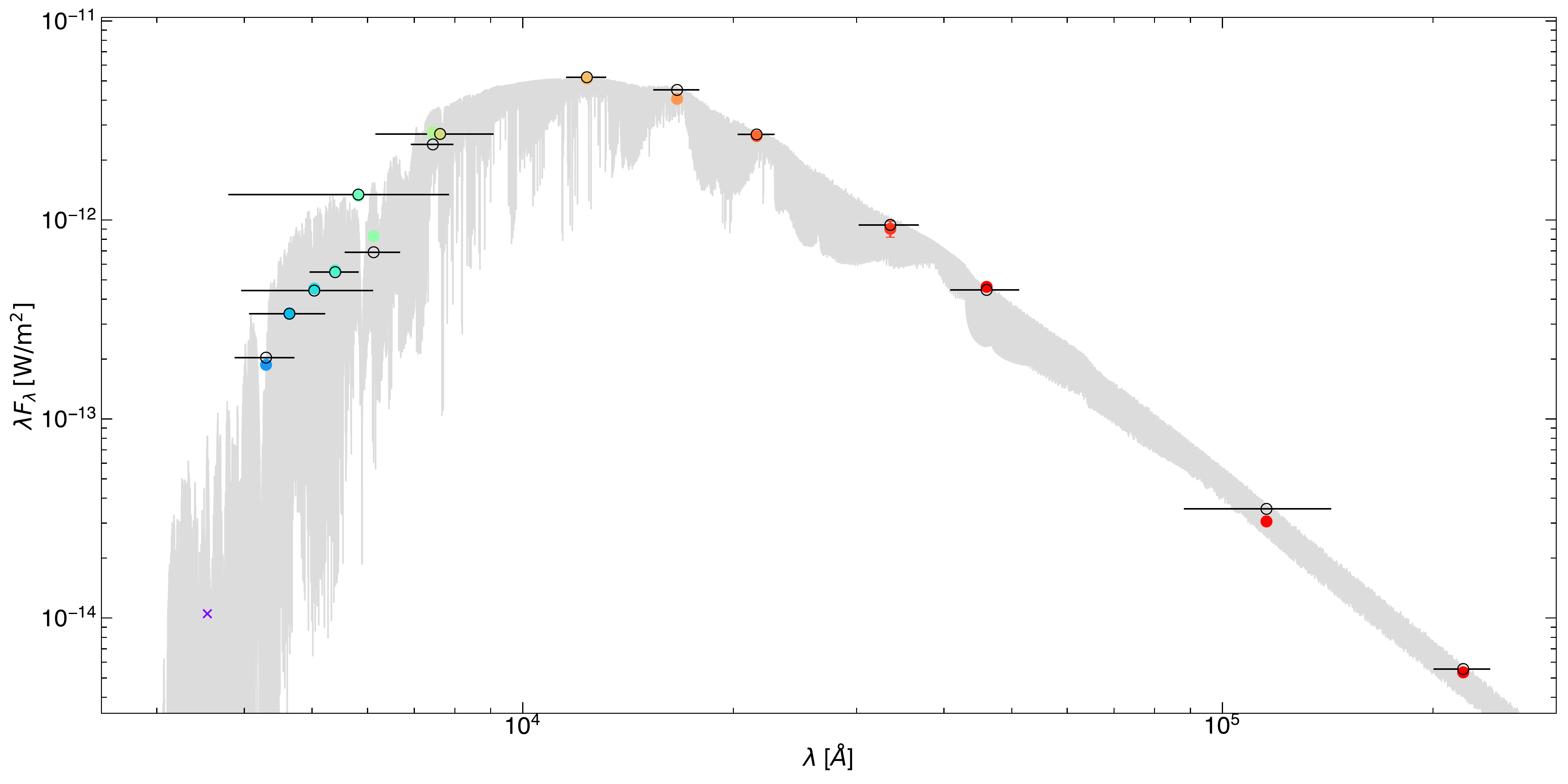}
 	\caption{Spectral energy distribution of Gl\,486.
 	The apparent fluxes (coloured filled circles) are imposed on a BT-Settl CIFIST spectrum (grey; $T_{\rm eff}$ = 3200\,K and $\log{g}$ = 5.5).
 	The modelled fluxes are depicted as black empty circles.
 	The photometric datum in $u'$ (purple cross) was not considered for integrating the bolometric luminosity.
 	Horizontal bars represent the effective widths of the bandpasses (equivalent to the horizontal size of a rectangle with height equal to maximum transmission and with the same area that the one covered by the filter transmission curve), while vertical bars (visible only for relatively large values) represent the flux uncertainty derived from the magnitude and parallax errors.
 	See details in \citet{Cifuentes2020}.}
     \label{fig:SED}
\end{figure}


\begin{table}[!ht]
\centering
\small
\caption{Published effective temperatures of Gl\,486$^a$.} 
\label{tab:teff}
\begin{tabular}{cr}
\hline
\hline
\noalign{\smallskip}
$T_{\rm eff}$ & Reference \\
{[K]} & \\
\noalign{\smallskip}
\hline
\noalign{\smallskip}
$3095$              & Cas08$^b$ \\
$3240$              & Mor08 \\
$3086$              & Jen09$^c$ \\
$3290$              & Lep13 \\
$3300$              & Raj13 \\
$3241$              & Ste13 \\
$3270 \pm 74$       & Gai14 \\
$3240 \pm 17$       & Fou18 \\
$3384 \pm 51$       & Pas18 \\
$3384$              & Raj18 \\
$3313$              & Hou19 \\
$3239 \pm 92$       & Hoj19 \\
$3218 \pm 110$      & Kuz19 \\
$3389 \pm 51$       & Sch19 \\
$3340 \pm 54$       & Pas19 \\
$3096 \pm 27$       & AK20 \\
$3200 \pm 100$      & Cif20 \\
$3408 \pm 45$       & Mar21 \\
\noalign{\smallskip}
\hline
\end{tabular}
\tablebib{
Cas08: \citet{Casagrande2008};
Mor08: \citet{Morales2008};
Jen09: \citet{Jenkins2009};
Lep13: \citet{Lepine2013};
Raj13: \citet{Rajpurohit2013};
Ste13: \citet{Stelzer2013};
Gai14: \citet{Gaidos2014};
Fou18: \citet{Fouque2018};
Pas18: \citet{Passegger2018};
Raj18: \citet{Rajpurohit2018};
Hou19: \citet{Houdebine2019};
Hoj19: \citet{Hojjatpanah2019};
Kuz19: \citet{Kuznetsov2019};
Sch19: \citet{Schweitzer2019};
Pas19: \citet{Passegger2019};
AK20: \citet{AntoniadisKarnavas2020};
Cif20: \citet{Cifuentes2020};
Mar21: \citet{Marfil2021}.
}
\tablefoot{
\tablefoottext{a}{The $T_{\rm eff}$ determined by us from the stellar bolometric luminosity, interferometric radius, and Stefan-Boltzman law is $3291 \pm 75$\,K.}
\tablefoottext{b}{\citet{Cifuentes2020} demonstrated that \citet{Casagrande2008} $T_{\rm eff}$ of M dwarfs, in contrast to FGK stars, are unreliable.}
\tablefoottext{c}{\citet{Jenkins2009} $T_{\rm eff}$ were determined using the $V-K_s$ relations taken from \citet{Casagrande2008}, while \citet{Cifuentes2020} again demonstrated that in the {\em Gaia} era the Johnson photometry should not be used for deriving parameters of M dwarfs.}
}
\end{table}


\begin{table}[!ht]
\centering
\small
\caption{Published relative iron abundances of Gl\,486.} 
\label{tab:Fe/H}
\begin{tabular}{cr}
\hline
\hline
\noalign{\smallskip}
[Fe/H] & Reference \\
{[dex]} & \\
\noalign{\smallskip}
\hline
\noalign{\smallskip}
$+0.03 \pm 0.13$     & New14 \\
$+0.06 \pm 0.16$     & Pas18 \\
$+0.01 \pm 0.10$     & Fou18$^a$ \\ 
$+0.12 \pm 0.09$     & Kuz19$^b$ \\
$+0.03 \pm 0.16$     & Sch19 \\
$+0.07 \pm 0.19$     & Pas19$^c$ \\
$+0.12 \pm 0.05$     & Hoj19 \\
$-0.15 \pm 0.13$     & Mar21$^d$ \\
\noalign{\smallskip}
\hline
\end{tabular}
\tablebib{
New14: \citet{Newton2014};
Pas18: \citet{Passegger2018};
Fou18: \citet{Fouque2018};
Kuz19: \citet{Kuznetsov2019};
Sch19: \citet{Schweitzer2019};
Pas19: \citet{Passegger2019};
Hoj19: \citet{Hojjatpanah2019};
Mar21: \citet{Marfil2021}.
}
\tablefoot{
\tablefoottext{a}{Tabulated uncertainty is the quadratic sum of the systematic and the {\tt MCAL} uncertainties.}
\tablefoottext{b}{From low-S/N X-Shooter spectra.}
\tablefoottext{c}{From CARMENES VIS+NIR spectra.}
\tablefoottext{d}{After $\alpha$-enhancement correction.}
}
\end{table}


\begin{table}[!ht]
\centering
\small
\caption{Emission measure distribution of Gl\,486.}
\label{tab:emd}
\begin{tabular}{cc}
\hline 
\hline
\noalign{\smallskip}
$\log{T}$ (K) & EM (cm$^{-3}$) \\
\noalign{\smallskip}
\hline
\noalign{\smallskip}
4.0 & 48.00: \\
4.1 & 47.90: \\
4.2 & 47.80: \\
4.3 & $47.65 \pm 0.35$ \\
4.4 & $47.55 \pm 0.20$ \\
4.5 & $47.40 \pm 0.25$ \\
4.6 & $47.15 \pm 0.30$ \\
4.7 & $47.00 \pm 0.20$ \\
4.8 & $46.90 \pm 0.25$ \\
4.9 & $46.85 \pm 0.20$ \\
5.0 & $46.80 \pm 0.15$ \\
5.1 & $46.70 \pm 0.30$ \\
5.2 & $46.60 \pm 0.20$ \\
5.3 & $46.50 \pm 0.20$ \\
5.4 & $46.50 \pm 0.20$ \\
5.5 & 46.35: \\
5.6 & 46.10: \\
5.7 & 46.00: \\
\noalign{\smallskip}
\hline
\end{tabular}
\end{table}


\begin{figure}[]
 	\centering
 	\includegraphics[width=0.49\textwidth]{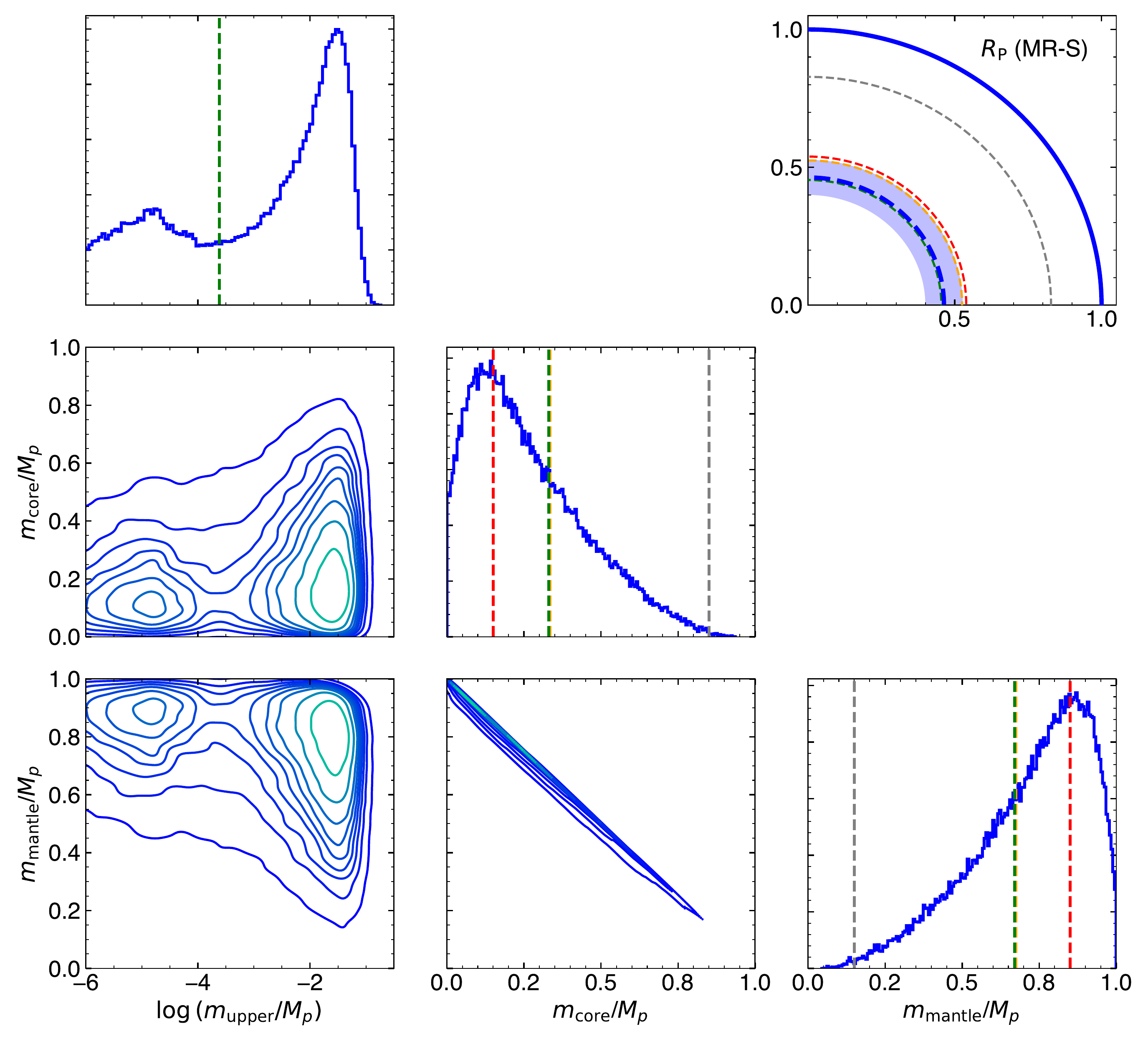}
 	\includegraphics[width=0.49\textwidth]{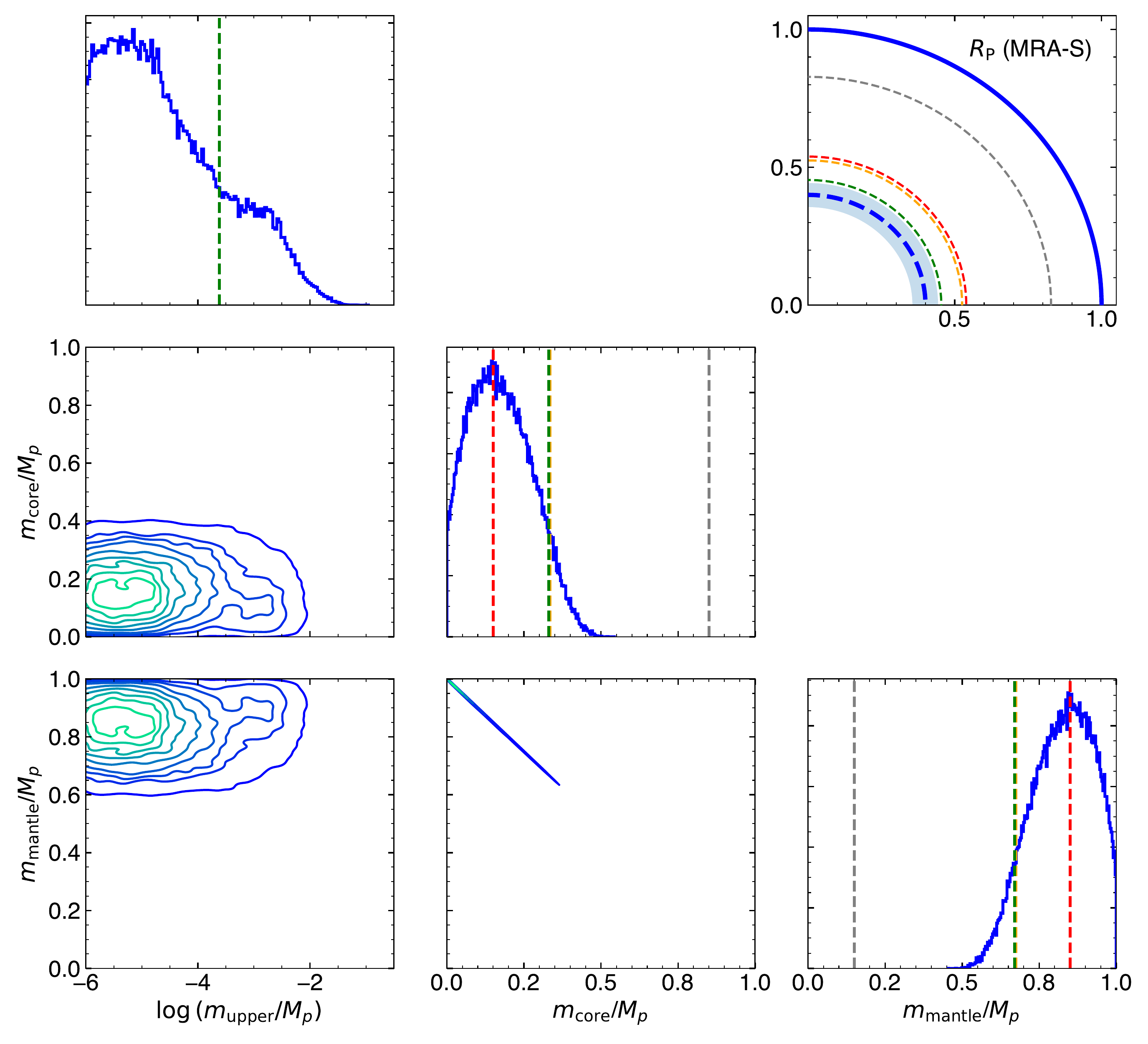}
 	\caption{Same as Fig.~\ref{fig:interior} but for the MR-S (baseline, top) and MR-SA (baseline plus relative stellar abundance constraints, bottom) planet interior scenarios.
 	}
    \label{fig:interior-appendix}
\end{figure}
 


\begin{figure*}[!ht]
    \centering
    \includegraphics[width=0.92\textwidth]{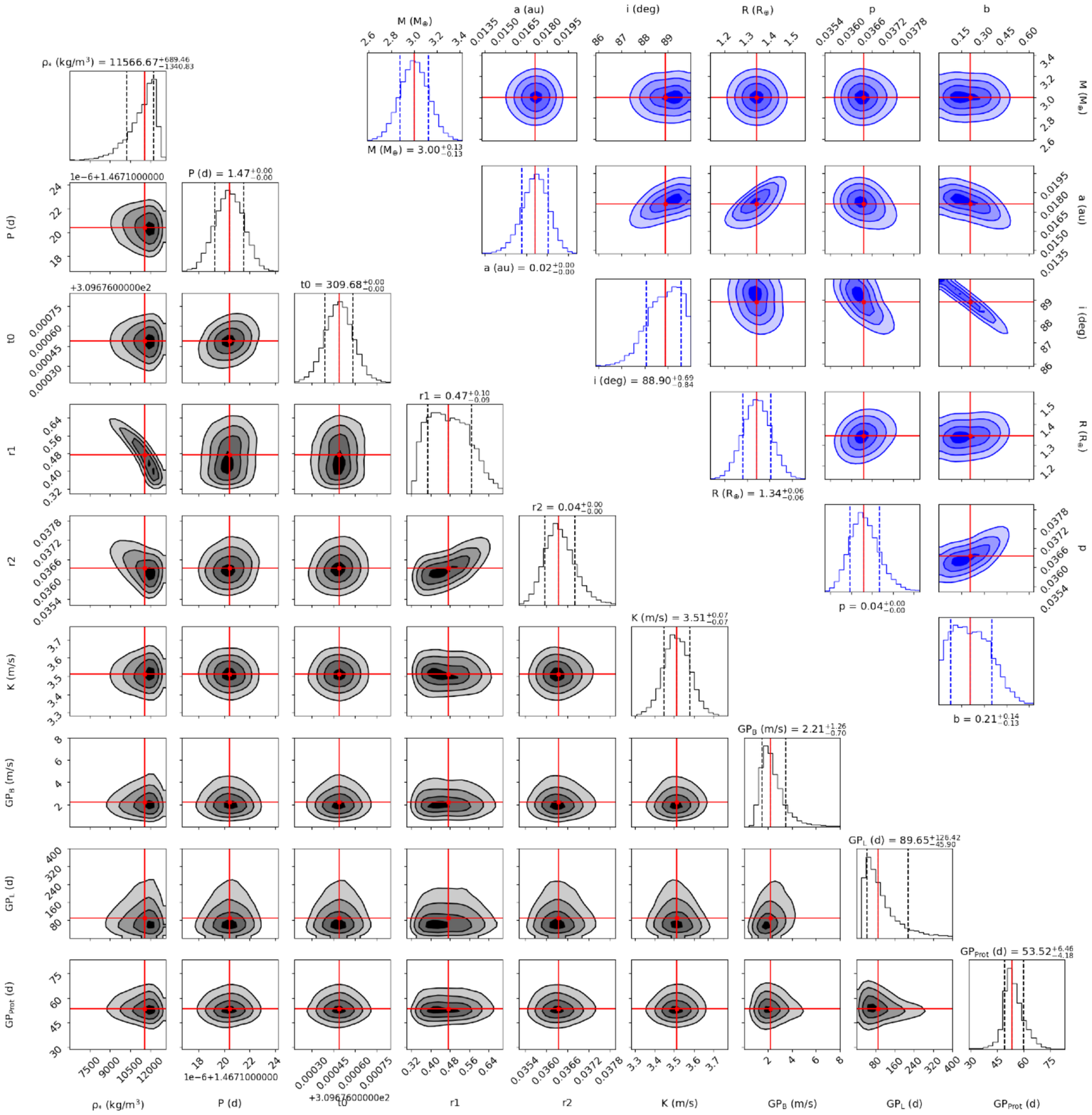}  
    \caption{Nested samples distribution of fitted (black) and derived (blue) parameters of Gl\,486~b with CARMENES and MAROON-X RV data and {\em CHEOPS} and {\em TESS} light curves.
    The position of the median from the posterior is marked with red grid lines. 
    The contours on the 2D panels represent the $1\sigma$, $2\sigma$, and $3\sigma$ confidence levels of the overall posterior samples.
    The top (black) and bottom (blue) panels of every column represent the 1D histogram distribution of each parameter.}
    \label{fig:cornerplot}
\end{figure*} 

\begin{table*}[!ht]
\centering
\caption{Priors and posteriors of the joint transit and RV fit of Gl\,486$^a$.}
\label{tab:priors+posteriors}
\begin{tabular}{l c c c r}
\hline
\hline
\noalign{\smallskip}
Parameter & Prior & Posterior & Unit & Description \\
\noalign{\smallskip}    
\hline  
\noalign{\smallskip}

\multicolumn{5}{c}{\textit{Stellar parameter}} \\
\noalign{\smallskip}

$\rho_{\star}$                          & $\mathcal{N}$(12106, 2593) & $11570 ^{+690} _{-1340}$ & kg\,m$^{-3}$ & Stellar density \\

\noalign{\smallskip}
\multicolumn{5}{c}{\textit{Photometry instrumental parameters}} \\
\noalign{\smallskip}

$\mu_{TESS, {\rm T23}}$      & $\mathcal{N}$(0, 10$^5$)     & $- 0.5 ^{+6.9} _{-7.0}$   & ppm    & Relative flux offset \\
$\mu_{CHEOPS, {\rm C01}}$    & $\mathcal{N}$(0, 10$^5$)     & $+1 ^{+15} _{-14}$   & ppm    & Relative flux offset \\
$\mu_{CHEOPS, {\rm C02}}$    & $\mathcal{N}$(0, 10$^5$)     & $+15 ^{+15} _{-16}$   & ppm    & Relative flux offset \\
$\mu_{CHEOPS, {\rm C03}}$    & $\mathcal{N}$(0, 10$^5$)     & $+1 ^{+17} _{-17}$   & ppm    & Relative flux offset \\
$\mu_{CHEOPS, {\rm C04}}$    & $\mathcal{N}$(0, 10$^5$)     & $+2 ^{+17} _{-17}$   & ppm    & Relative flux offset \\
$\mu_{CHEOPS, {\rm C05}}$    & $\mathcal{N}$(0, 10$^5$)     & $-17 ^{+16} _{-16}$   & ppm    & Relative flux offset \\
$\mu_{CHEOPS, {\rm C06}}$    & $\mathcal{N}$(0, 10$^5$)     & $-2 ^{+18} _{-19}$   & ppm    & Relative flux offset \\
$\mu_{CHEOPS, {\rm C07}}$    & $\mathcal{N}$(0, 10$^5$)     & $+6 ^{+21} _{-20}$   & ppm    & Relative flux offset \\

\noalign{\smallskip}                                                        

$\sigma_{TESS, {\rm T23}}$      & $\mathcal{LU}$(1, 8498) & $58 ^{+1118} _{-53}$     & ppm  & Transit extra jitter \\ 
$\sigma_{CHEOPS, {\rm C01}}$    & $\mathcal{LU}$(1, 3322) & $75 ^{+728} _{-69}$      & ppm  & Transit extra jitter \\ 
$\sigma_{CHEOPS, {\rm C02}}$    & $\mathcal{LU}$(1, 3320) & $77 ^{+806} _{-71}$      & ppm  & Transit extra jitter \\ 
$\sigma_{CHEOPS, {\rm C03}}$    & $\mathcal{LU}$(1, 3323) & $44 ^{+555} _{-40}$      & ppm  & Transit extra jitter \\ 
$\sigma_{CHEOPS, {\rm C04}}$    & $\mathcal{LU}$(1, 3322) & $40 ^{+495} _{-36}$      & ppm  & Transit extra jitter \\ 
$\sigma_{CHEOPS, {\rm C05}}$    & $\mathcal{LU}$(1, 3336) & $40 ^{+528} _{-36}$      & ppm  & Transit extra jitter \\ 
$\sigma_{CHEOPS, {\rm C06}}$    & $\mathcal{LU}$(1, 3388) & $38 ^{+511} _{-35}$      & ppm  & Transit extra jitter \\ 
$\sigma_{CHEOPS, {\rm C07}}$    & $\mathcal{LU}$(1, 3329) & $62 ^{+704} _{-58}$      & ppm  & Transit extra jitter \\ 

\noalign{\smallskip}                                                        

$q_{1, TESS}$                     & $\mathcal{U}$(0, 1.0)             & $0.57 ^{+0.23} _{-0.21}$                  & ...   & $u_1$ quadratic limb-darkening \\
$q_{1, CHEOPS}$                   & $\mathcal{U}$(0, 1.0)             & $0.51 ^{+0.16} _{-0.15}$                  & ...   & $u_1$ quadratic limb-darkening \\

\noalign{\smallskip}                                                        

$q_{2, TESS}$                     & $\mathcal{U}$(0, 1.0)             & $0.19 ^{+0.20} _{-0.12}$                  & ...   & $u_2$ quadratic limb-darkening \\
$q_{2, CHEOPS}$                   & $\mathcal{U}$(0, 1.0)             & $0.119 ^{+0.119} _{-0.078}$               & ...   & $u_2$ quadratic limb-darkening \\

\noalign{\smallskip}                                                        

$D_{TESS}$                      & \multicolumn{2}{c}{1.0 (fixed)}                                               & ...   & Dilution factor \\
$D_{CHEOPS}$                    & \multicolumn{2}{c}{1.0 (fixed)}                                               & ...   & Dilution factor \\

\noalign{\smallskip}
\multicolumn{5}{c}{\textit{RV instrumental parameters}} \\
\noalign{\smallskip}

$\gamma_{\rm CARMENES}$         & $\mathcal{U}$(--10.0, +10.0)     & $-0.05 ^{+0.40} _{-0.44}$      & m\,s$^{-1}$       & Relative RV offset \\
$\gamma_{\rm MAROON-X,Blue,1}$  & $\mathcal{U}$(--10.0, +10.0)     & $+1.33 ^{+0.52} _{-0.56}$      & m\,s$^{-1}$       & Relative RV offset \\
$\gamma_{\rm MAROON-X,Red,1}$   & $\mathcal{U}$(--10.0, +10.0)     & $+1.32 ^{+0.51} _{-0.53}$     & m\,s$^{-1}$       & Relative RV offset \\
$\gamma_{\rm MAROON-X,Blue,2}$  & $\mathcal{U}$(--10.0, +10.0)     & $+0.31 ^{+0.85} _{-0.80}$      & m\,s$^{-1}$       & Relative RV offset \\
$\gamma_{\rm MAROON-X,Red,2}$   & $\mathcal{U}$(--10.0, +10.0)     & $+0.28 ^{+0.84} _{-0.82}$      & m\,s$^{-1}$       & Relative RV offset \\
$\gamma_{\rm MAROON-X,Blue,3}$  & $\mathcal{U}$(--10.0, +10.0)     & $+0.1 ^{+1.2} _{-1.1}$      & m\,s$^{-1}$       & Relative RV offset \\
$\gamma_{\rm MAROON-X,Red,3}$   & $\mathcal{U}$(--10.0, +10.0)     & $+0.1 ^{+1.2} _{-1.1}$      & m\,s$^{-1}$       & Relative RV offset \\

\noalign{\smallskip}                                                        

$\sigma_{\rm CARMENES}$         & $\mathcal{LU}$(0.001, 5.0)       & $0.035 ^{+0.245} _{-0.031}$       & m\,s$^{-1}$       & RV extra jitter \\ 
$\sigma_{\rm MAROON-X,Blue,1}$  & $\mathcal{LU}$(0.001, 5.0)       & $0.66 ^{+0.16} _{-0.16}$       & m\,s$^{-1}$       & RV extra jitter \\ 
$\sigma_{\rm MAROON-X,Red,1}$   & $\mathcal{LU}$(0.001, 5.0)       & $0.136 ^{+0.097} _{-0.124}$    & m\,s$^{-1}$       & RV extra jitter \\ 
$\sigma_{\rm MAROON-X,Blue,2}$  & $\mathcal{LU}$(0.001, 5.0)       & $0.023 ^{+0.106} _{-0.019}$    & m\,s$^{-1}$       & RV extra jitter \\ 
$\sigma_{\rm MAROON-X,Red,2}$   & $\mathcal{LU}$(0.001, 5.0)       & $0.026 ^{+0.152} _{-0.022}$    & m\,s$^{-1}$       & RV extra jitter \\ 
$\sigma_{\rm MAROON-X,Blue,3}$  & $\mathcal{LU}$(0.001, 5.0)       & $0.017 ^{+0.093} _{-0.014}$       & m\,s$^{-1}$       & RV extra jitter \\ 
$\sigma_{\rm MAROON-X,Red,3}$   & $\mathcal{LU}$(0.001, 5.0)       & $0.24 ^{+0.28} _{-0.23}$       & m\,s$^{-1}$       & RV extra jitter \\ 

\noalign{\smallskip}
\multicolumn{5}{c}{\textit{RV quasi-periodic GP parameters}} \\
\noalign{\smallskip}

$B_{\rm GP}$                          & $\mathcal{U}$(0.01, 50.0) & $2.21 ^{+1.26} _{-0.70}$ & m\,s$^{-1}$       & GP kernel amplitude in Eq.~\ref{eq:GP_BLProt} \\
$C_{\rm GP}$                & \multicolumn{2}{c}{0.0 (fixed)} & m\,s$^{-1}$       & GP kernel amplitude in Eq.~\ref{eq:GP_BLProt} \\
$L_{\rm GP}$                          & $\mathcal{LU}$(20, 10$^4$) & $90 ^{+126} _{-46}$ & d & GP modulation time scale \\
$P_{\rm rot,GP}$                      & $\mathcal{N}$(50.9, 10.0) & $53.5 ^{+6.5} _{-4.2}$ & d & GP quasi-periodic rotation period \\

\noalign{\smallskip}                                                        
\multicolumn{5}{c}{\textit{Planet $b$ fitted parameters} }\\   
\noalign{\smallskip}

$P$                     & $\mathcal{N}$ (1.467, 0.010)          & $1.4671204 ^{+0.0000011} _{-0.0000011} $ & d                 & Orbital period \\
$t_0$                   & $\mathcal{U}$ (2459309.0, 2459311.0)  & $2459309.676549 ^{+0.000097} _{-0.000096} $ & d                & Time of periastron passage \\
$K$                     & $\mathcal{U}$ (0, 10.0)               & $3.512 ^{+0.069} _{-0.065}$               & m\,s$^{-1}$       & RV semi-amplitude \\   
$r_1$                   & $\mathcal{U}$ (0, 1.0)                & $0.473 ^{+0.097} _{-0.087}$              & ...               & Parametrisation for $p$ and $b$  \\   
$r_2$                   & $\mathcal{U}$ (0, 1.0)                & $0.03635 ^{+0.00046} _{-0.00039}$          & ...               & Parametrisation for $p$ and $b$    \\   
$e$                     & \multicolumn{2}{c}{0.0 (fixed)}                                                   & ...               & Orbital eccentricity \\
$\omega$                & \multicolumn{2}{c}{90.0 (fixed)}                                                  & deg               & Periastron angle\\

%

\noalign{\smallskip}    
\hline  
\end{tabular}
\tablefoot{
\tablefoottext{a}{Median and upper and lower 68.3\,\% posterior credibility intervals (1$\sigma$).
The prior labels of $\mathcal{N}(\mu,\sigma)$, $\mathcal{U}(a,b)$,  
and $\mathcal{LU}(a,b)$ represent normal (mean $\mu$ and variance $\sigma^2$), uniform, 
and log-uniform distributions (minimum $a$ and maximum $b$), respectively.
The unit symbol ``ppm'' stands for part per million.}
}
\end{table*}


\begin{table}[]
\centering
\caption{CARMENES and MAROON-X Red and Blue RVs of Gl\,486$^a$.} 
\label{tab:CARM_data} 
\begin{tabular}{c c l}
    \hline 
    \hline
	\noalign{\smallskip}
Epoch [BJD] & RV [m\,s$^{-1}$] & Instrument \\
	\noalign{\smallskip}
	\hline
	\noalign{\smallskip}

2457400.7408 & 4.54 $\pm$ 1.07 & CARMENES \\
2457401.7424 & 0.30 $\pm$ 1.31 & CARMENES \\
2457418.7185 & --2.23 $\pm$ 1.14 & CARMENES \\
2457421.7051 & --2.64 $\pm$ 0.98 & CARMENES \\
2457426.6930 & 0.95 $\pm$ 1.13 & CARMENES \\
 & ... & \\
2459364.9516 & 3.05 $\pm$ 0.94 & MAROON-X Blue \\
2459365.9122 & --3.64 $\pm$ 0.69 & MAROON-X Blue \\
2459365.9122 & --3.26 $\pm$ 0.32 & MAROON-X Red \\
2459367.8489 & 1.75 $\pm$ 0.47 & MAROON-X Blue \\
2459367.8489 & 1.88 $\pm$ 0.26 & MAROON-X Red \\

	\noalign{\smallskip}
	\hline
\end{tabular}
\tablefoot{
\tablefoottext{a}{\em Ask the first author for the full table.}
}
\end{table}


\newpage

\section{Sources of error and propagation of uncertainty}
\label{sec:errors}

\subsection{Planet radius and mass}

\begin{figure*}
    \centering
    \includegraphics[height=0.39\textwidth]{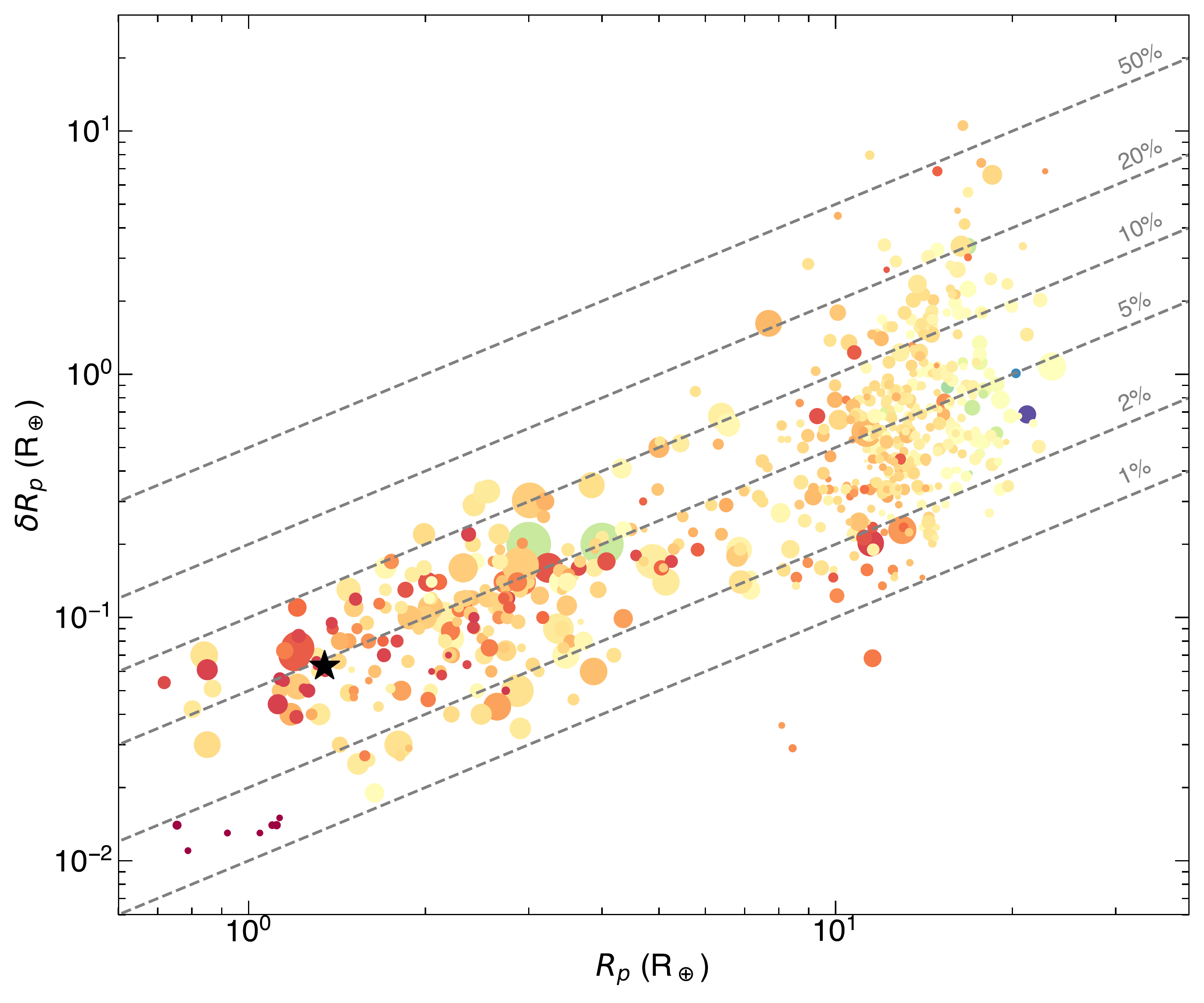}
    \includegraphics[height=0.39\textwidth]{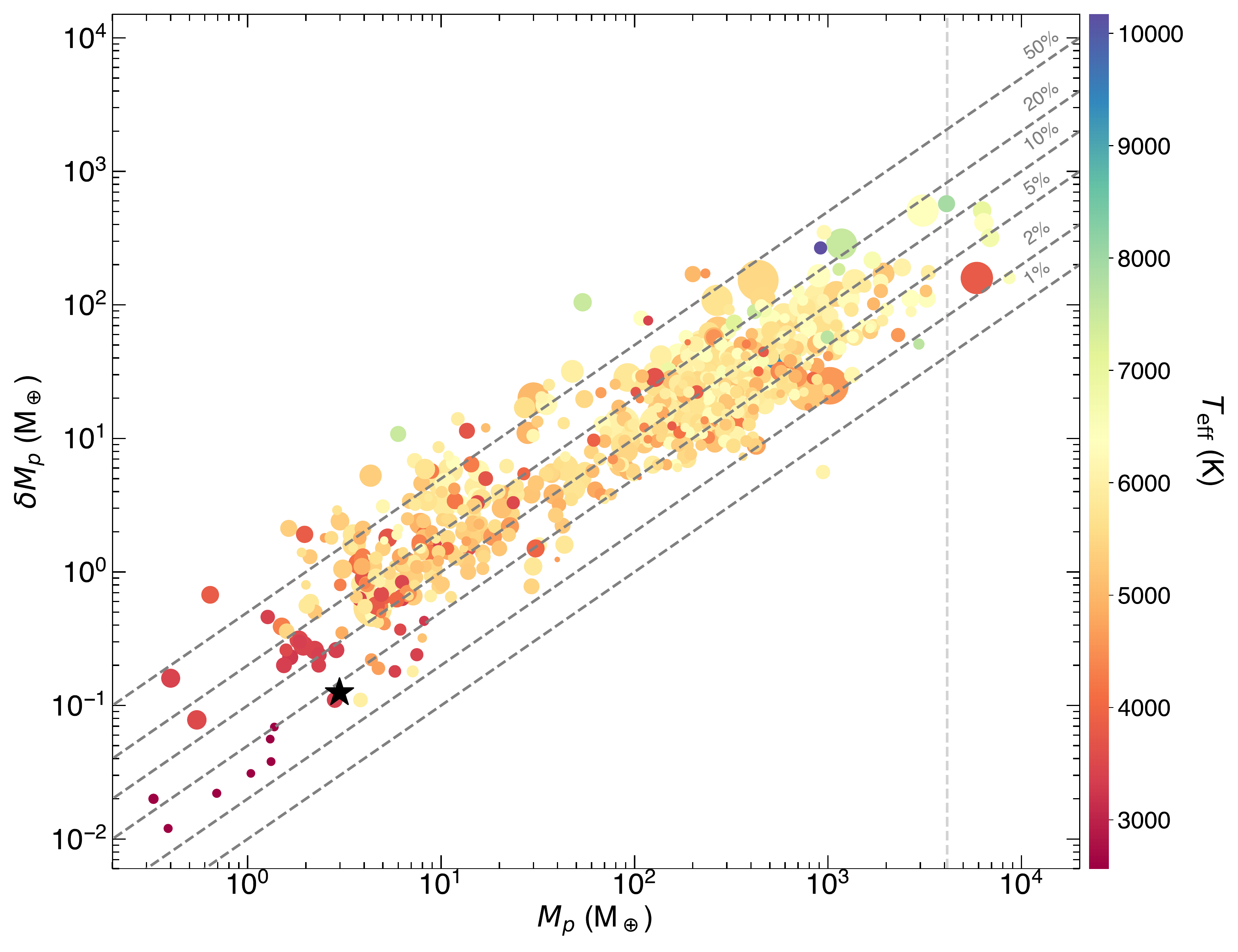}
         \caption{Uncertainty in radius versus radius ({\em left}) and uncertainty in mass versus mass ({\em right}) of all transiting exoplanets with mass determination (from RV or transit time variations).
         The symbol colour denotes the stellar host effective temperature, while the symbol size is proportional to mass and radius, respectively.
         Diagonal dashed lines indicate lines of constant relative uncertainty (1, 2, 5, 10, 20, 50\,\%).
         Gl~486\,b is marked with a black star in both panels.
         Only in the right panel, the grey vertical dotted line marks the deuterium burning mass limit.
The planets with $\delta R_{\rm p} / R_{\rm p}$ ratios less than 1\,\% (left panel) are
\object{Kepler-16}\,b ($R_{\rm p} = 8.448 ^{+0.028} _{-0.026}\,R_\oplus$), 
a Saturn-size circumbinary planet around a eclipsing binary with both precise and accurate stellar mass and radius determination \citep{Doyle2011},
\object{HATS-72}\,b ($R_{\rm p} = 8.097 \pm 0.036\,R_\oplus$), whose stellar host is a K dwarf with a questionable tabulated relative radius uncertainty of merely 0.29\,\% \citep{Hartman2020},
and \object{Wendelstein-1}\,b ($R_{\rm p} = 11.561 \pm 0.068\,R_\oplus$), for which the uncertainty in the stellar radius was not correctly propagated
\citep{Obermeier2020}.
The planet with $\delta M_{\rm p} / M_{\rm p}$ ratio less than 1\,\% (right panel) is \object{TIC~172900988}\,b ($M_{\rm p} = 942.0 \pm 5.6\,M_\oplus$ for one of the six possible model solutions), a Jupiter-size circumbinary planet around another eclipsing binary \citep{Kostov2021}.
         }
        \label{fig:dRMvsRM}
\end{figure*}

Gl~486\,b is among the smallest and least massive transiting planets, and among the ones with the smallest relative uncertainties in both radius and, especially, mass.
However, there are even smaller and less massive planets (e.g. in the TRAPPIST-1 system), which have smaller reported relative uncertainties.
Planets with very small reported uncertainties orbit eclipsing binary stars (with precise and accurate $R_\star$ and $M_\star$ determinations) or suffer from incorrect error propagation.
This is illustrated in the $\delta R_{\rm p}$-$R_{\rm p}$ and $\delta M_{\rm p}$-$M_{\rm p}$ diagrams in Fig.~\ref{fig:dRMvsRM}.

With $(R_{\rm core} / R_{\rm p})_{\rm MRA-SH} = 0.404 ^{+0.040} _{-0.045}$ in Gl~486\,b, we reached the 10\,\% precision boundary on the core-to-mantle ratio of a telluric exoplanet.
Breaking this precision boundary will give rise to planet internal composition studies not done before.
However, in order to obtain a precision better than 10\,\% on the core-to-mantle ratio of a terrestrial exoplanet (actually, the ``core radius fraction''), it is required a measurement of the mass better than 11\,\% and a measurement on the radius better than 3\,\% \citep{Suissa2018}.
The approximate uncertainty ratio of three-to-one is also applied to the planet bulk density.
From the sphere density formula, the uncertainty in planet radius $R_{\rm p}$ contributes three times more to the final error of planet bulk density $\rho_{\rm p}$ than the uncertainty in planet mass $M_{\rm p}$: 

\begin{equation}
    \rho_{\rm p} = \frac {3 M_{\rm p}} {4 \pi R_{\rm p}^3},
\end{equation}

\noindent and:

\begin{equation}
    \left( \frac {\delta \rho_{\rm p}} {\rho_{\rm p}} \right)^2 = 
    \left( \frac {\delta M_{\rm p}} {M_{\rm p}} \right)^2 + 
    \left( \frac {3 \delta R_{\rm p}} {R_{\rm p}} \right)^2. 
\end{equation}

For Gl~486b, the measured relative uncertainties in planet radius and mass translate into a bulk density relative uncertainty of 9.3\,\%, which allowed to determine its internal composition and structure.
In our case, we fitted the $r_1$ and $r_2$ parameters, from which we derived the radius ratio, $p$, and impact parameter, $b$ \citep{Espinoza2019}.
The planet radius comes from the simple definition of $p$:

\begin{equation}
R_{\rm p} = p ~ R_\star, 
\end{equation}

\noindent and its uncertainty is, therefore:

\begin{equation}
    \left( \frac {\delta R_{\rm p}} {R_{\rm p}} \right)^2 = 
    \left( \frac {\delta p} {p} \right)^2 +
    \left( \frac {\delta R_\star} {R_\star} \right)^2.
\end{equation}

\noindent As a result, the relative uncertainty in $R_{\rm p}$ can never be smaller than that in $R_\star$.
For Gl~486\,b, $p = 0.03644^{+0.00048}_{-0.00042}$, the relative error in $p$ is very small, and, therefore, $\delta R_{\rm p} / R_{\rm p} \approx \delta R_\star / R_\star$.
Actually, from Sect.~\ref{sec:planetradiusandmass}, the derived planet radius is $R_{\rm b} = 1.343^{+0.063}_{-0.062}\,R_\oplus$, which represents a relative uncertainty of 4.7\,\% and is very similar to the relative uncertainty of the stellar radius, at 4.5\,\%.
The small difference of 0.2\,\% through the $p$ parameter is mostly due to the quality of the {\em TESS} and {\em CHEOPS} transit data.

The derivation of the relative uncertainty in $M_{\rm p}$ is, a priori, more complicated and highly non-linear.
From the definition of the RV semi-amplitude:

\begin{equation}
K = \frac{1}{(1-e^2)^{1/2}} 
    \frac{M_{\rm p} \sin{i}}{(M_{\rm p} + M_\star)^{2/3}}
    \left( \frac{2 \pi G}{P} \right)^{1/3},
\end{equation}

\noindent it can be deduced the following:

\begin{equation}
\frac{M_{\rm p}}{(M_{\rm p} + M_\star)^{2/3}} = \mathcal{H}(P,K,i,e),
\end{equation}

\begin{equation}
\mathcal{H}(P,K,i,e) = (1-e^2)^{1/2}
    \frac{K}{\sin{i}}
    \left( \frac{P}{2 \pi G} \right)^{1/3},
\end{equation}

\noindent where $P$, $K$, and $e$ are fitted parameters, $i$ is a derived parameter, and $M_\star$ is an input parameter.
Nevertheless, in most cases it happens that $M_\star \gg M_{\rm p}$.
For example, for Gl~486\,b, the ratio between planet and star masses is about $2.7 \cdot 10^{-5}$, which justifies the following approximation:

\begin{equation}
M_{\rm p} \approx M_\star^{2/3} ~ \mathcal{H}(P,K,i,e).
\end{equation}

\noindent As a result, the uncertainty in planet mass becomes:

\begin{equation}
  \begin{aligned}    
(\delta M_{\rm p})^2 & \approx 
    \left( \frac{2 \mathcal{H}}{3 M_\star^{1/3}} \right)^2 \delta^2 M_\star \\
    & + \left( \frac{\partial \mathcal{H}}{\partial P} \right)^2 \delta^2 P + 
    \left( \frac{\partial \mathcal{H}}{\partial K} \right)^2 \delta^2 K + 
    \left( \frac{\partial \mathcal{H}}{\partial i} \right)^2 \delta^2 i + 
    \left( \frac{\partial \mathcal{H}}{\partial e} \right)^2 \delta^2 e. 
  \end{aligned}    
\end{equation}

\noindent When the uncertainty in $M_\star$ dominates the global error contribution with respect to $P$, $K$, $i$, and $e$, as in the case of typical observations with CARMENES+MAROON-X or ESPRESSO, the latter equation remains just:

\begin{equation}
    \frac {\delta M_{\rm p}} {M_{\rm p}} \approx 
    \frac {2 \delta M_\star} {3 M_\star}.
\end{equation}

From Sect.~\ref{sec:planetradiusandmass}, the derived planet mass is $M_{\rm b} = 3.00^{+0.13}_{-0.13}\,M_\oplus$, which represents a relative uncertainty of 4.2\,\%, and is about two thirds of the relative uncertainty of the stellar mass, at 5.6\,\%.
The small difference of 0.4\,\% through the $P$, $K$, and $i$ (and $e$) parameters is, in this case, mostly due to the quality of the CARMENES and MAROON-X RV data, some of them with sub-metre-per-second precision.

\begin{figure*}[]
 	\centering
 	\includegraphics[height=0.39\textwidth]{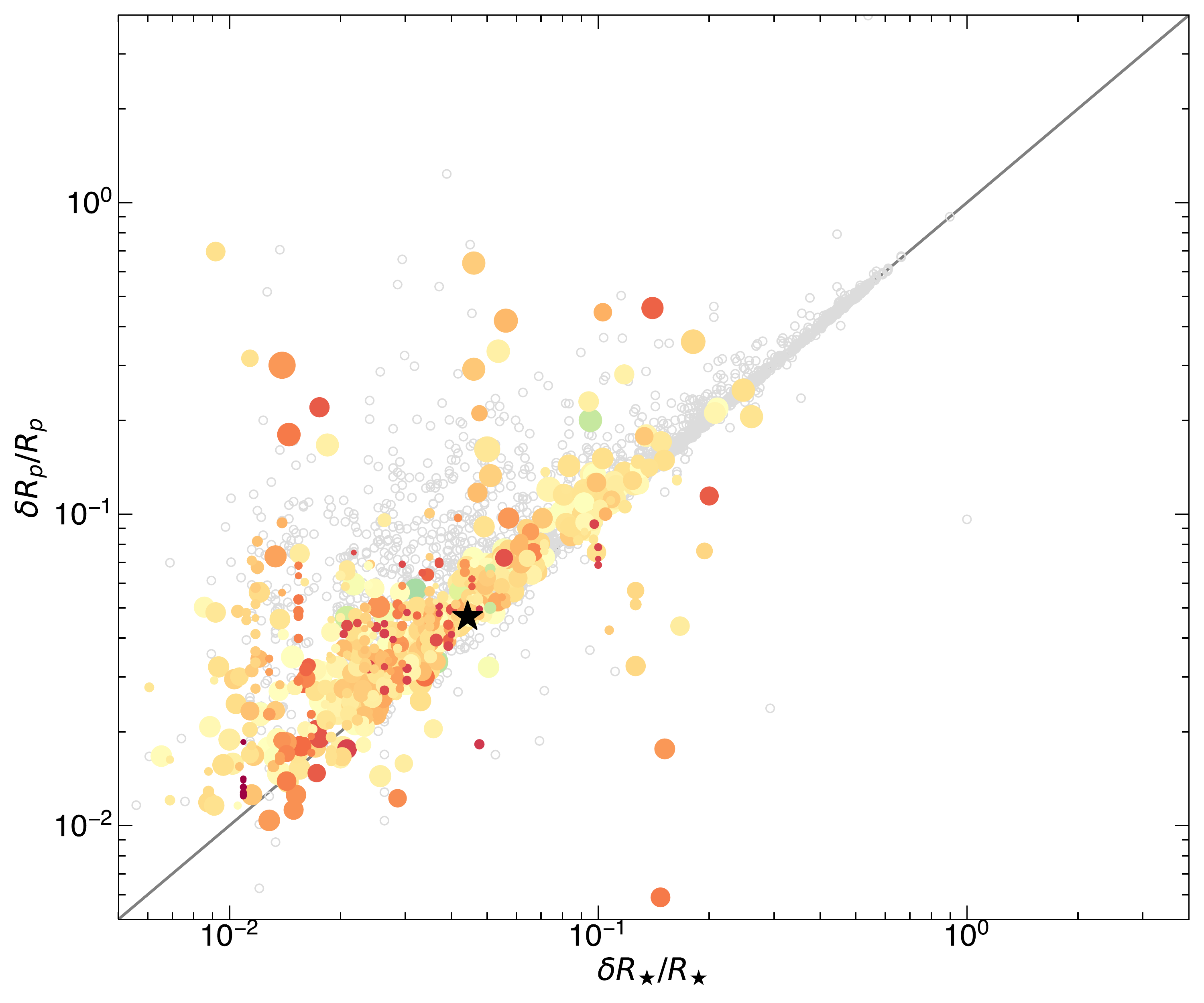}
 	\includegraphics[height=0.39\textwidth]{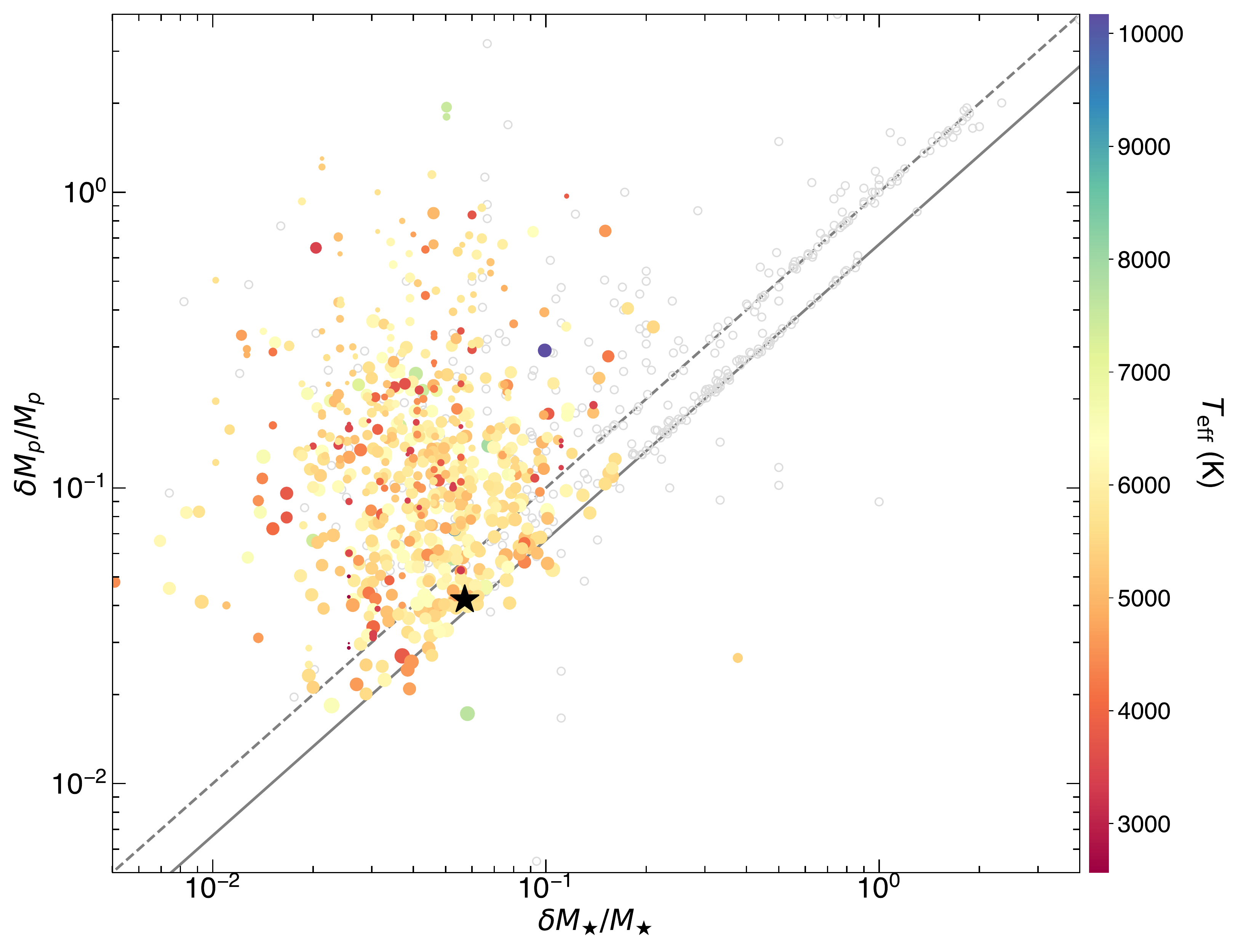} 	
 	\caption{
 	Relative errors in star (X axis) and planet (Y axis) radius ({\em left}) and mass ({\em right}).
 	Grey open circles: all exoplanet candidates in {\tt exoplanets.org}.
 	Coloured filled symbols: planets with both mass and radius determination (from RV and transits or transit time variations); the symbol colour denotes the stellar host effective temperature, while the symbol size is log-proportional to the planet radius ({\em left}) and mass ({\em right}).
    Black star: Gl~486 system.
 	Solid diagonal lines: 1:1 radius ({\em left:}) and 2:3 mass ({\em right:}) relative error ratios.
 	Only in the {\em right panel}, we also plot a dashed diagonal line at the 1:1 mass relative error ratio, which corresponds to microlensing planets.
    }
    \label{fig:relativeerrors} 
 \end{figure*}

Fig.~\ref{fig:relativeerrors} illustrates these computations.
The majority of confirmed planets fall on or above the 1:1 radius relative error ratio in the $\delta R_{\rm p} / R_{\rm p}$ vs. $\delta R_\star / R_\star$ diagram (left panel) and on or above the 2:3 mass relative error ratio in the $\delta M_{\rm p} / M_{\rm p}$ vs. $\delta M_\star / M_\star$ diagram.
Gl~486\,b, displayed in both panels of Fig.~\ref{fig:relativeerrors} with a black filled star, has radius and mass determinations near, but slightly above, the empirical boundaries at $\delta R_{\rm p} / R_{\rm p} \sim \delta R_\star / R_\star$ and  $\delta M_{\rm p} / M_{\rm p} \sim 2 \delta M_\star / 3 M_\star$.
The outliers with $\delta R_{\rm p} / R_{\rm p} \ll \delta R_\star / R_\star$ 
    (Wendelstein~1\,b and~2\,b, \citealt{Obermeier2020}; 
 	\object{CoRoT-27}\,b, \citealt{Parviainen2014};
 	and \object{Kepler-30}\,b, c and~d, \citealt{SanchisOjeda2012})
and $\delta M_{\rm p} / M_{\rm p} \ll \delta M_\star / M_\star$ 
    (\object{BD+46~2629A}\,b = Kepler-13\,b, \citealt{Esteves2015}; and 
    \object{LTT~9779}\,b, \citealt{Jenkins2020}) 
suffered again from incorrect error propagation.
For example, \citet{SanchisOjeda2012} tabulated $3.9 \pm 0.2$, $12.3 \pm 0.4$, and $8.8 \pm 0.5\,R_\oplus$ for planets b, c, and d in the Kepler-30 system but, from their $R_\star$ and $\Delta = (R_{\rm p} / R_\star)^2$ and the SI values for $R_\odot$ and $R_\oplus$, we determined instead $4.21 \pm 0.54$, $13.2 \pm 1.7$, and $9.4 \pm 1.2\,R_\oplus$, respectively\footnote{We used $R_\odot = 6.957 \cdot 10^8$\,m and $R_\oplus = 6.3781 \cdot 10^6$\,m.}.

\subsection{Stellar radius and mass}

While we break by a wide margin the planet mass boundary of \citet{Suissa2018} for internal composition studies at 11\,\%, we get very close to break the planet radius boundary at 3\,\%.
The space photometry and RV spectroscopy add only 0.2\,\% and 0.4\,\% extra uncertainty to the fit, respectively, so we are in the case of planet parameters limited by the stellar parameter uncertainties, especially the stellar radius.
In our case, this problem is partially alleviated by our accurate interferometric measurements.
We analyse below the sources of limiting errors in determining stellar parameters, apart from improvements in RV and transit photometry precision and systematics correction.

We computed the stellar mass $M_\star$ from the stellar radius $R_\star$ with the linear mass-radius relation of \citet{Schweitzer2019}:

\begin{equation}
    M_\star = \alpha + \beta ~ R_\star,
\end{equation}


\noindent where $\alpha = -0.0240 \pm 0.0076$\,$M_\odot$, $\beta = 1.055 \pm 0.017$\,$M_\odot \,R_\odot^{-1}$, and $R_\star$ is expressed in solar units (Sect.~\ref{sec:star+planet} -- we use $\alpha,~\beta$ instead of $a,~b$ as in the original work for avoiding confusion with $p,~b$ {\tt juliet} parameters).
From this relation, 
the uncertainty in stellar mass is, therefore:

\begin{equation}
    \left( \delta M_\star \right)^2 =
    \left( \delta \alpha \right)^2 +
    \left( \delta \beta ~ R_\star \right)^2 +
    \left( \beta ~ \delta R_\star \right)^2.
\end{equation}

In our case, we determined the stellar radius with near-infrared interferometric measurements with MIRC-X at the CHARA Array (Sects.~\ref{sec:chara} and~\ref{sec:Rstar}). 
In particular, $R_{\star,{\rm interf}}$ is a simple function of the stellar angular diameter, $\theta$, and the parallax, $\varpi$, or, alternative, the parallactic distance, $d$:

\begin{equation}
    R_{\star,{\rm interf}} = \frac {\theta} {2 \varpi} = \frac {\theta} {2} d,
\end{equation}

\noindent and the corresponding relative uncertainty is:

\begin{equation}
    \left( \frac {\delta R_{\star,{\rm interf}}} {R_{\star,{\rm interf}}} \right)^2 = 
    \left( \frac {\delta \theta} {\theta} \right)^2 + 
    \left( \frac {\delta \varpi} {\varpi} \right)^2 =
    \left( \frac {\delta \theta} {\theta} \right)^2 + 
    \left( \frac {\delta d} {d} \right)^2.
\end{equation}

To sum up, the uncertainty in stellar mass from an interferometric stellar radius becomes:

\begin{equation}
  \begin{aligned}    
    \left( \delta M_{\star,{\rm interf}} \right)^2 & =
    \left( \delta \alpha \right)^2 +
    \beta^2 R_{\star,{\rm interf}}^2 \\
    & \times \left( 
        \left( \frac {\delta \beta} {\beta} \right)^2 + 
        \left( \frac {\delta \theta} {\theta} \right)^2 + 
        \left( \frac {\delta d} {d} \right)^2
    \right),
  \end{aligned}    
\end{equation}

In most cases, especially for stars within 10\,pc of the Sun such as Gl~486, the uncertainty in distance is much smaller than those in angular diameter and slope of the mass-radius linear relation.
In Sect.~\ref{sec:chara} we determined a stellar limb-darkened disc diameter $\theta_{\rm LDD} = 0.390 \pm 0.018$\,mas, which error propagated from the scatter of the squared visibility, $V^2$, as a function of spatial frequency, $B'/\lambda$, and the uncertainties in all fit parameters (i.e., $T_{\rm eff}^{\tt LDTK}$, $V_0^2$, $\mu_H$).
The uncertainty in $\theta_{\rm LDD}$, of $\sim$4.5\,\%, is about three times larger that of $\beta$, of $\sim$1.6\,\%, and almost 20 times larger than that of $d$, of barely $\sim$0.3\,\%.
As a result, $\delta R_{\star,{\rm interf}} / R_{\star,{\rm interf}} \approx \delta \theta_{\rm LDD} / \theta_{\rm LDD}$, which agrees with the stellar radius relative uncertainty of $\sim$4.5\,\% as in Table~\ref{tab:star}.
After including the $\alpha$ and $\beta$ contributing errors, the stellar mass relative uncertainty becomes the nominal 5.6\,\%.

When there is no interferometric determination of the stellar radius, \citet{Schweitzer2019} proposed deriving $R_\star$ through the Stefan-Boltzmann law from bolometric luminosity and an equilibrium temperature derived from spectral synthesis: 

\begin{equation}
    R_{\star,{\rm synth}}^2 = \frac {L_\star} {4 \pi \sigma T_{\rm eff}^4},
\end{equation}

\noindent and its uncertainty is, thus:

\begin{equation}
    \left( \frac {\delta R_{\star,{\rm synth}}} {R_{\star,{\rm synth}}} \right)^2 = 
    \left( \frac {\delta L_\star} {2 L_\star} \right)^2 + 
    \left( \frac {2 \delta T_{\rm eff}} {T_{\rm eff}} \right)^2. 
\end{equation}

\noindent As a result, the uncertainty in $T_{\rm eff}$ to $R_{\star,{\rm synth}}$ contributes four times that in $L_\star$.
In turn, the uncertainty in luminosity, which is computed from the distance and observed flux after integrating the stellar spectral energy distribution from the blue optical to the mid infrared ($L_\star = 4 \pi d^2 F_{\rm obs}$), is relatively small for nearby stars with precise {\em Gaia} EDR3 parallactic distance and a wealth of well-calibrated multiband photometry:

\begin{equation}
    \left( \frac {\delta L_\star} {L_\star} \right)^2 = 
    \left( \frac {\delta F_{\rm obs}} {F_{\rm obs}} \right)^2 +
    \left( \frac {2 \delta d} {d} \right)^2.
\end{equation}

\noindent As a result, the error in the determination of $R_{\star,{\rm synth}}$ is dominated by that of $T_{\rm eff}$, which can be 50--200\,K in M dwarfs \citep{Passegger2022}.

The following equation summarises all the contributions to the uncertainty in stellar mass in absence of interferometric observations:

\begin{equation}
  \begin{aligned}    
  \left( \delta M_{\star,{\rm synth}} \right)^2 & =
    \left( \delta \alpha \right)^2 +
    \beta^2 R_{\star,{\rm synth}}^2 \\
    & \times \left( 
        \left( \frac {\delta \beta} {\beta} \right)^2 + 
        \left( \frac {2 \delta T_{\rm eff}} {T_{\rm eff}} \right)^2 + 
        \left( \frac {\delta F_{\rm obs}} {2 F_{\rm obs}} \right)^2 + 
        \left( \frac {\delta d} {d} \right)^2
    \right).
  \end{aligned}    
\end{equation}

There are different ways of reducing the uncertainties in stellar radius and mass:

\begin{itemize}
    \item Acquiring more and better interferometric data.
    There are, however, technical and logistics limitations to this, as observing at wide baselines and dense ranges of spatial frequencies with up to six CHARA Array telescopes or any other interferometer is time consuming.
    Besides, although the scenario is not as serious as for the AstroLAB site, Gl~486 culminates at an altitude of only $\sim$65.5\,deg as seen from Mount Wilson.
    
    \item If there are no interferometric data, improving the $T_{\rm eff}$ determination. 
    We refer to \citet{Marfil2021} and \citet{Passegger2022} for recent and exhaustive comparisons of methodologies for determining $T_{\rm eff}$ of M dwarfs.
    
    \item Improving the mass-radius relation.
    {\em TESS} is discovering new detached, M-dwarf, eclipsing binaries \citep[e.g.,][]{Lendl2020, Prsa2012}, some of them with relatively large orbital periods that lack enhanced magnetic activity and, thus, stellar inflation as in the ones with the shortest periods \citep{Kraus2011}.
    Special attention must also be given to not including in the fit young eclipsing binaries in stellar kinematic groups that are still on the Hayashi track \citep{Schweitzer2019}.
    
    \item Measuring a more precise bolometric observed flux.
    Differences between $F_{\rm obs}$ computed by us with {\tt VOSA} or by other teams are in the details, such as origin of the photometry, template choice, handling of zero-points and transmission profiles \citep{Cifuentes2020}.
    It is however difficult to get better than 1.1\,\% as measured by us because, even if the photometry has tiny errors and the spectral templates are perfect, almost all photometry is calibrated to the same set of standard stars. 
    Those standards only have their true $F_{\rm obs}$ measured to about 1--2\,\% based on STIS/{\em Hubble Space Telescope} spectro-photometric calibrations \citep{Bohlin2004,Bohlin2007,MaizApellanizWeiler2018}.
    {\em Webb} and its extended wavelength range towards the near- and mid-infrared can soon be used for improving the bolometric flux of standard stars.
    
    \item Improving the parallactic distance determination.
    There is strong evidence that {\em Gaia} EDR3 parallax errors are underestimated, especially for bright stars \cite[e.g.,][]{Luri2018, ElBadry2021, Fabricius2021, MaizApellaniz2022}. 
    At $G \sim$ 10\,mag the underestimation can be up to 60\,\%.
    Since {\em Gaia} DR3 parallactic distances will be those already published in EDR3, we will have to wait for DR4\footnote{\url{https://www.cosmos.esa.int/web/gaia/release}} for having uncertainties so small that they will in general be negligible with respect to the errors in other parameters.
\end{itemize}

\subsection{Element abundances}

Slightly different values of $A$(X) and, therefore, [X/H] can be obtained if other input $T_{\rm eff}$ are used.
Different $T_{\rm eff}$ at a fixed $L_{\rm bol}$ translates into distinct $R_\star$ and, therefore, $M_\star$, $R_{\rm p}$, and $M_{\rm p}$.
However, as discussed in Sect.~\ref{sec:star+planet}, our $T_{\rm eff}$ from our interferometric radius and bolometric luminosity matches most literature values (Table~\ref{tab:teff}).
The abundances and, therefore, the planet interior models, are also sensitive in a lower degree to the used $\log{g}$ and [Fe/H] values.
While the iron abundances of \citet{Marfil2021} seem to be the most reliable ones published to date in M dwarfs \citep{Passegger2022}, their surface gravities are a matter of concern.
For example, the Gl~486 surface gravity from our interferometric radius and the mass-radius relation of \citet{Schweitzer2019} is $\log{g} \approx$~5.4\,dex, 
which contrasts $\log{g_{\rm spec}} = 4.82 \pm 0.12$\,dex from \citet{Marfil2021}.
Following \citet{Passegger2022}, ``further in-depth investigations of the employed methods [to determine  effective temperatures, surface gravities, and metallicities in M dwarfs] would be necessary in order to identify and correct for the discrepancies that remain''.
In any case, the Fe, Mg, and Si abundances derived by us, which are solar within generous uncertainties, can be applicable to current and future planet interior structure and composition models.

\end{document}